\documentclass[article]{aa}
\usepackage{graphicx} 
\usepackage[colorlinks=true,     linkcolor=blue, citecolor=blue, filecolor=blue, urlcolor=blue]{hyperref}
\graphicspath{ {plots/} }
\usepackage[]{natbib}
\usepackage{amsmath}
\usepackage{amssymb}
\usepackage{soul}
\usepackage{sidecap}
\usepackage[dvipsnames]{xcolor}
\usepackage{tikz}
\usepackage{multirow}

\bibpunct{(}{)}{;}{a}{}{,} 

\titlerunning{Exploration of the Galactic Bulge}
\authorrunning{Nepal et al.}

\def\kms{km~s$^{-1}$}
\def\teff{$\textit{T}_{\text{eff}}$}

\def\logg{$\text{log}(\textit{g})$}
\def\mh{$[\text{M}/\text{H}]$}
\def\feh{[\text{Fe}/\text{H}]}
\def\alpham{$[\alpha/\text{M}]$}
\def\alphafe{$[\alpha/\text{Fe}]$}

\defcitealias{rvs_cnn_2023}{G24}
\defcitealias{Queiroz2021}{Q21}

\title{The Spheroidal Bulge of the Milky Way:\\ Chemodynamically Distinct from the Inner-Thick Disc and Bar}

\author{S.~Nepal \inst{1, 2},
C.~Chiappini \inst{1},
A. P\'erez-Villegas \inst{3}, 
A. B. Queiroz\inst{4,5}, 
S. Souza\inst{6}, 
M. Steinmetz\inst{1, 2}, 
F. Anders\inst{7,8,9},
A. Khalatyan\inst{1},
B. Barbuy\inst{10}, 
G. Guiglion\inst{6,11}
}
\institute{Leibniz-Institut f\"ur Astrophysik Potsdam (AIP), An der Sternwarte 
16, 14482 Potsdam, Germany
\and
Institut f\"ur Physik und Astronomie, Universit\"at Potsdam, Karl-Liebknecht-Str. 24/25, 14476 Potsdam, Germany
\and
{Instituto de Astronom\'ia, Universidad Nacional Aut\'onoma de M\'exico, A. P. 106, C.P. 22800 Ensenada, B. C., M\'exico}
\and
{Instituto de Astrof\'{\i}sica de Canarias, E-38205 La Laguna, Tenerife, Spain}
\and
{Universidad de La Laguna, Departamento de Astrof\'{\i}isica, 38206 La Laguna, Tenerife, Spain}
\and
{Max Planck Institute for Astronomy, K\"onigstuhl 17, 69117 Heidelberg, Germany}
\and
{Departament de Física Qu\`antica i Astrof\'isica (FQA), Universitat de Barcelona, Mart\'i i Franqu\`es 1, 08028 Barcelona, Spain}
\and
{Institut de Ciències del Cosmos (ICCUB), Universitat de Barcelona, Martí i Franqués, 1, 08028 Barcelona, Spain}
\and
{Institut d'Estudis Espacials de Catalunya (IEEC), Esteve Terradas, 1, Edifici RDIT, 08860 Castelldefels, Spain}
\and
{Universidade de S\~{a}o Paulo, IAG, Departamento de Astronomia, 05508-090 S\~{a}o Paulo, Brazil}
\and
{Zentrum f\"ur Astronomie der Universit\"at Heidelberg, Landessternwarte, K\"onigstuhl 12, 69117 Heidelberg, Germany}
}

\date{Received 9 July 2025 / Accepted 30 November 2025}

\begin{document}

\abstract{Studying the composition and origin of the inner region of our Galaxy—the “Galactic bulge”—is crucial for understanding the formation and evolution of the Milky Way and other galaxies. We present new observational constraints based on a sample of around 18,000 stars in the inner Galaxy, combining {\it Gaia} DR3 RVS and APOGEE DR17 spectroscopy. {\it Gaia}-RVS complements APOGEE by improving the sampling of the metallicity, \feh\,, in the $-$2.0 to $-$0.5 dex range. This work marks the first application of {\it Gaia}-RVS spectroscopy to the bulge region, enabled by a novel machine learning approach ({\tt hybrid-CNN}) that derives stellar parameters from intermediate-resolution spectra with precision comparable to APOGEE’s infrared data. We performed full orbit integrations using a barred Galactic potential and applied orbital frequency analysis to disentangle the stellar populations in the inner Milky Way. For the first time, we are able to robustly identify the long-sought pressure-supported bulge traced by the field stars. We show this stellar population to be chemically and kinematically distinct from the other main components co-existing in the same region. The spheroidal bulge has a metallicity distribution function (MDF) peak at around $-$0.70 dex extending to solar values. It is dominated by a high-\alphafe\ population with almost no dependency on metallicity, consistent with very rapid early formation predating the thick disc and the bar. We find evidence that the bar has influenced the dynamics of the spheroidal bulge, introducing a mild triaxiality and radial extension. We identify a group of stars on $X_4$ orbits, likely native to the early spheroid, as this population mimics the chemistry of the spheroidal bulge, with a minor contamination from the more metal-poor (\feh<$-$1.0) halo. We find the inner thick disc to be kinematically hotter ($\overline{V_{\phi}}\approx$125 \kms) than the local thick disc. The disc, chemically distinct from the spheroidal bulge and bar, is predominantly metal-poor with an MDF peak at $\feh\approx-0.45$ dex and includes a high fraction of stars with sub-solar \feh\ and intermediate \alphafe. In contrast to the spheroidal bulge, the \alphafe\ disc shows a steeper decline with [Fe/H], consistent with smaller star formation efficiency than that of the spheroidal bulge. Both the thick disc and the spheroidal bulge present vertical metallicity gradients. We find that the Galactic bar contains both metal-rich and metal-poor stars, as well as high and low \alphafe\ in nearly equal measure. However, their relative contributions vary significantly across different orbital families. The bar shows no strong metallicity trends in orbital extent or velocity dispersions and maintains a consistent elongated shape across all metallicities, indicating that it is a well-mixed dynamical structure. Despite their spatial overlap, the spheroidal bulge, thick disc, and bar occupy distinct regions in both phase space and chemical abundance, indicating separate formation pathways. The stars with [Fe/H]$<-$1.0 and crossing the Galactic bulge are comprised by accreted populations primarily (70\%) belonging to the Gaia-Enceladus/Sausage (GES) merger with an MDF peak at $-$1.30 dex and possibly a secondary merger remnant with an MDF peak at $-$1.80 dex.}


\maketitle

\clearpage
\newpage
\section{Introduction}

The Galactic bulge—dense and ancient—holds the fossil record of the formative epochs of the Milky Way (MW): its collapse, dynamical awakening, and emergence as a barred spiral galaxy. Only in the MW bulge is it possible to obtain a detailed, star-by-star view of stellar populations, providing critical constraints on models of bulge formation and galaxy evolution. The complexity of this region arises from the interplay between gas flows, bar dynamics, as well as rapid and efficient star formation. As such, it remains a major challenge to model the bulge reliably in cosmological simulations \citep[e.g.][]{vanDonkelaar2025}. In spite of these challenges, our understanding of the MW bulge has evolved substantially in recent years \citep[see][for reviews]{Barbuy2018, Zoccali2024} thanks to large photometric and spectroscopic surveys (most notably VISTA Variables in the Via Lactea (VVV) -- \citealt{Minniti2010} and Apache Point Observatory Galactic Evolution Experiment (APOGEE) -- \citealt{Majewski2017}). 

Although large, age-dated stellar samples in the bulge are still lacking, it is well established that this component hosts some of the oldest known globular clusters (GCs) (\citealt{Bica2024, Souza2024a, Ortolani2025}, and references therein). The ages of these systems, approaching 13 Gyr, suggest that the oldest stellar populations in the MW bulge may represent the local counterparts of the compact bulges observed at very high redshifts by James Webb Space Telescope (JWST). This is supported by recent findings, such as \citet{XiaoM2025}, who identified an ultra-massive grand-design spiral galaxy at z $\simeq$ 5.2, exhibiting a classical bulge structure. 

Moreover, the bulge contains a reservoir of the most metal-rich stars in the Galaxy \citep[e.g.][]{RojasArriagada2020, queiroz2020, Queiroz2021, Rix2024,Horta_Knot2025}. Indeed, thanks to {\it Gaia}'s precise parallaxes and proper motions, combined with extensive spectroscopic and photometric datasets, precise distances have become available for field stars in the innermost regions of the MW. These have been derived using Bayesian spectro-photometric tools such as the {\tt StarHorse} code \citep{queiroz2018, queiroz2020, Queiroz2023}. Recent comparisons with independent tracers such as Mira variables and horizontal branch stars \citep{ZhangH2024, Arentsen2024} have further validated the precision of these distances, even in highly extinguished and distant inner-Galaxy regions. This progress has allowed bulge studies to move beyond the classical sky-projected ($l, b$) analyses and towards chemo-orbital mapping, which is crucial for disentangling the multiple, co-spatial stellar populations existing in the bulge.

\citet{Queiroz2021} (Q21 hereafter) pioneered this approach, revealing distinct chemodynamical populations in the bulge, including a metal-rich bar component alongside a significant metal-poor contribution, in addition to a kinematically hotter population. Their orbital parameter maps linked chemical properties with the maximum height from the galactic midplane and eccentricity, showing that at low heights a transition occurs towards very metal-rich stars with large eccentricities consistent with bar-supporting orbits. \citetalias{Queiroz2021} further showed that among these inner stars on hot orbits, a high fraction is moderately metal poor. These populations fade rapidly at lower metallicities (below the typical metallicity of bulge GCs; see \citealt{Bica2024}) and are scarcely found in samples targeting very metal-poor stars in the inner MW (e.g. \citealt{Arentsen2024}). Importantly, \citetalias{Queiroz2021} has also shown that a substantial portion of both metal-rich and intermediate-metallicity stars are supported by the bar, and that the expectation that the bar was mainly composed of metal-rich stars was too simplistic. 

Another way to convey a similar message was shown by \citet{Rix2024}, who confirmed the dramatic shift among extremely metal-rich stars (with metallicities above [Fe/H] $ > +$0.5 dex) from a flattened, rotating bar to a more compact, less-rotating component with intermediate metallicities. \citet{Han2025} complemented these findings through the combined analysis of RR Lyrae stars and APOGEE giants, confirming that kinematically hotter, centrally concentrated stars do not align with bar dynamics, whereas stars farther from the centre exhibit bar-like rotation. Their orbital classifications based on apocentre distances effectively reduce disc and halo contamination, underscoring the need for orbital analysis to disentangle bulge populations.

However, the new data also brought some inconsistencies with previous interpretations based solely on photometry. Although photometric studies suggest the bar is dominated by metal-rich stars (expected to possess low-alpha-over-iron ratios; see \citealt{Lim2021}), spectroscopic evidence reported by \citepalias{Queiroz2021} reveals a persistent alpha-enhanced and moderately metal-poor population within the bar, demonstrating its chemical diversity. The recent review by \citet{Zoccali2024} synthesises these findings, emphasising the bulge’s bimodal metallicity distribution function (MDF), with metal-rich ([Fe/H] $= +$0.25) and metal-poor ([Fe/H] $=-$0.3) populations differing in spatial distribution and kinematics. While metal-rich stars dominate intermediate latitudes with cylindrical rotation, metal-poor stars are dynamically hotter, increasing in relative importance towards the bulge’s centre and higher latitudes. However, although the photometry traces millions of stars, it cannot resolve the different stellar populations co-existing in the innermost regions of the MW. In addition, the transition of the MDF in the central regions is very dependent on position \citep[e.g.][]{Zoccali2014,Barbuy2018,RojasArriagada2020}, and the inner five degrees play a crucial role if one wants to sample all the co-existing stellar populations in the bulge.

This complex interplay calls for tighter chemodynamical constraints combining chemistry and orbital dynamics. Further theoretical insight into the complexity of the bulge structure comes from simulations by \citet{Saha2012}, who showed that even a low-mass, initially non-rotating classical bulge can be dynamically transformed by angular momentum exchange with the Galactic bar. In their high-resolution N-body models, this process spins up a spheroidal bulge through resonant and non-resonant interactions, reshaping it into a fast-rotating, radially anisotropic, and triaxial structure with cylindrical rotation. This 'classical bulge-bar', embedded within the larger boxy bulge formed from the disc, may leave distinct dynamical signatures observable today. These results, along with many others in the literature \citep[e.g.][and references therein]{BlandHawthorn2016,Zoccali2024}, reinforce the need for robust, orbit- and chemistry-based constraints in bulge studies, as purely positional analyses (e.g. using only Galactic coordinates) are insufficient to disentangle the diverse, co-spatial populations now known to occupy the inner Galaxy.

In summary, collectively, these studies reveal the MW bulge as a chemically and dynamically intricate system where multiple populations co-exist within the same spatial volume but differ in orbital and chemical properties. Disentangling these requires comprehensive chemo-orbital analyses -- the objective of our ongoing work leveraging {\it Gaia}, APOGEE, distances, and {\it Gaia} RVS data to place new constraints on the bulge’s formation and evolution within a unified chemodynamical framework.

In this work, we extend the findings of \citetalias{Queiroz2021}, by complementing the bulge sample with stars from the {\it Gaia} data release 3 (DR3) Radial Velocity Spectrograph (RVS; \citealt{gaiadr3_survey_properties}). Thanks to the analysis by \citet{rvs_cnn_2023}, who used a hybrid convolutional neural network ({\tt hybrid-CNN}), precise atmospheric parameters and chemical abundances were obtained even for low signal-to-noise spectra. This has significantly expanded the number of stars with high-quality stellar parameters. This advancement has enabled the construction of statistically robust and chemically precise samples, particularly within the 1–2 kpc volume around the Sun. Crucially, it has allowed the derivation of accurate stellar ages, which has opened new windows into the temporal dimension of Galactic structure studies. For instance, \citet{Nepal2024_bar} used this enhanced dataset to study super-metal-rich (SMR) stars, showing that a recent star formation burst around 3–4 Gyr ago could be associated with bar-driven dynamics. Furthermore, the same data reveal a surprising population of stars on thin disc-like orbits with ages exceeding 12 Gyr (appearing to be the local counterpart of the large discs now observed at the high redshifts quoted above).

Here we use the brightest giant stars in the improved RVS dataset to study the galactic bulge (although, in this case, the data lack age information). The aim is to extend the orbital analysis first introduced in \citetalias{Queiroz2021}, now incorporating APOGEE data release 17 (DR17) {\tt StarHorse} distances \citep{Queiroz2023}. This allows for a more detailed characterisation of the multiple stellar populations in the bulge, with a focus on the stars that remain confined in the inner 5 kpc of the MW. Crucially, we perform, for the first time, a comprehensive chemo-orbital mapping that links distinct orbital families with their chemical signatures. As we show, this approach reveals a population of stars on $X_4$ orbits (retrograde), likely tracing a roughly spheroidal bulge component that predates the formation of the bar. We also identify a thick disc structure in the bulge region, which contribute to a wide MDF. This refined orbital framework and the availability of alpha abundances are essential for properly interpreting the MDFs in the innermost regions of the MW and, ultimately, for providing robust constraints to models of chemical evolution (see \citealt{Matteucci2021} and references therein) and chemodynamical formation scenarios \citep[e.g.][]{Debattista2017,Fragkoudi2020, Orkney2022}.

This paper is organised as follows. In Section \ref{sec:data}, we describe the two stellar samples adopted in this study and highlight their complementarity. We demonstrate, for the first time, how {\it Gaia} RVS data can be effectively used to study stars in the bulge region, showing strong consistency between optical and near-infrared datasets in terms of stellar parameters, distances, metallicities, and \alpham\ abundances.
In Section \ref{sec:chemdyn}, we present maps of global kinematics and on-sky MDFs, before proceeding to a detailed orbital analysis aimed at disentangling the multiple co-existing populations in the studied region. We map the stars supported by the bar to their chemical properties and isolate the thick disc and the spheroidal bulge. Finally, we present the main chemo-kinematical properties of the spheroidal bulge, Galactic bar, and the inner thick disc, summarising their main properties also with respect to their [$\alpha$/Fe] enhancement. Discussion and conclusions are presented in Section \ref{sec:conclusion}.

\begin{figure*}[!ht]
    \centering
    \includegraphics[width=0.99\linewidth]{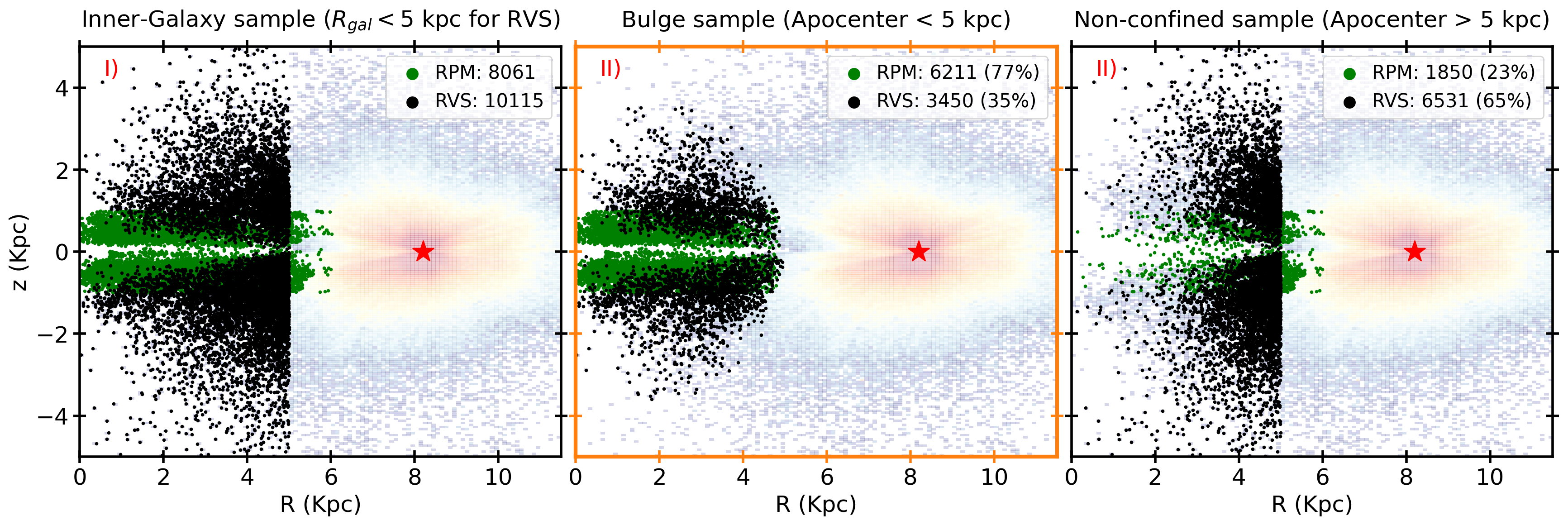}
    \caption{Sample selection schema. I) Distance from the Galactic midplane (Z) vs the galactocentric radius (R) distribution of the parent inner-galaxy sample. The green (RPM) and black (RVS) colours represent the two spectroscopic sources. The background shows the 2D density distribution for the full RVS data, which is mostly concentrated at the solar neighbourhood. The red stars represents the current position of the Sun. II) Z vs R for the bulge sample stars ~[\,Total\,(9661)\,=\,RPM(6211)\,+\,RVS\,(3450)\,] confined to within 5 kpc. Unless otherwise stated, \textit{“bulge sample”} refers to this group. The fraction of stars that are in the bulge sample from the parent inner-galaxy sample are also shown. III) Similar to panel I), but for stars that do not remain confined within 5 kpc.}
    \label{fig:sample_selection}
\end{figure*}

\section{Data}\label{sec:data}

This work utilises spectroscopic samples from two different sources. The first corresponds to the reduced proper motion (RPM) sample of approximately 8,000 stars from \citetalias{Queiroz2021}, which, in turn, is based on the spectroscopic parameters released during DR17 (\citealt{Abdurrouf2022}) of the Sloan Digital Sky Survey (SDSS)-III/IV for the stars observed by the APOGEE survey \citep{Majewski2017}. Using the very precise proper motions of {\it Gaia} EDR3 and {\tt StarHorse}\footnote{\texttt{StarHorse} is a Bayesian isochrone-fitting code \citep{queiroz2018, Queiroz2023} that combines spectroscopic parameters with broadband photometry and {\it Gaia} parallaxes to derive the distances, extinctions, and astrophysical parameters for stars.} distances, \citetalias{Queiroz2021} employed the RPM technique to obtain stars cleaned from the foreground disc and confined to the inner Galaxy (see \citetalias{Queiroz2021} for details).

Our second sample was extracted from the catalogue by \citet[][G24]{rvs_cnn_2023}. \citetalias{rvs_cnn_2023} analysed about one million spectra observed with the RVS\footnote{The RVS spectra were originally analysed during {\it Gaia} DR3 (10.17876/gaia/dr.3) by the General Stellar Parametriser for spectroscopy (GSP-Spec, \citealt{recioblanco2023}) module of the Astrophysical parameters inference system (Apsis, \citealt{Creevey2023}).} of the {\it Gaia} spacecraft. \citetalias{rvs_cnn_2023} used the {\tt hybrid-CNN} method in order to derive precise atmospheric parameters (\teff, \logg, and overall \mh) and chemical abundances (\feh\, and \alpham ). The training sample for this machine learning method was based on the high-quality parameters from APOGEE DR17. \citetalias{rvs_cnn_2023} significantly increased the number of reliable targets with {\tt hybrid-CNN} and made improvements over GSP-Spec thanks to the novel method and inclusion of the additional information from {\it Gaia} magnitudes \citep{Riello2021}, parallaxes \citep{Lindegren2021}, and the Blue and Red Prism Photometer (XP) coefficients \citep{denageli_2023}.  

For the \citetalias{rvs_cnn_2023} sample, the distances, extinctions, and stellar ages for a subset of main-sequence turnoff (MSTO) and sub-giant (SGB) stars have also been computed using the {\tt StarHorse} code. This catalogue has been previously used to study the description of the \alphafe\, bimodality from the inner to the outer disc of the MW (\citetalias{rvs_cnn_2023}), the distribution of SMR stars in the Solar neighbourhood, hints on the formation epoch of the Galactic bar (\citet{Nepal2024_bar}), and the MW's early disc history (\citet{Nepal2024_disc}). We refer to these papers for a detailed validation and description of the catalogue. The final catalogue with {\tt StarHorse} products for the \citetalias{rvs_cnn_2023} sample will be published in Nepal et al. (in prep.).

To calculate positions and velocities in the galactocentric rest-frame and to integrate the orbits of the stars, we used the 6D phase-space vector (sky positions, parallaxes, proper motions, and radial velocities) from {\it Gaia} DR3 \citep{gaiadr3_survey_properties}, along with the {\tt StarHorse} distances. The radial velocities for the RPM sample are from APOGEE DR17. We used {\tt Astropy} \citep{astropy2022} for position and velocity transformations, assuming the Sun is located at radius ~$\mathrm{R_{0}}$\,=\,8.2\,kpc and the circular velocity of local standard of rest (LSR) is ~$\mathrm{V_{0}}$\,=\,233.1\,\kms\,\citep{galpy2015, McMillan2017}. The peculiar velocity of the Sun with respect to the LSR is $\mathrm{(U, V, W)_\odot} = (11.1,\,12.24,\,7.25)$\,\kms\ \citep{Schonrich2010}. 

For the calculation of orbital parameters, we adopted the MW model from \citet{Portail2017} with the analytical approximation given by \citet{Sormani2022}. The Galactic model is composed of the X-shaped boxy-peanut, a long bar, a central concentration mass, a disc, and a dark-matter halo. The bar rotates with a pattern speed of $\mathrm{39\,km\,s^{-1}\,kpc^{-1}}$ and is oriented at an angle of $25^\circ$ with the Sun–Galactic centre line of sight. Using the Action-based Galaxy Modelling Architecture (AGAMA; \citealt{Agama2019}), a Python package for orbital integration\footnote{\url{http://github.com/GalacticDynamics-Oxford/Agama}}, we performed forward orbital integrations, for both the \citetalias{Queiroz2021} and \citetalias{rvs_cnn_2023} samples, for a 3 Gyr period and saved each orbit's trajectory every 1 Myr. The adopted Galactic potential, bar pattern speed, and orientation follow commonly used models in the literature. While variations in these parameters could affect individual orbits, simple change of pattern speed would be dynamically inconsistent with the assumed potential. A full, self-consistent re-computation is beyond the scope of this work. Our analysis focuses on the broad orbital properties of the main stellar populations, which remain robust within current model uncertainties. For instance, we recover the trends reported by \citepalias{Queiroz2021} despite using a different potential, supporting the reliability of our results.

\begin{figure*}[!ht]
    \centering
    \includegraphics[width=0.7\linewidth]{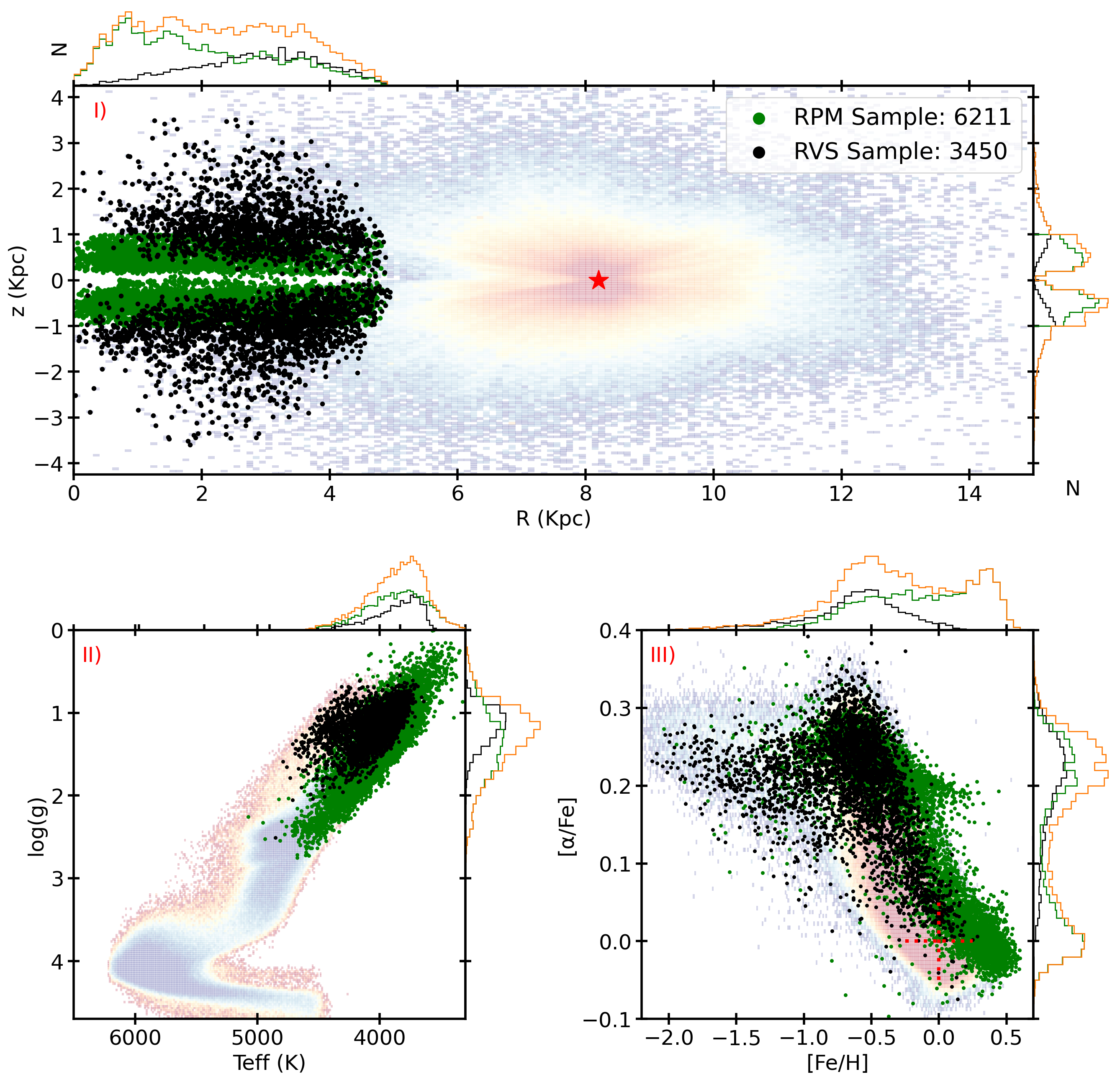}
    \caption{General properties of the bulge sample stars: I) Distance from the Galactic midplane (Z) vs the galactocentric radius (R) of the bulge sample, along with histograms showing the individual distributions for the RPM (green) and RVS (black) stars; II) Kiel diagram (\teff\, vs \logg\,); III) \alphafe\,vs \feh\, diagram for the bulge sample plus the individual histograms.}
    \label{fig:data intro}
\end{figure*}

\subsection{The sample definition}

The current work aims to explore the inner Galactic region (i.e. within the inner $\sim$5 kpc), with a focus on the complementarity of the RVS sample. For this reason, we chose to begin from the \citetalias{Queiroz2021} well-studied RPM sample of 8,061 stars. This study will be expanded in the future to include APOGEE data from SDSS-V, which will be available with the 20th SDSS-V data release. From the \citetalias{rvs_cnn_2023} catalogue, we selected the stars reliably parametrised by setting `flag\_boundary'=`00000000'. We then cleaned any spurious measurements by applying the following uncertainty limits: `sigma\_teff'$<$100 K, `sigma\_logg'<0.1, `sigma\_feh'<0.2, and `sigma\_alpham'<0.1. We removed any stars with poor astrometric solutions by limiting `RUWE' to <1.4, and we also removed known variable stars using {\it Gaia} flag `phot\_variable\_flag'$\neq$`VARIABLE' (see \citealt{gaiadr3_survey_properties}). For the {\tt StarHorse} parameters, we employed distance uncertainty < 10\% and extinction uncertainty < 0.25 Mag., giving us a high-quality set of 585,790 stars.

We then selected the {\it Gaia}-RVS stars with current galactocentric radii less than 5 kpc, which yielded a parent sample of 18,176 stars (RPM = 8,061 and RVS = 10,115), dubbed the “inner-galaxy sample”. From this sample, we then selected stars confined to 5 kpc from the galactic centre by selecting Apocenter < 5 kpc. We named this the “bulge sample”. It contains a total of 9,661 stars (RPM = 6,211 and RVS = 3450). We considered a limit of R = 5 kpc to study the stellar populations and dynamical structures in the inner Galaxy, including the galactic bar, which is expected to have a radius of about 4 kpc \citep{BlandHawthorn2016, Portail2017, ZhangH2024}. A limit of 5 kpc should also allow us to characterise the inner bulge-bar-disc interface. A similar radius has widely been adopted in various recent work studying the bulge region \citep[e.g.][]{Xiaojie2024, Arentsen2024,Horta_Knot2025, Han2025}. In addition, we repeated the analysis with a relaxed apocentre limit (5.5 kpc vs 5.0 kpc) and found negligible impact on the spheroidal bulge and bar, while primarily adding more thick disc-like stars with similar chemo-kinematical properties as well as a contamination from thin disc stars. This confirms that our adopted apocentre $<$5 kpc selection effectively isolates the genuine inner-Galaxy population, minimising contamination from the thin disc and halo while preserving the core bulge sample.

Our sample selection schema is illustrated in Fig. \ref{fig:sample_selection}. Panel I) shows the Z versus R projected distribution for the full inner-galaxy sample, panel II) for the bulge sample, and panel III) for the stars not confined to the inner 5 kpc. We note here that, since the authors in \citetalias{Queiroz2021} did not apply any distance cuts to select the RPM sample, it includes a low number of ($\sim$350) stars with $\mathrm{R_{gal}>5\,kpc}$ that are not in the bulge sample. Interestingly, we find that a high fraction of RPM stars (77\%) are retained in the bulge sample, while only 23\% RVS stars remain confined to 5 kpc. This difference is explained by the two different sample selection methods adopted for the two sources and the fact that most of the RVS stars lie farther from the galactic plane. The “non-confined” group consists mostly of stars belonging to the thin disc for the RPM and to the thick disc and halo for RVS sample -- we refer to Appendix \ref{sec:nonConfined} for details on the characterisation of this group. Throughout this paper, we use the term, `Bulge Sample', except in Sec. \ref{sec:bulge_halo} when discussing the link between the bulge and the halo.

In Fig. \ref{fig:data intro} we present the general properties of our bulge sample. The stars have uniformly distributed radii within 4 kpc with a decreasing tail out to 5 kpc. The RVS stars have a wider coverage in the Z direction with a peak at 1 kpc and extend to about 3 kpc, while the RPM sample, by design, is confined to within |Z| = 1 kpc. At the innermost region, i.e. R < 1 kpc and |Z| < 1 kpc, we mostly have RPM stars. Our sample mostly includes M and K giants with a larger log(g) range for the APOGEE stars. We do not have the brightest giants in RVS due to the training sample limits adopted in \citetalias{rvs_cnn_2023}. The sample covers a wide range of \feh with a clear bimodality in \alphafe. RVS provides a high fraction of metal-poor stars compared to RPM; namely, less than 3\% of RPM (164 stars) have $\feh \leq -1.0$, while it is about 15\% for RVS (551 stars). Conversely, RPM provides most of the super-solar metallicity stars, which are known to reside mostly close to the galactic plane.

\begin{figure*}[!ht]
    \centering
    \includegraphics[width=0.9\linewidth]{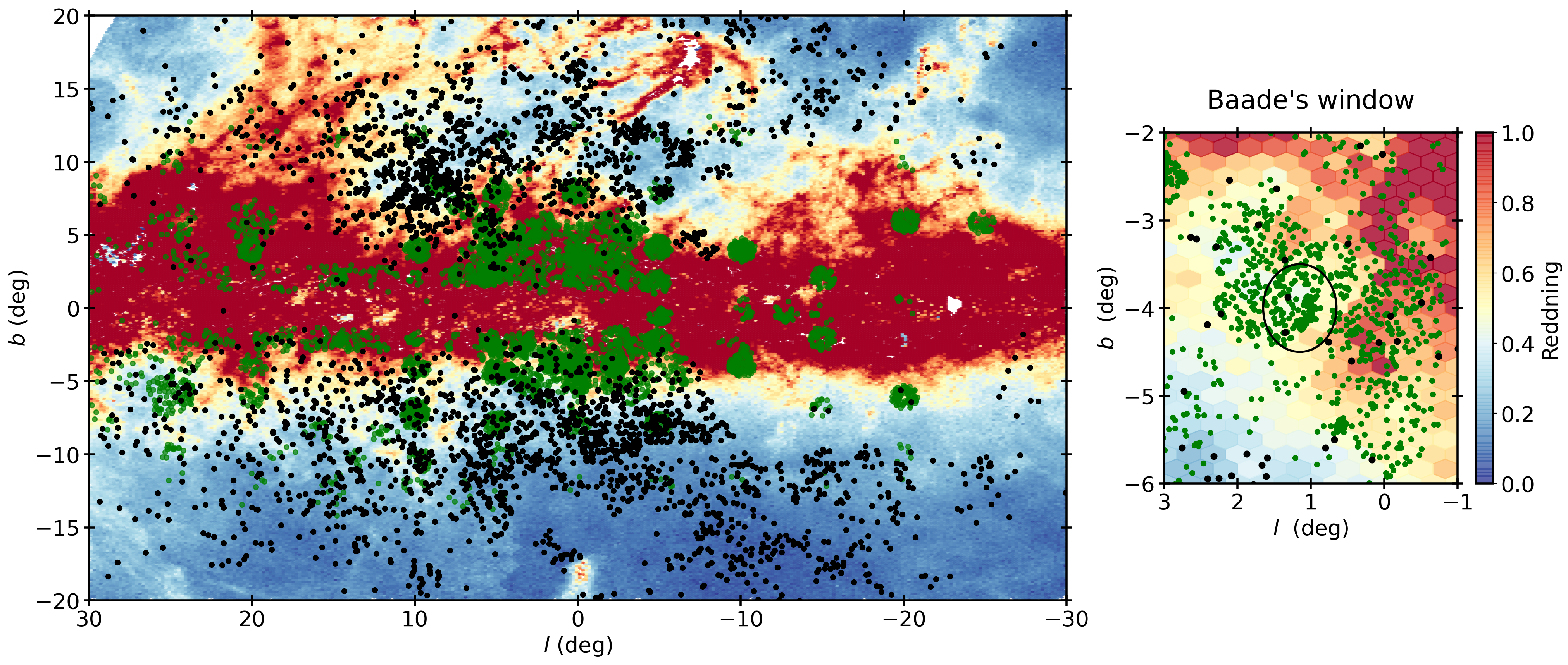}
    \caption{Sky distribution (in Galactic coordinates) of the bulge sample stars, RPM (green), and RVS (black), above the extinction map covering the Bulge region obtained with results from \citet{Anders2022}. Right: Zoom-in view of the Baade's window (black circle). The pencil-beam-like observation of the APOGEE survey is evident in the sky distribution of the targets, while the {\it Gaia} spacecraft observes all-sky and covers a much larger area.}
    \label{fig:lb_extinction}
\end{figure*}

In Figure \ref{fig:lb_extinction} we present the sky distribution (in Galactic coordinates) of our bulge sample stars above an extinction map. Complementary to the distribution in the Z versus R plane shown in Fig. \ref{fig:data intro}, the RPM stars are mostly confined to $|b| < 10^{\circ}$ with a significant fraction within $5^{\circ}$. APOGEE infrared spectroscopy clearly has an advantage over {\it Gaia} spectroscopy in the high extinction regions. The RVS sample mostly avoids high extinction regions and is hence absent close to the galactic plane, i.e. around $|b| = 0^{\circ}$ -- they have a wider $b$ distribution with most of the stars between $+5^{\circ}$ to $+15^{\circ}$ and $-5^{\circ}$ to $-15^{\circ}$. A complicated survey selection is evident for both surveys in this challenging region of the sky. As a validation of our RVS stars covering low extinction regions close to the galactic midplane, we present a zoom-in of the Baade's Window in the smaller panel on the right. While a lot of observations are available with APOGEE spectra in this region, we managed to retain a few stars with the RVS spectra.

\begin{figure}[!ht]
    \centering
    \includegraphics[width=0.99\linewidth]{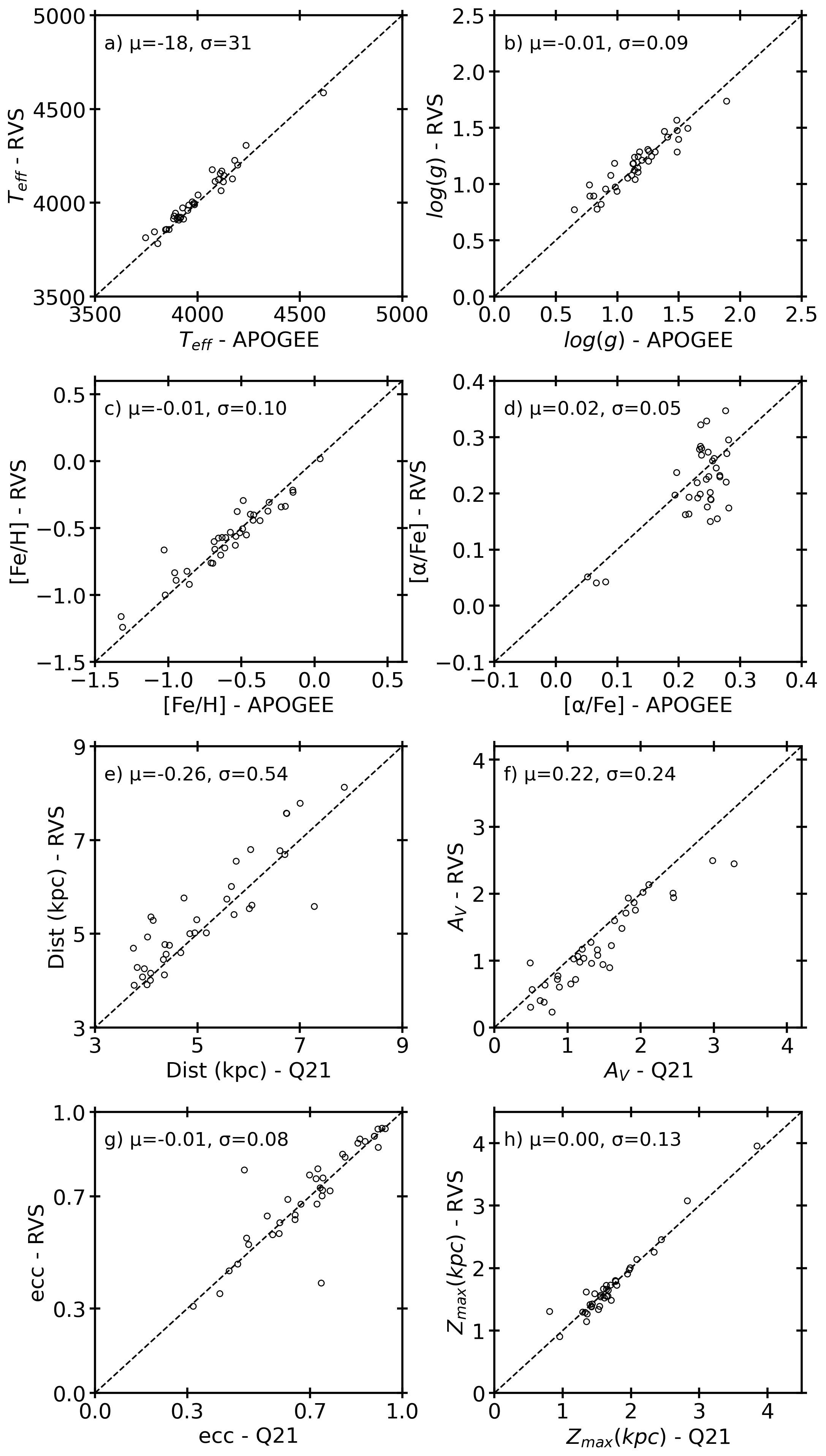}
    \caption{Comparison of the \teff, \logg, \feh, \alphafe, distances, extinctions, ecc, and $\mathrm{Z_{max}}$\, for the 36 common stars in the RPM and RVS bulge samples. For the RPM bulge sample, \teff, \logg, \feh\,, and \alphafe\, are from APOGEE DR17. The distances and extinctions correspond to the {\tt StarHorse} estimates from \citetalias{Queiroz2021}, and the orbital parameters, ecc and $\mathrm{Z_{max}}$, were re-estimated using new galactic potential. For the RVS bulge sample, \feh\, and \alphafe\, are from \citetalias{rvs_cnn_2023}; the distances and extinctions correspond to the StarHorse estimates from current work; and the orbital parameters, ecc and $\mathrm{Z_{max}}$, were estimated using the new galactic potential.}
    \label{fig:RPM_RVS_bulge_common}
\end{figure}

It is beyond the scope of the present work to account for sample selection effects. The reason is that simple corrections based on extinction maps or Gaia/APOGEE magnitude limits are insufficient to characterise the selection function of our sample. Additional and hard-to-quantify biases arise from proper-motion availability, the convergence behaviour of the RVS-CNN and APOGEE pipelines in different stellar-parameter regimes, and {\tt StarHorse} convergence. Our primary goal is to demonstrate the relevance of Gaia-RVS data—so far used mainly in the nearby volume—for bulge studies. For these reasons, a full treatment of selection effects lies beyond the scope of this work.

\subsection{Comparison and complementarity of the APOGEE-RPM and the {\it Gaia}-RVS samples}\label{sec:rvs_apo}

In Fig. \ref{fig:RPM_RVS_bulge_common} we present a one-to-one comparison for our 36 bulge sample stars common between RVS and RPM. These 36 stars span well the range of S/N ($\sim$20 to $\sim$50) and G magnitudes (12 to 14) covered by the RVS-CNN bulge sample, which indicates that they are representative of the full sample. Additionally, we also checked the distributions of uncertainties in [Fe/H] and [$\alpha$/Fe] as a function of S/N, \textit{Gaia} G magnitudes, and [Fe/H] to find similar trends, confirming that the comparison sample is not biased towards higher-quality spectra. Panels a to d show excellent match for the atmospheric parameters \teff\, and \logg\, with negligible biases and a scatter of $\sim$30 K and $\sim$0.1 dex, respectively. Also, for \feh\, no apparent bias is seen even at metallicities below $-$1.0 dex; there is a small overall scatter of 0.1 dex. The {\tt hybrid-CNN} exhibits larger uncertainties towards the metal-poor end ([Fe/H]$\lesssim-$1.5 dex), due to the limitations in the training sample (\citetalias{rvs_cnn_2023}). However, this bias affects only a low fraction of stars and does not impact our main conclusions, which concern the bulk of the spheroidal bulge, bar, and inner-thick disc populations at higher metallicities. \alphafe\, shows a visible spread for the high-$\alpha$ stars with a mean scatter of 0.05 dex; however, it is worth noting that none of the stars with high-$\alpha$ values from APOGEE have traditionally ‘low-$\alpha$’ values (i.e. < 0.1 dex) in the RVS sample. These results -- especially for \teff\,, \logg\,, and \feh\, from such narrow wavelength and low signal-to-noise RVS spectra -- highlight the amazing precision (and accuracy) achieved even for these bulge stars with the {\tt hybrid-CNN} machine learning application of \citetalias{rvs_cnn_2023}. 

In panels e) and f) we show the comparison for the {\tt StarHorse} estimates of distances and extinctions for the two samples. The distances are mildly overestimated for the RVS with the opposite for the extinctions as indicated by the biases of $-$0.26 kpc and 0.22 mag, respectively. However, these stars, for both samples, have distance uncertainties of less than 10\%. In panels g) and h), we show the orbital parameters of eccentricity and $\mathrm{Z_{max}}$. We find well-matched orbital parameters despite the respective uncertainties in distances and radial velocities for the two sources, which were taken into consideration during the orbit integrations. Only two out of 36 stars have a difference of more than 0.3 for orbital eccentricities.

\begin{figure}[!ht]
    \centering
    \includegraphics[width=0.8\linewidth]{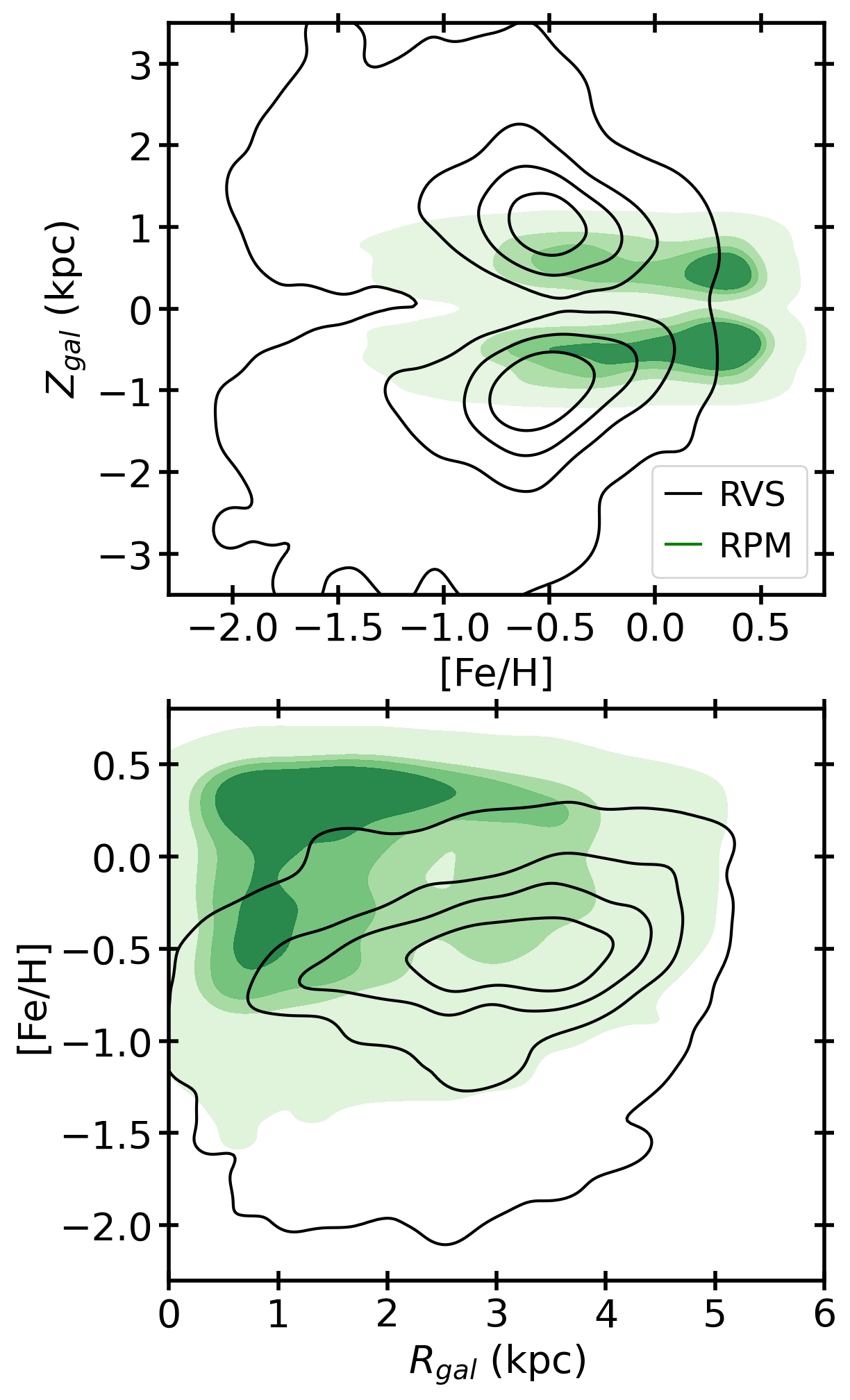}
    \caption{\feh\ vs spatial distribution of the bulge sample. Top: $\mathrm{Z_{gal}}$ as a function of \feh\, shown as a kernel density estimate (KDE) for the RVS (black) and RPM (green) samples. Bottom: \feh\ as a function of $\mathrm{R_{gal}}$.}
    \label{fig:R_Z_FeH}
\end{figure}

In Fig. \ref{fig:R_Z_FeH} we present \feh\, as a function of $\mathrm{Z_{gal}}$ and $\mathrm{R_{gal}}$ for the bulge sample. We find that the RPM sample maps the regions close to the galactic plane, along with the innermost radii, well and hence provides most of the super-solar metallicity stars. The RVS complements with a larger fraction of metal-poor stars. 

\begin{figure*}[!ht]
    \centering
    \includegraphics[width=0.8\linewidth] {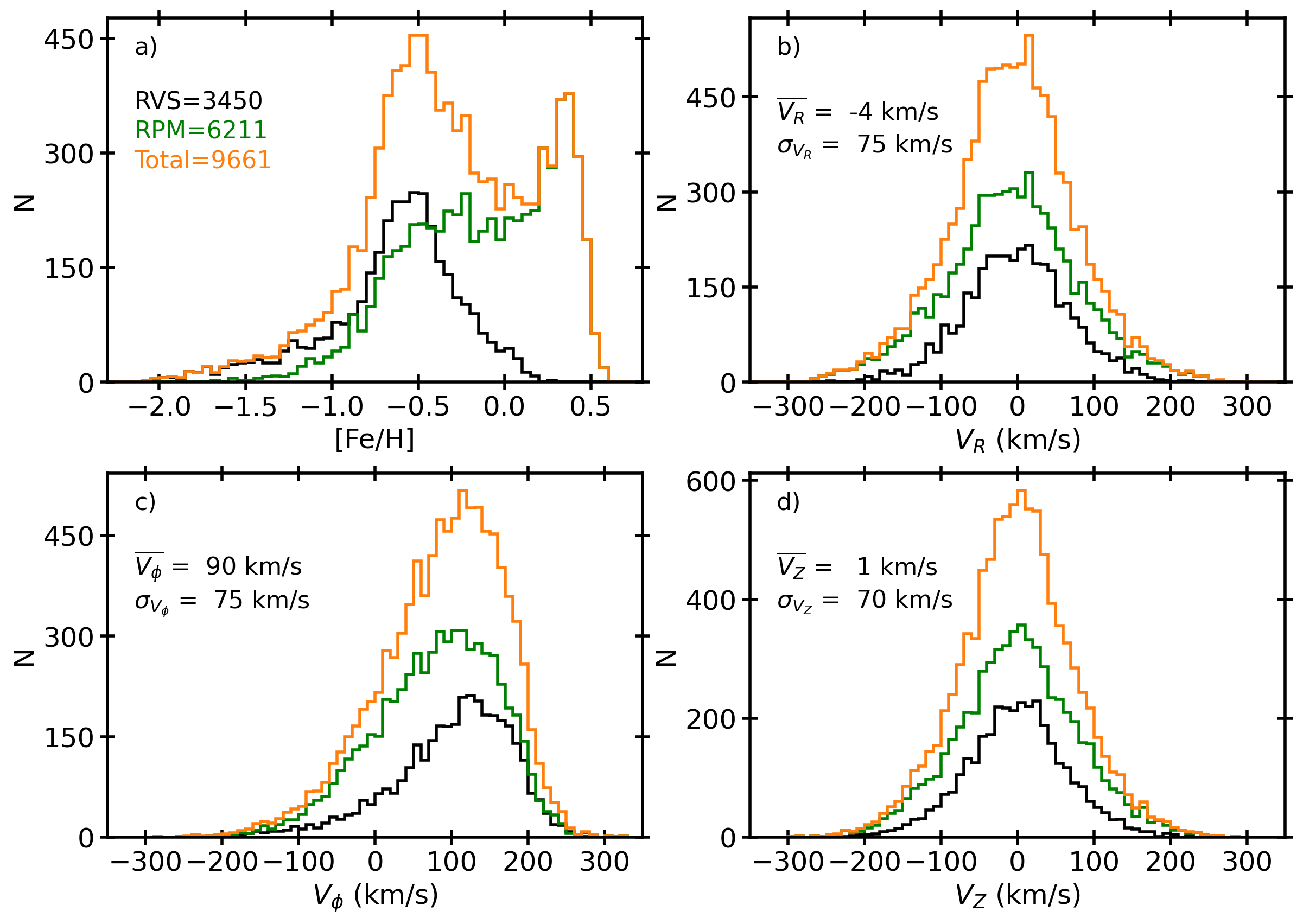}
    \caption{General MDF and kinematics of the combined bulge sample. Top: a) MDF for the combined bulge sample (orange), MDFs for the RVS (black), and RPM (green) samples shown separately; b) Distribution of the galactocentric radial velocity ($V_{R}$ in \kms). Bottom: c) Galactocentric azimuthal velocity ($V_{\phi}$); d) Galactocentric vertical velocity ($V_{Z}$). The mean and dispersion for the three velocity components, for the total, are also shown and were calculated via bootstrapping resampling.}
    \label{fig:full_mdf_kine}
\end{figure*}

\section{Stellar populations in the Galactic Bulge} 
\label{sec:chemdyn}

\subsection{Overall kinematics and the on-sky MDFs}

\begin{figure*}[!ht]
    \centering
    \includegraphics[width=0.99\linewidth]{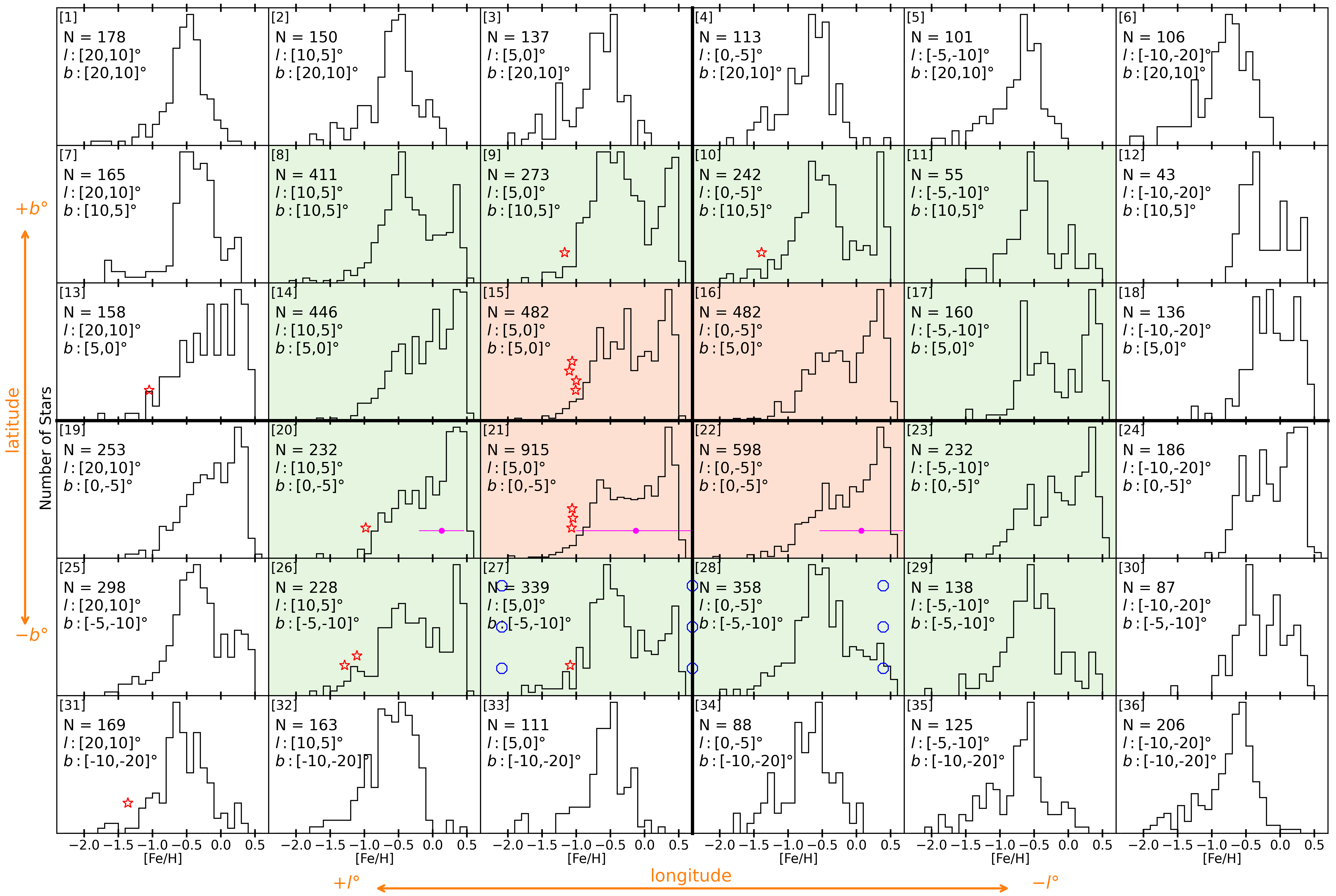}
    \caption{MDF of the combined bulge sample for different longitude-latitude ($l$, $b$) bins, in point-of-view of sky distribution (in Galactic coordinates). The thick lines represent the axes for $l$ and $b$, i.e. $0^\circ$ lines with positive (+ve) longitude on the left and negative (-ve) on the right, and -ve latitude downwards and +ve upwards. The red panels represent the central 5 degrees, and the green panels represents the central 10 degrees. Each panel shows the respective $lb$ range, number of stars, and the MDF. Red stars, placed according to their on-sky position, show the metallicity for the Bulge GCs from \citet{Souza2024a}. Magenta circles show the median \feh\, for the micro-lensed dwarf stars from \citet{Bensby2013}. The error bars represent the spread of the \feh\, in each $lb$ bins. The blue hexagons show the positions of the fields from the study of \citet{Lim2021}, using BDBS bulge red-clump stars}
    \label{fig:mdf_in_lb_main}
\end{figure*}

We begin the chemodynamical exploration of the galactic bulge by first looking at the general chemo-kinematics of our bulge sample. In Fig. \ref{fig:full_mdf_kine} we present the MDF and the galactocentric radial, tangential, and vertical velocities. We show distributions from both sources to highlight the individual contributions from RVS and RPM but here discuss the properties of the combined bulge sample. The MDF has, in general, a bimodal shape with a long tail towards the metal-poor end extending to $\sim$ --2.0 dex. We observe a super-solar peak at $\sim$0.3 dex and a metal-poor peak at $\sim$ --0.5 dex. The metal-poor peak is close to the local thick disc MDF peak, while the metal-rich peak is at a much higher \feh\, value compared to the local thin disc. Later in the paper, we show that this bimodal MDF is a superimposition of multiple components (see Fig. \ref{fig:mdf_in_lb_main} and Sec. \ref{sec:zmaxecc}). We note that the true fraction of metal-poor versus metal-rich stars strongly depends on the sample selection. For instance, the highest metallicity stars are expected to be close to the galactic plane, which is the region most affected by extinction and hence difficult to obtain spectroscopy. However, the relative behaviour is informative, and a correction for the selection function is beyond the scope of the present work. 

Our sample has relatively high velocity dispersions for all three components with ($\sigma_{V_R}$, $\sigma_{V_{\phi}}$, $\sigma_{V_Z}$) = (75, 75, 70) \kms. Interestingly, but as expected, we observe an important property: unlike galactocentric R and Z velocities, which are normally distributed with a mean $\sim$0 \kms\,, the azimuthal velocity has a mean of 90 \kms. At the higher end, $V_{\phi}$ shows a near-sharp cut at $\sim$200 \kms\, , while the distribution is negatively skewed with a tail towards the negative velocities. A small bump around 0 \kms\, is also visible. 

These velocity distributions hint at the presence of a rotating structure, such as the bar or the disc, reflected by the major component with positive $V_{\phi}$. The left-skewed nature of the $V_{\phi}$ with a bump around 0 \kms\, also hints at the presence of a pressure-supported component in the Galactic bulge. Similarly, the peaked nature of the MDF suggests the presence of multiple stellar populations with possibly distinct kinematics. Several previous studies using different photometric and spectroscopic data have hinted at the presence of distinct galactic components, such as the bar, disc, and a pressure-supported component, and linked it to distinct metallicity regimes. Our data also reflect the presence of a combination of such chemo-kinematic components.

To further understand the general nature of our sample, we present in Fig. \ref{fig:mdf_in_lb_main} the MDFs in different longitude-latitude ($l, b$) bins (on-sky distribution). An obvious but remarkable trend clearly appears when we look at different latitude bins for a fixed longitude range. For example, for the outermost longitude bins of $l \in [20,10]^\circ$ or $[-10,-20]^\circ$, we observe the MDF peak shift from metal-poor to metal-rich and back to metal-poor when we move from top to bottom. The plane of the MW, i.e. closer to $l = 0^\circ$, clearly has more metal-rich stars. At the outermost latitudes (i.e. $b > 10^\circ$ or $b < -10^\circ$), the MDFs appear similar for the full range of longitudes -- they lack super-solar stars and the metal-poor stars dominate here with a major peak at $\sim$ --0.5 along with multiple peaks at lower values. Moving inwards, at the fixed latitude bins of $b \in [10,5]^\circ$ or $[-5,-10]^\circ$, from left to right we observe that the broad metal-poor MDF gain an additional, well-separated, super-solar component for the innermost longitudes, i.e. for $l \in [10,-5]^\circ$. Close to the galactic plane, i.e. for the latitude bins of $b \in [5,-5]^\circ$, the metal-rich stars are present at all longitudes and not so well separated from the metal-poor component as in the case of the higher latitudes. The outer longitudes show a higher fraction of metal-rich stars, while at the innermost region, i.e. the central $5^\circ$ shown in red panels, we observe an increase in the metal-poor stars with a tail to $\feh \approx - 1.5$. The MDF components appear to be well mixed in the central regions.

These on-sky distributions highlight the fact that the MDF significantly varies with the on-sky location of the sample used. The MDF in the MW bulge shows the chemodynamical complexity of this central Galactic structure, reflecting a composite origin from multiple stellar populations. Early indications of a metallicity gradient were reported by \citet{Rodgers1986} and \citet{Terndrup1988} through photometric colour analysis of Reg Giant Branch (RGB) stars across different fields. The first MDF based on high-resolution spectroscopy came from \citet{Terndrup1988}, who observed 11 K giants in Baade’s Window, revealing a broad metallicity spread. Subsequent spectroscopic studies \citep[e.g.][]{Fulbright2006,Rich2007,Zoccali2008,Johnson2011,Hill2011,Gonzalez2011,Rich2012} consistently found that the bulge spans a metallicity range of approximately $-1.5 \leq \feh \leq +0.5$\,dex, typically peaking at solar or slightly super-solar metallicity. The advent of large spectroscopic surveys such as ARGOS \citep{Ness2012}, BRAVA \citep{Kunder2012}, GIBS \citep{Zoccali2014}, Gaia-ESO \citep{Gilmore2022}, and APOGEE \citep{Majewski2017} significantly expanded spatial coverage and sample sizes, enabling MDF measurements across the full bulge extent. These surveys converge on the presence of at least two dominant components: a metal-poor population with $\feh \approx -0.3$ dex and a metal-rich one near +0.3 dex, together producing a bimodal MDF nearly ubiquitous across the bulge \citep[e.g.][]{Ness2013a, Zoccali2014, RojasArriagada2014,RojasArriagada2020}. More detailed analyses, particularly using APOGEE data, further resolve this structure into three overlapping components with peaks at $\feh \approx$ $+$0.32, $-$0.17, and $-$0.66 dex \citep{RojasArriagada2020}, consistent with earlier bimodal interpretations once observational uncertainties and sample sizes are considered. Some studies (e.g. ARGOS, \citealt{Ness2013a}; micro-lensed dwarfs, \citealt{Bensby2017} and APOGEE, \citealt{GarciaPerez2018}) have even suggested four or five components, hinting at an extended and continuous chemical evolution.

Vertical metallicity gradients in the bulge are now understood to arise from varying contributions of these components with Galactic latitude: the metal-poor population shows a spheroidal distribution, while the metal-rich one follows the bar/boxy shape morphology \citep{Zoccali2018}, a pattern reproduced by N-body simulations of thin and thick disc components with differing kinematics \citep{Debattista2017, Fragkoudi2018}. These studies suggest the metal-poor spheroidal population may originate from an in situ thick disc, although an early accretion scenario remains plausible \citep{Athanassoula2017}. In contrast, the MDF of bulge RR Lyrae stars (RRLs) -- tracing exclusively older, metal-poor populations -- differs significantly. While based largely on photometric metallicity estimates, these studies \citep[e.g.][and references therein]{Pietrukowicz2015, Zoccali2024} show broader, unimodal distributions with peaks between $-$1.5 and $-$1.0 dex, and extensions down to $-$2.5 dex, which nonetheless includes contamination by halo stars. Spectroscopic MDF for bulge RRLs \citep{Savino2020} confirms a wide range ($-2.5 \leq \feh \leq +0.5$) with a peak at $\feh \approx -1.4$\,dex. Although differences across studies may reflect systematics in photometric estimates, they collectively indicate that the bulge RRL population is chemically distinct from the K- and M-giant dominated MDF, likely tracing an older and more metal-poor stellar component (similarly to that traced by bulge GCs). The recent results of \citet{Kunder2024} further clarify the nature of metal-poor stars in the inner Galaxy, showing that their orbital properties depend strongly on metallicity and location. While stars with metallicities above [Fe/H] $\approx -$1.0 are dominated by in situ bulge-disc populations, those below this threshold include both halo and possible accreted stars. 

Taken together, these results demonstrate that the bulge MDF is shaped by multiple overlapping populations with distinct formation histories, spatial distributions, and ages—underscoring the complex evolutionary history of the inner MW. Therefore, studying the MDF of the full population alone is not very informative. A more insightful approach requires examining the orbital properties and additional chemical abundances, which is the goal of the next sections.

\begin{figure*}[!ht]
    \centering
    \includegraphics[width=0.99\linewidth]{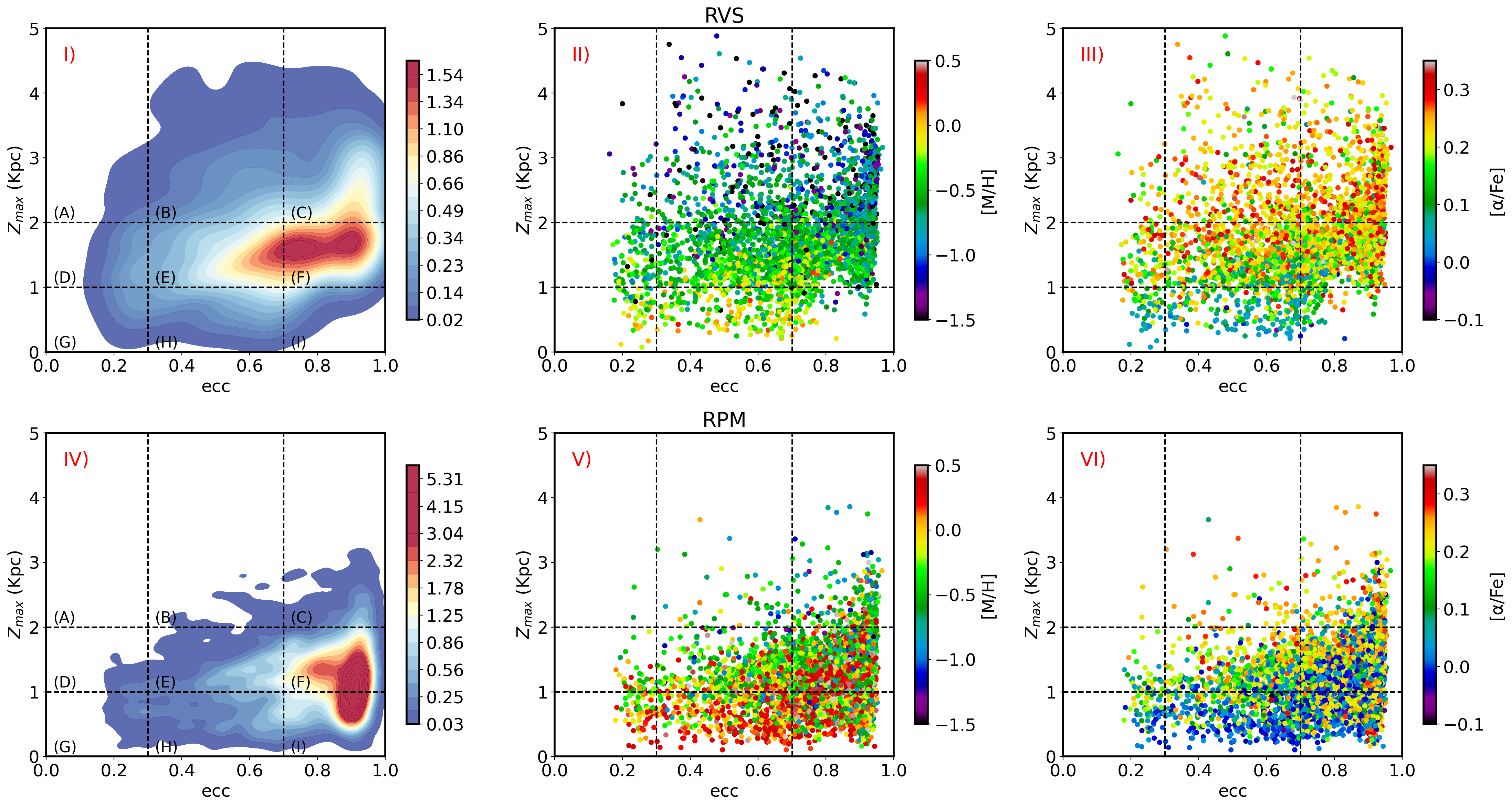}
    \caption{$\mathrm{Z_{max}}$--eccentricity diagram for the bulge sample (Top row: RVS; Bottom row: RPM). Left (Panels I and IV): kernel density estimate (KDE) plot. The dashed lines divide the plane into nine cells, which are labelled alphabetically from A to I. Middle (II and V): Scatter plot colour-coded by individual \feh. Right (III and VI): Scatter plot colour-coded by individual \alphafe.}
    \label{fig:ecc_zmax}
\end{figure*}

\subsection{The $\mathrm{Z_{max}}$-- eccentricity plane} \label{sec:zmaxecc}

The $\mathrm{Z_{max}}$--eccentricity plane has proven to be an important diagnostic tool for the study of stellar populations and the respective galactic components. \citet{Feltzing2008} first used it to analyse thin and thick disc stars for a sample of local dwarfs. \citet{Boeche2013} employed it extensively in their study of the galactic disc with the RAVE survey (see also \citealt{Steinmetz2020b}) and established it as an alternative way of identifying stellar populations based on the similarity of orbits. The eccentricity gives information on the shape of the orbits, and $\mathrm{Z_{max}}$ informs about the oscillation of the star perpendicular to the Galactic plane. In \citetalias{Queiroz2021}, the authors used this parameter space to disentangle the Galactic Bulge region into the thin disc, thick disc, bar, and possibly a spheroidal bulge component, and study their distinct chemical and kinematical properties. According to \citetalias{Queiroz2021}: \textit{'}This parameter space offers a powerful way to disentangle the co-existing populations in the region (avoiding the use of predefined Galactic populations based on properties of the more local samples)'. In this section, similar to \citetalias{Queiroz2021}, we analyse our bulge sample in the $\mathrm{Z_{max}}$--eccentricity plane.

\begin{figure*}
    \centering
    \includegraphics[width=0.99\linewidth]{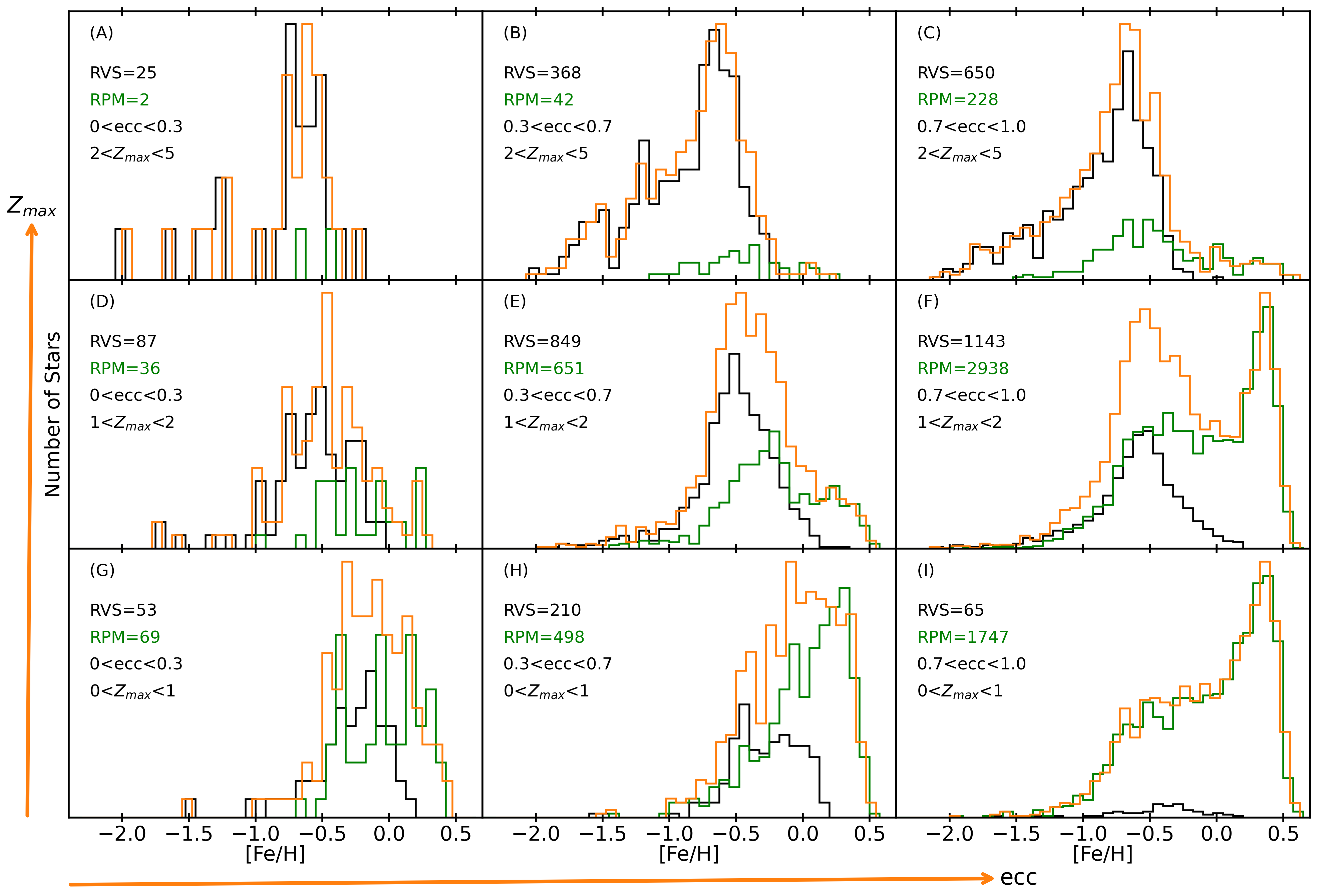}
    \caption{MDFs across the $\mathrm{Z_{max}}$--eccentricity plane (see Fig. \ref{fig:ecc_zmax}) for the bulge sample (full in orange; RVS in black; RPM in green).}
    \label{fig:mdf_ecc_zmax}
\end{figure*}

Figure \ref{fig:ecc_zmax} shows the distribution of our stars in the $\mathrm{Z_{max}}$--eccentricity plane. For the RPM sample, as discussed in \citetalias{Queiroz2021}, most of the stars have high eccentricity and low $\mathrm{Z_{max}}$, i.e. ecc > 0.7 and $\mathrm{Z_{max}}$ < 2 kpc (cells I, F). However, we note here that there are two differences compared to the \citetalias{Queiroz2021} analysis. First, as described in Sec. \ref{fig:data intro}, the adopted Galactic potential, although very similar, is slightly different. Second, we only selected stars confined within 5 kpc, which excludes a high fraction of disc and halo stars (see also Appendix \ref{sec:nonConfined}). For the RVS sample, most of the stars have high eccentricity and low $\mathrm{Z_{max}}$, with ecc > 0.5 and 1 < $\mathrm{Z_{max}}$ < 3 kpc (cells C, E, and F). In the middle and right panels, we show our stars in the plane colour-coded by their individual \feh\, and \alphafe\, , respectively. A general trend is visible, for both \feh\, and \alphafe, showing metal-rich and low-\alphafe\, stars closer to the Galactic plane, while higher $\mathrm{Z_{max}}$ is populated by metal-poor and high-\alphafe\, stars. However, interestingly, the well-populated cells also show a clear mixture of high and low-\alphafe\, as well as metal-poor and metal-rich stars in these cells. In \citetalias{Queiroz2021}, the authors showed that the \alphafe\, bimodality is strongly present in all their well-populated cells (i.e. with $\mathrm{Z_{max}}$ < 2 kpc and ecc > 0.3). These cells also had a high fraction of stars with a high bar-support probability. Next, we analyse the MDFs and the velocities for each cell to better understand the constituent populations.

Figure \ref{fig:mdf_ecc_zmax} shows the MDFs for the different cells across the $\mathrm{Z_{max}}$--eccentricity plane. The orange histogram represents the full bulge sample, while individual RVS and RPM are shown in black and green, respectively. Our analysis is based on the full sample, but here it is important to note the contribution from the RVS sample, particularly from the cells sparsely populated by the RPM (i.e. cells A, B, C, D, and E). The authors in \citetalias{Queiroz2021} find hints of a non-negligible contribution from a spheroid-like component exhibiting a distinct shape in the \alphafe\,vs \feh\,plane; however, they refrain from any strong conclusions due to low statistics in these cells. With the RVS sample, we significantly increased the statistics for these cells.

\begin{figure*}[!ht]
    \centering
    \includegraphics[width=0.99\linewidth]{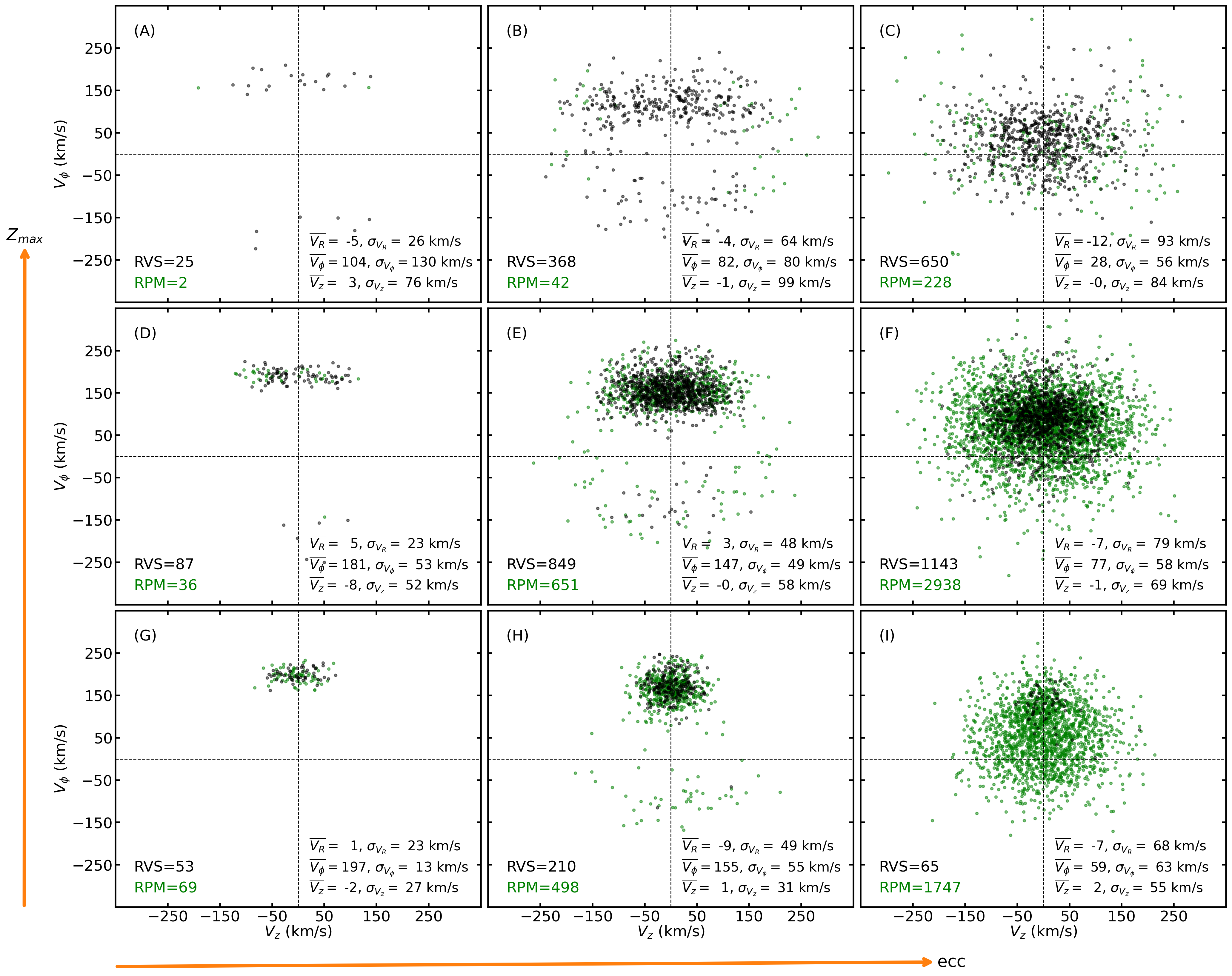}
    \caption{$V_{Z}$ vs $V_{\phi}$ across the $\mathrm{Z_{max}}$--ecc plane (see Fig. \ref{fig:ecc_zmax}). For each cell, the mean and dispersion of the galactocentric radial ($V_{R}$), azimuthal ($V_{\phi}$), and vertical ($V_{Z}$) velocity are also shown and have been calculated via bootstrap resampling.}
    \label{fig:vz_vphi}
\end{figure*}

In Figure \ref{fig:mdf_ecc_zmax}, the MDF in cell G -- corresponding to low eccentricity (ecc < 0.3) and low $\mathrm{Z_{max}}$ (< 1 kpc) -- shows that nearly all stars lie within $-0.5 < \feh < 0.5$, with very few below $-0.5$, and the distribution peaks below solar \feh. At similarly low eccentricity but higher $\mathrm{Z_{max}}$ (cells D and A), the MDF displays a sharp decline in metal-rich stars and a significant increase in metal-poor stars, extending down to $-2.0$ dex. As $\mathrm{Z_{max}}$ increases, the peak of the distribution gradually shifts towards lower metallicity. For intermediate eccentricity (0.3 < ecc < 0.7) and low $\mathrm{Z_{max}}$, in cell H, we see a high fraction of super-solar-metallicity stars and the distribution shows a peak at solar value. At higher $\mathrm{Z_{max}}$ (cells E and B), we find a similar trend of increase in metal-poor stars and the overall shift of distribution towards lower metallicity. However, even with only a visual inspection, the MDFs clearly reflect multiple components. Cell E shows a major component with the peak at $\sim-0.5$, in addition to a minor metal-poor one extending down to $-2.0$ dex and a super-solar component. In cell B, we find that nearly all stars have metallicities below the sub-solar values and roughly three components with peaks at $\sim-0.6$, $\sim-1.2$, and $\sim-1.5$ dex. Finally, for the high eccentricity bin (ecc > 0.7), in the low $\mathrm{Z_{max}}$ cell I, we observe a high fraction of super-solar stars with a peak at $\sim+0.3$ dex, followed by a plateau towards lower metallicity up to $\sim-0.7$ and a tail down to $-2.0$ dex. For the intermediate $\mathrm{Z_{max}}$ cell F, again with only visual inspection, we observe a very clear bimodal distribution -- a metal-rich component with a peak at $\sim+0.3$ dex and a metal-poor component with a peak at $\sim-0.5$ dex, along with a tail down to $-2.0$ dex. At the highest $\mathrm{Z_{max}}$ (cell C), the distribution is mostly metal-poor, similar to cell B, with a peak at $\sim-0.6$ dex and a smooth, gradually declining tail towards lower metallicities, reflecting a steadily decreasing number density.

In Fig. \ref{fig:vz_vphi} we present the velocities ($V_{Z}$ vs $V_{\phi}$) across the $\mathrm{Z_{max}}$--ecc plane. We describe the figure in a similar order to Fig. \ref{fig:mdf_ecc_zmax} in the previous paragraph. In cell G, we observe local, thin disc-like velocities with $\overline{V_{\phi}} = 200$ km/s and a very small dispersion of 13 km/s. This cell also shows the lowest vertical and radial dispersions. As we go to higher $\mathrm{Z_{max}}$ in the same low ecc range, we continue to find disc-like distributions, while the mean $V_{\phi}$ gradually decreases and the dispersions for all three velocity components increase. Additionally, in cells D and A, we also find a small group of stars with negative $V_{\phi}$. This affects the estimates of velocity mean and dispersion for these cells. These 13 stars show strong counter-rotation in the disc with $\overline{V_{\phi}} = -180$ km/s and $\sigma_{V_{\phi}} = 34$ km/s and are all metal-poor ($\feh<-0.5$) with a median of $\sim-1.2$ dex. We refer to Sec. \ref{sec:X4} for discussion on a possible origin of this counter-rotating group. Excluding these counter-rotating stars, we get $\overline{V_{\phi}}$ and $\sigma_{V_{\phi}}$ = 190 km/s and 14 km/s for cell D, and  $\overline{V_{\phi}}$ and $\sigma_{V_{\phi}}$ = 175 km/s and 18 km/s for cell A. For the intermediate eccentricity range, in all three cells (H, E, and B), we observe a distinct disc-like rotational component that shows a decreasing mean azimuthal velocity with increasing $\mathrm{Z_{max}}$ range. Cells H and E, both with $\overline{V_{\phi}} \approx$ = 150 km/s and $\sigma_{V_{\phi}}$ = 50 km/s, appear to include a thick disc-like component. In cell B, the disc-like component with $\overline{V_{\phi}}$ = 80 km/s depicts a much slower rotation and could comprise the hotter extension of the thick disc; however, the MDF of this cell is more metal-poor than that of cells E and H. Additionally, we observe a significant number of stars in the three cells with negative $V_{\phi}$. These stars have a wide range of metallicity ($-2<\feh<0.45$) with a median of -0.7 dex. We would like to remind readers of the multi-component MDF for these cells. Finally, for the high eccentricity bin (ecc > 0.7) in cells I, F, and C, the velocity distribution plots show kinematically hot populations in all three cells reflected by the large values for velocity dispersions. Despite the large $\sigma_{V}$ and the eccentricities, ample azimuthal rotation is still observed in cells I ($\approx60$ km/s) and F ($\approx80$ km/s). For the highest ecc and $\mathrm{Z_{max}}$ cell (C), we observe a small rotational contribution.

Figs. \ref{fig:mdf_ecc_zmax} and \ref{fig:vz_vphi} together reveal the complex interplay of kinematics and chemistry in the Galactic bulge, highlighting the presence of multiple stellar populations occupying overlapping regions of the eccentricity--$\mathrm{Z_{max}}$ plane. At low eccentricity and low $\mathrm{Z_{max}}$ (cell G), stars exhibit thin disc-like characteristics, with a narrow MDF peaking just below solar metallicity and cold kinematics ($\overline{V_{\phi}} \sim 200$ km/s, minimal velocity dispersion). As $\mathrm{Z_{max}}$ increases (cells D and A), metal-poor stars become more prominent, the MDF broadens and shifts to lower metallicities, and velocity dispersions rise, even though a significant disc-like rotation is still retained. These cells also host a subset of strongly counter-rotating, metal-poor stars. In the intermediate eccentricity range (cells H, E, and B), the MDFs become increasingly complex, with evidence of both thick disc-like and metal-poor halo-like populations, accompanied by reduced mean azimuthal velocities and increasing dispersions. Notably, retrograde stars are found across this regime, spanning a wide range in metallicity. At high eccentricity (cells I, F, and C), the velocity distributions are dominated by kinematically hot populations, yet moderate prograde rotation persists in cells I and F, alongside MDFs with multiple peaks and a high fraction of super-solar stars. The high eccentricity and super-solar metallicity stars possibly constitute the galactic bar. In \citetalias{Queiroz2021}, the authors find cells E, F, H, and I to contain a high fraction of stars with bar-supporting orbits. Additionally, the high fraction of metal-poor stars seen in the MDF for the innermost $(l,b)$ bins in Fig. \ref{fig:mdf_in_lb_main} and the metal-poor MDF component with hot kinematics seen in cells I, C, and F across the $\mathrm{Z_{max}}$--ecc plane in Figs. \ref{fig:mdf_ecc_zmax} and \ref{fig:vz_vphi} hint at the possible existence of a metal-poor hot component. Overall, the data clearly illustrate a superposition of chemically and kinematically distinct components—thin disc, thick disc, and bar-supporting stars, and possibly a spheroidal pressure-supported component co-existing within the bulge region. Finally, in \citetalias{Queiroz2021}, a counter-rotating population is identified. As discussed in more detail in Section \ref{sec:X4}, this feature is now more evident—thanks to the complementary information provided by the RVS sample.

As a first attempt to disentangle the bulge components, we explored our data in the $\mathrm{Z_{max}}$--eccentricity plane. However, to fully identify the bulge stellar populations and to characterise their main properties, it becomes necessary to employ the full orbital parametrisation. In the next section, we utilise the orbital frequency analysis methods to identify and characterise the stars supporting the Galactic bar. This is a crucial step towards better isolating the old spheroidal bulge and thick disc components.

\subsection{Elements of the Galactic Bar}\label{sec:bar}

The MW's Galactic bar is a fundamental structure formed from dynamical instabilities in the disc \citep[see e.g.][]{Binney2008}, playing a central role in shaping Galactic dynamics and sculpting the bulge into its characteristic boxy-peanut or X shape. This prominent bar traps stars on specific resonant orbits, such as the bar-supporting $X_1$ family and associated box orbits, which define the bar’s internal phase-space structure \citep[e.g.][]{Contopoulos1980, Pfenniger1984, Voglis2007, Patsis2018, Patsis2019, Portail2015, Valluri2016, Parul2020, Beraldo2023}. Crucially, these distinct orbital families are expected to be linked to chemically distinct stellar populations that make up the bulge and inner disc \citep[e.g.][]{Debattista2017, Fragkoudi2018}. Therefore, understanding the orbital structure and stellar composition of the bar is vital to fully characterise the chemo-kinematic complexity of the bulge and to trace the formation and evolution of the MW.

In this section, we use full orbital analysis to identify stars that support the bar and characterise their chemo-kinematic properties with the goal of isolating the bar population to subsequently investigate the remaining components of the bulge population. 

Orbital frequency analysis is a crucial tool for deciphering the building blocks of barred galaxies (e.g. \citealt{Binney1982, Carpintero1998, Voglis2007, Valluri2010, Vasiliev2013, Portail2015, Valluri2016, Parul2020, Beraldo2023, Tahmasebzadeh2024}; see also \citealt{Koppelman2021, Queiroz2021, Nieuwmunster2024, ZhangH2024} for studies, with observational data, ranging from MW's nuclear stellar disc to the solar neighbourhood). The family of $X_1$ orbits is considered to be the bar-supporting orbit family \citep[e.g.][]{Contopoulos1980, Athanassoula1983}. The $X_1$ are the orbits that close after one revolution and two radial oscillations. We calculated the fast Fourier transform to obtain the orbital frequencies $f_x$, $f_z$ (in Cartesian coordinates), and $f_R$ (the cylindrical radius). We considered an orbit to be bar-supporting if $f_{R}/f_{x} = 2\pm0.1$ \citep[see e.g.][]{Portail2015} and also if the orbit's extent along bar's major-axis is greater than along the minor-axis (i.e. if $X_{\mathrm{max}}/Y_{\mathrm{max}}>1$). For each star in our bulge sample, we obtained a bar-support probability ($0\%<\mathrm{P_{bar}}<100\%$) following our Monte Carlo method for orbital integration (see data section). We note here that our analysis of the bar is limited to the main $X_1$ family of orbits, and we did not consider other near-resonance or higher-multiplicity orbits shown to play a secondary role in the bar’s structure (see \citealt{Patsis2019} and also \citealt{Voglis2007}).

\begin{figure}
    \centering
    \includegraphics[width=0.8\linewidth] {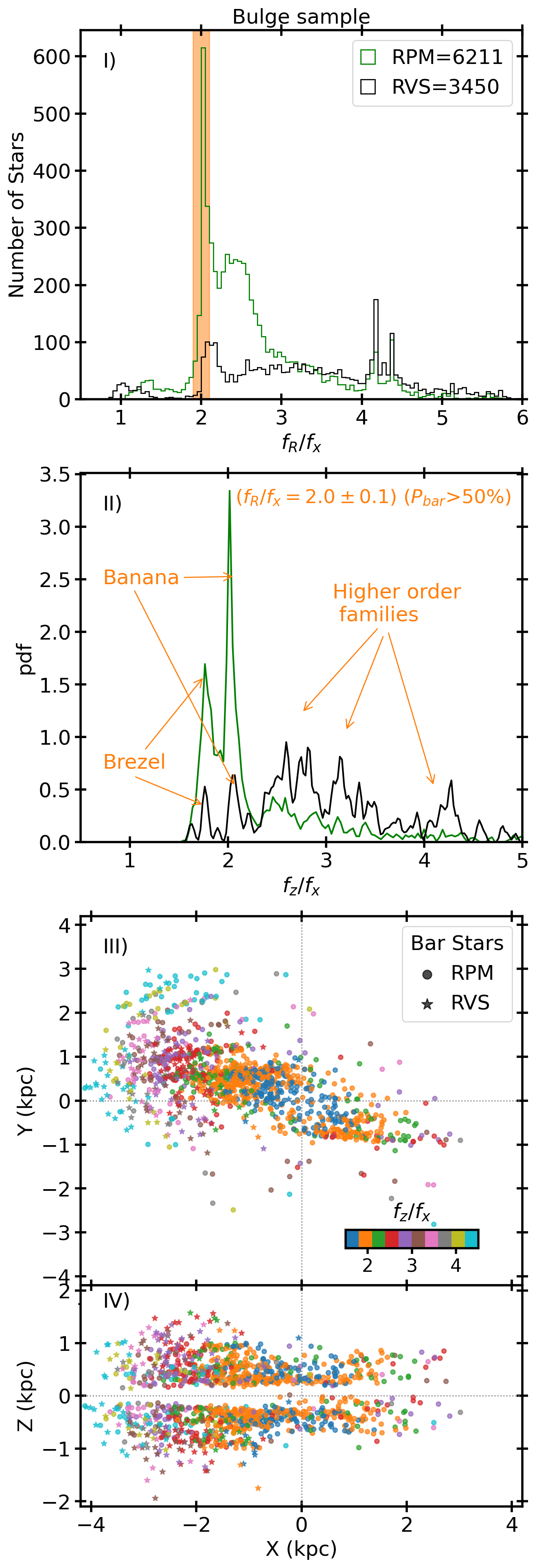}
    \caption{Identification of the bar-supporting stars: I) Distribution of the $f_{R}/f_{x}$ ratio for the bulge sample (RVS-black, RPM-green); II) Distribution of the $f_{z}/f_{x}$ frequency ratios for the bar supporting stars with $\mathrm{P_{bar}}\geq50\%$ (see text for details); III) Current galactocentric positions of the bar stars, in face-on (XY - top) and edge-on (XZ - bottom) view, colour-coded by their $f_{z}/f_{x}$ values.}
    \label{fig:freq_distros}
\end{figure}

In Fig. \ref{fig:freq_distros} panel I), we show the distribution of $f_{R}/f_{x}$ for our bulge sample stars - the orange highlights a large peak around $2\pm0.1$, corresponding to $X_1$ family of orbits. In Fig. \ref{fig:freq_distros} panel II), we present the distribution of $f_{z}/f_{x}$ for stars that support the bar with a condition of $\mathrm{P_{bar}}\geq50\%$. We obtain a total of 1379 (RVS=232, RPM=1147) stars with $\mathrm{P_{bar}}>50\%$ (see Appendix. See \ref{sec:lowPbar} for a discussion of the stars with low probabilities to be on bar orbits, i.e. $0\%< \mathrm{P_{bar}}<50\%$). The $f_{z}/f_{x}$ ratio, i.e. the ratio of frequency along the vertical and bar-major-axis direction, is useful for identifying the different members of the $X_1$ orbit family. The two major members, the banana-like and brezel-like orbits, are prominent and clearly identified, along with a wide range of higher-order members. In panels III) and IV), we present the current galactocentric positions X, Y, and Z of our bar-supporting stars, colour-coded by their respective $f_{z}/f_{x}$ ratios. The identified stars clearly manifest the elongated bar and the $X$ shape in our observed data. The orbits with lower $f_{z}/f_{x}$ ratios reside in the innermost regions of the bar, while those with higher $f_{z}/f_{x}$ ratios constitute the outer regions. We note here that the contribution of the {\it Gaia}-RVS (black) is mostly to the near-side of the bar and the higher-order families, while the RPM (green) -- again owing to APOGEE's H-band spectroscopy -- samples the inner bar regions and captures the building blocks of inner $X$-shape (see also below). Given the complex selection function of APOGEE and {\it Gaia}-RVS in this region, we cannot conclusively quantify the fractional contribution of an orbital family to the overall $X/BP$ shape. In the next section, we analyse the orbital shapes contributed by the different families via face-on and side-on orbital projections and examine their chemical make-up.

\subsubsection{Bar orbital families and their chemistry}\label{sec:bar_chem}

\begin{figure*}[!ht]
    \centering
    \includegraphics[width=0.99\linewidth]{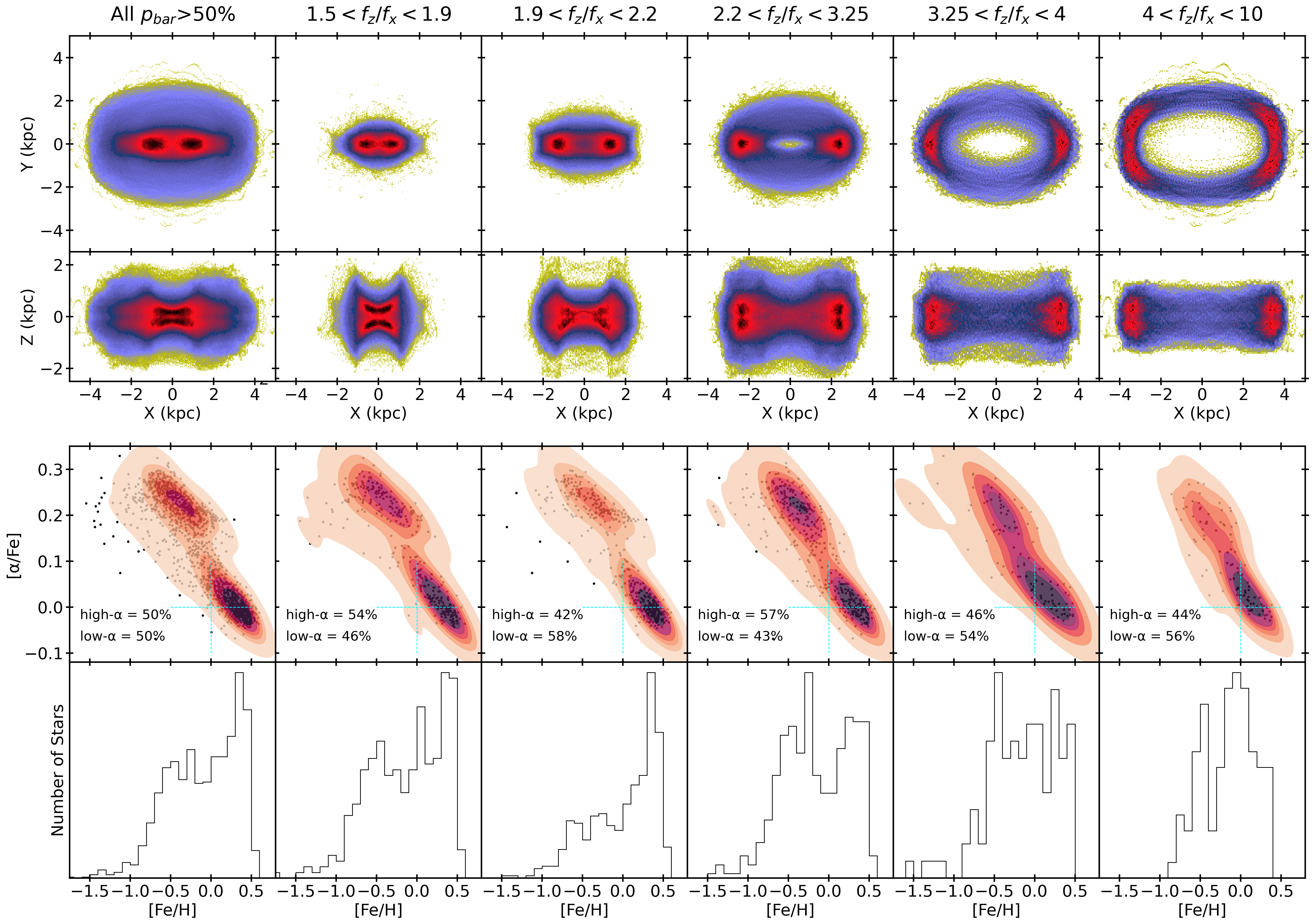}
    \caption{Orbital density projections and chemistry of bar stars in $f_{z}/f_{x}$ ratio bins. First and second rows: Orbital density projections in face-on (XY) and edge-on (XZ) views for different orbital family groups in the $f_{z}/f_{x}$ bins. Third panel: Corresponding \alphafe\, vs \feh\ distribution for each of orbital family groups, along with the fraction of high-\alphafe\ and low-\alphafe\ stars estimated using a limit of \alphafe\=0.1 dex. Fourth row: Corresponding MDFs. This figure highlights the link between orbital dynamics and chemical composition for the galactic bar.}
    \label{fig:chem_orbit_families}
\end{figure*}

In this section, using the $f_{z}/f_{x}$ ratio, we study the contribution of different orbital members to the overall bar shape and the chemical properties of these orbital family members. In Figure \ref{fig:chem_orbit_families}, in the first and second row, we present the orbital density projections in XY and XZ planes for our bar-supporting stars. The left-most panel shows the full bar sample, while the consecutive panels are arranged in increasing order of $f_{z}/f_{x}$ ratio bins.

In the first column of Fig. \ref{fig:chem_orbit_families}, the bar-supporting stars together show strong bar morphology in both the XY and XZ planes. Stars with frequency ratios $1.5<f_{z}/f_{x}<1.9$, belonging to the brezel group of orbits, show a strong central X-shaped structure. The brezel orbits are stars in 5:3 vertical resonances ($f_{z}/f_{x}\approx5/3$) that require ten oscillations in z and six oscillations in x in order to close. This group includes about 23\% of the bar stars. Next, with $1.9<f_{z}/f_{x}<2.2$, we have stars belonging to the x1v1 or the banana group of orbits—identified by their unique projection in the XZ plane. These orbits follow the 2:1 vertical resonance and actually comprise two populations mirroring on the z-axis, i.e. the banana and anti-banana. This group is the largest and includes one-third of the bar stars. Contrary to the brezel, the banana and anti-banana groups appear centrally thin and contribute to an outer $X$-shape. For the range of $2.2<f_{z}/f_{x}<3.25$, we clearly see a boxy-peanut-like shape in the orbital projection. This `boxy-peanut' group has 29\% of our bar stars and is vertically the thickest. Finally, for the ranges $3.25<f_{z}/f_{x}<4.0$ and $4.0<f_{z}/f_{x}<10$, the orbits get vertically thinner and wider in the XY plane, giving a boxy shape in appearance. These orbits support the edges of the bar and include about 8\% and 7\% of our bar stars.

We now turn to the chemical properties of these bar-supporting orbits. In the third row of Fig. \ref{fig:chem_orbit_families}, we present the \alphafe\ versus \feh, and in the fourth row, we show the corresponding MDFs, for both the full bar sample and each orbital group defined above. We use a limit of \alphafe\ = 0.1 dex to separate the high-\alphafe\ and low-\alphafe\ stars. We find that the bar has an equal contribution (50/50\%) of the high- and low-\alphafe\ stars. The MDF shows a high fraction of metal-rich stars with a narrow peak at $\sim+0.4$ dex with a plateau towards lower metallicities up to $\sim-$0.5 and a tail down to $-1.5$ dex. The brezel orbits ($1.5<f_{z}/f_{x}<1.9$) show an 8\% higher contribution from the high-\alphafe\ stars, while the MDF shows three clear peaks at around +0.4, 0.0, and $-$0.5 dex. The `banana' orbits ($1.9<f_{z}/f_{x}<2.2$) have 16\% more low-\alphafe\ stars, and the MDF shows a massive contribution from super-solar metallicity stars. Half of stars in this group have \feh>0 dex. The `boxy-peanut' group ($2.2<f_{z}/f_{x}<3.25$) have 14\% more high-\alphafe\ stars, with the MDF now showing a higher number of metal-poor stars. The higher-order orbit groups show an inversion from the `peanut' group with increasing contribution from the low-\alphafe\ stars. For these groups, most of the stars have $-0.5<\feh<0.5$. The metal-rich peak shifts towards the solar metallicity, while the fraction of both very metal-rich and metal-poor stars is diminished. Interestingly, among the inner bar families, we also find a few metal-poor stars with lower \alphafe\ (around 0.1-0.15 dex) in the more metal-poor range, abundance akin to the GC stars captured by the Galactic bar, as discussed in \citet{Souza2024b}. In Sec. \ref{sec:SpBDB} we discuss a possible reservoir of such stars in the bulge region.

\begingroup
\renewcommand{\arraystretch}{1.5}
\begin{table}[ht]
\caption{Mean azimuthal velocity ($\overline{V_{\phi}}$) as well as azimuthal ($\sigma_{V_{\phi}}$), radial ($\sigma_{V_{R}}$), and vertical ($\sigma_{V_{Z}}$) velocity dispersions for bar-supporting stars across different orbital family groups defined by $f_z/f_x$ frequency ratios.}
\centering
\resizebox{0.48\textwidth}{!}{
\begin{tabular}{|l|c|c|c|c|}
\hline
\textbf{Orbital group} & $\boldsymbol{\overline{V_{\phi}}}$ [km/s] & $\boldsymbol{\sigma_{V_{\phi}}}$ [km/s] & $\boldsymbol{\sigma_{V_{R}}}$ [km/s] & $\boldsymbol{\sigma_{V_{Z}}}$ [km/s] \\ [0.8ex]
\hline\hline
All ($\mathrm{P_{bar}} > 50\%$)      & $108 \pm 2$ & $57 \pm 1$ & $71 \pm 1$ & $63 \pm 1$ \\[0.5ex]
\hline
$1.5 < f_z/f_x < 1.9$  & $70 \pm 3$  & $53 \pm 2$ & $79 \pm 3$ & $71 \pm 2$ \\
$1.9 < f_z/f_x < 2.2$  & $85 \pm 2$  & $48 \pm 2$ & $79 \pm 2$ & $67 \pm 2$ \\
$2.2 < f_z/f_x < 3.25$ & $132 \pm 2$ & $40 \pm 1$ & $66 \pm 2$ & $58 \pm 2$ \\
$3.25 < f_z/f_x < 4.0$ & $165 \pm 2$ & $22 \pm 2$ & $38 \pm 2$ & $44 \pm 3$ \\
$4.0 < f_z/f_x < 10$   & $180 \pm 2$ & $17 \pm 1$ & $39 \pm 3$ & $35 \pm 2$ \\
\hline
\end{tabular}
}
\label{tab:orb_kinematics}
\end{table}
\endgroup

In addition to the morphological and chemical differences discussed above, we also examine the overall kinematic properties of the various orbital groups. As shown in Table \ref{tab:orb_kinematics}, we find that the velocity dispersion systematically decreases with increasing frequency ratio, while the mean azimuthal velocity ($V_{\phi}$) increases. This trend reflects the expected increase in rotational velocity with radius: orbits with lower frequency ratios are confined to the inner bar, whereas those with higher ratios contribute to the outer bar structure. 

While the total bar population shows roughly equal contributions from the high- and low-\alphafe\ populations, their relative proportions vary significantly across different orbital families. The MDF for the `banana' orbits, comprising the largest fraction of super-solar metallicity stars, is consistent with being generated via `vertical bifurcation' of the in-plane and 2D parent $X_1$ orbit \citep[e.g.][]{Contopoulos2013}. Similarly, the higher fraction of metal-poor and high-\alphafe\ stars in the `brezel' or `peanut' groups, coupled with the vertical metallicity gradient we observe in the inner-thick disc (see Sec. \ref{sec:SpBDB}, can provide observational constraints for the mechanisms responsible for bar thickening \citep[see][]{Quillen2002, Quillen2014, Sellwood2020}. \citet{Debattista2017} introduced the concept of `kinematic fractionation' where the Galactic bar's secular evolution segregates stellar populations based on their radial random motions. This process results in metal-rich stars tracing the X-shaped structure, while older, metal-poor stars populate a more boxy bulge morphology. Furthermore, \citet{Debattista2020} demonstrated that the vertical thickening of stellar populations increases monotonically with the radial action of stars from before the bar formed, reinforcing the connection between orbital dynamics and chemical properties. Looking ahead, the availability of larger and more comprehensive stellar samples will allow these chemodynamical observables to be used more effectively to constrain the specific processes involved in bar formation.

\subsubsection{The $X_4$ retrograde orbital family.}\label{sec:X4}

\begin{figure*}[!ht]
    \centering
    \includegraphics[width=.8\linewidth] {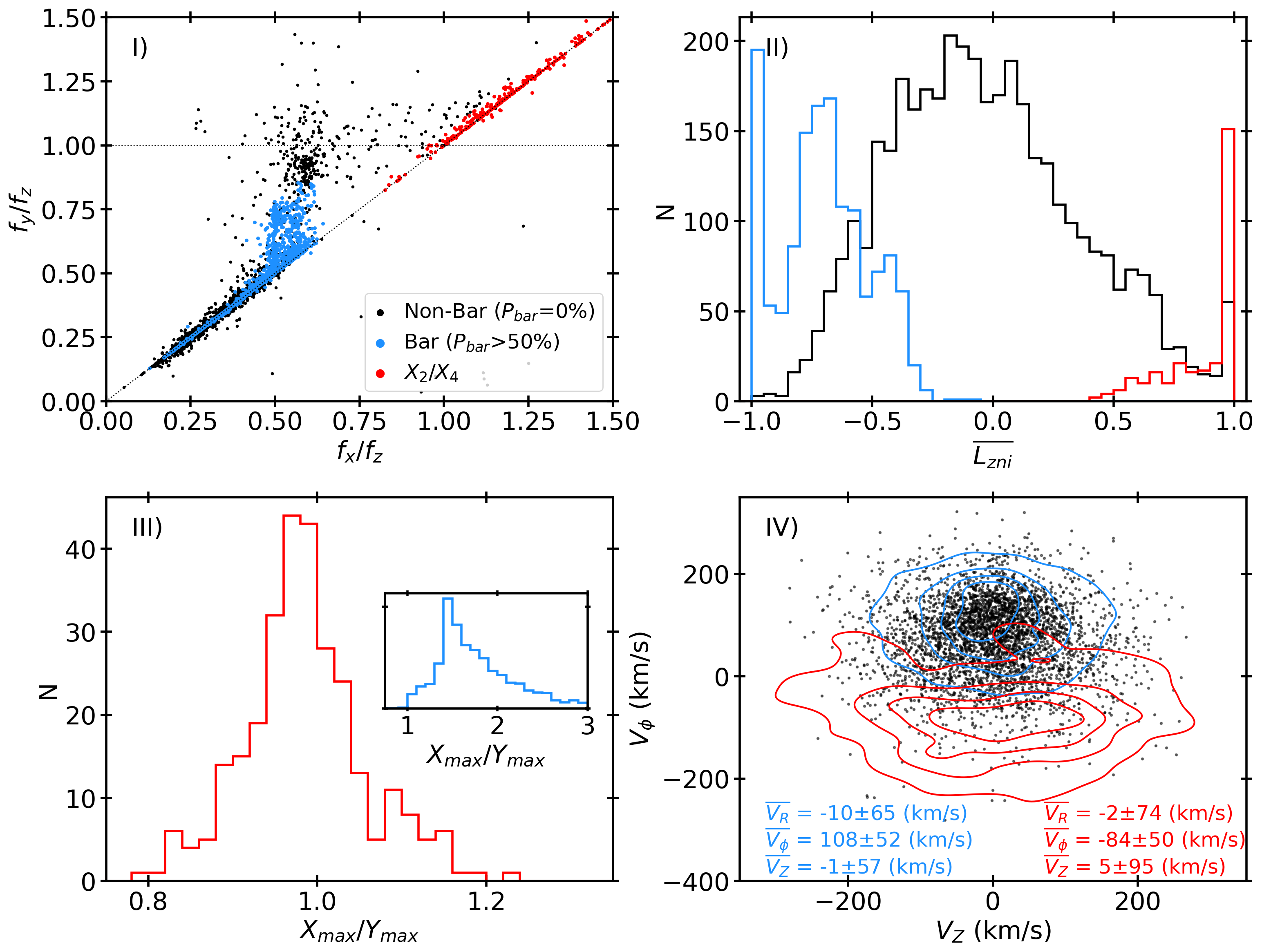}
    \caption{Identification of $X_4$ orbits: I) Frequency map, $f_{y}/f_{z}$ vs $f_{x}/f_{z}$, showing the bar-supporting stars (blue), the $X_4$ orbits (red), and the non-bar stars (black); II) Distribution of the net angular momentum parameter ($\overline{L_{\mathrm{zni}}}$) in the non-inertial reference frame; III) Distribution of the $X_{\mathrm{max}}/Y_{\mathrm{max}}$ ratio for the $X_4$ orbits. The inset shows the same for the bar stars. IV) $V_{\phi}$ vs $V_{Z}$ distribution for the $X_4$ and the bar stars shown as contour plots, along with mean velocities, galactocentric velocities, and their respective dispersions. The non-bar stars are shown as black points.}
    \label{fig:fmap_x4}
\end{figure*}

The $X_4$ family consists of short-axis (z) tube orbits parented by 1:1 periodic orbits in the x–y plane—similar to the $X_1$ family. However, they are retrograde around the z-axis in the bar's rotating frame and have their main axis perpendicular to the main axis of the bar \citep{Contopoulos2013, Valluri2016}. The $X_4$ family does not support the bar. Studies with N-body models of barred galaxies predict a low number of $X_4$ orbits ($\sim1.5\%$), while their prograde counterparts, the $X_2$ orbits, are expected to be absent \citep[e.g.][]{Voglis2007, Valluri2016}. Additionally, the $X_4$ orbits are expected to be elongated along the y-axis of the bar at smaller apocentres and get rounder outwards \citep{Sellwood1993}.

In Fig. \ref{fig:fmap_x4}, panel I, we present the Cartesian frequency map for our bulge sample stars\footnote{There is a difference in the distribution of $\overline{L_{\mathrm{zni}}}$ for RVS and RPM samples. This difference arises from the fact that the RPM sample is in the NIR - has a higher fraction of orbital families ($1.5<f_{z}/f_{x}<2.2$) supporting inner bar structure -- i.e. the $X$-shape orbits. The RVS sample, on the other hand, has a higher fraction of orbital families contributing to the outer bar component. This is also reflected in the mean distribution of $V_{\phi}$ of stars in bar-shaped orbits for the two samples -- i.e. a larger mean $V_{\phi}$ for RVS compared to that of the RPM sample.}. The blue points represent the bar-supporting stars identified and discussed in the previous section. Along the diagonal line, and at the upper right corner, the red points represent the $X_4$ orbits in our bulge sample. Despite a low fraction, these orbits are easily identified in the frequency maps \citep[see][]{Valluri2016}. The identified 287 stars constitute $\sim3\%$ of the bulge sample and $\sim1.5\%$ of the full inner-galaxy sample. This fraction is consistent with the estimation from the N-body models \citep[e.g.][]{Valluri2016}. 

In panel II, to confirm the retrograde nature of these $X_4$ orbits in the bar frame, we present the distribution of the net rotation parameter. The net rotation parameter is defined as \(\overline{L_{\mathrm{zni}}} = \frac{\Sigma_{i=1}^{i=N} sign(L_{\mathrm{zni}})_i}{N} \), where N is the number of time steps. For each time-step, for positive and negative values of $L_{\mathrm{z}}$, sign($L_{\mathrm{z}}$) is $+$1 and $-$1, respectively \citep[see][]{Tahmasebzadeh2024}. Here, $\overline{L_{\mathrm{zni}}}$ represents the net rotation parameter around the z-axis in the non-inertial frame, and $-$1 ($+$1) corresponds to full prograde (retrograde) rotation along the bar's rotation. We find that the bar-supporting stars of the $X_1$ family (blue) and the identified $X_4$ family (red) have opposite $\overline{L_{\mathrm{zni}}}$ distributions. As all our selected stars have net rotation opposite to the bar, we have no $X_2$ stars in our selection, as expected according to the models discussed above.

\begin{figure}[!ht]
    \centering
    \includegraphics[width=0.99\linewidth] {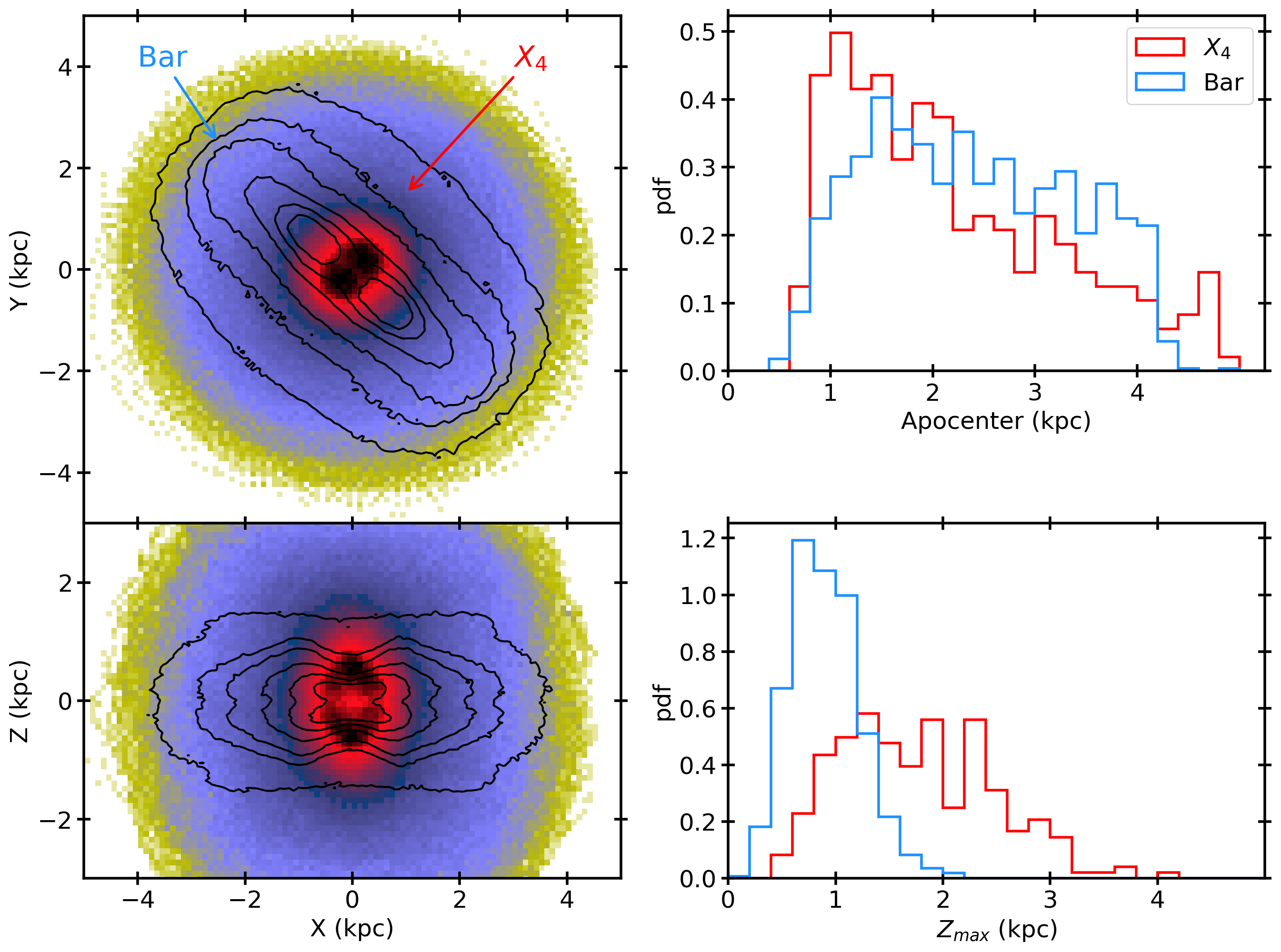}
    \caption{Morphology of $X_4$ orbit stars. Left: Orbital density projections in $XY$ and $XZ$ planes for the $X_4$ orbits compared to the $X_1$ orbits (black contours), tracing the bar. Right: Comparison of the apocenter and $\mathrm{Z_{max}}$ distributions for the two orbital families.}
    \label{fig:orbit_x4}
\end{figure}

In panel III we present the distribution for $X_{\mathrm{max}}/Y_{\mathrm{max}}$, where $X_{\mathrm{max}}$ and $Y_{\mathrm{max}}$ are the maximum extent of the orbit along bar major-axis and minor-axis, respectively. Most of the $X_4$ orbits are elongated along the bar minor-axis, while the bar-supporting stars have substantial elongation along the major-axis (see also Fig. \ref{fig:orbit_x4}). 

Finally, in panel IV of Fig. \ref{fig:fmap_x4}, we present the kinematic properties of the $X_4$ family and compare it with the bar. The galactic bar has a mean azimuthal velocity of $\approx110$\,\kms\ , while the $X_4$ family has $\overline{V_{\phi}}\approx-85$\,\kms\,, reflecting a strong counter-rotation in the kinematic space. The two groups show similar dispersions for the radial ($\approx70$ \kms) and azimuthal ($\approx50$ \kms) velocities, while the $X_4$ family has much higher vertical velocity dispersion of 95 \kms. These kinematical properties are similar to those of the counter-rotating stars seen in Fig. \ref{fig:vz_vphi} and in \citetalias{Queiroz2021}. The latter identified a counter-rotating population in the MW bulge, characterised by stars with negative azimuthal velocities ($V_{\phi} < -$50 km/s). This population was found to be predominantly metal-poor. However, the authors noted that the statistical significance of this result was marginal and suggested that larger data samples would be necessary to confirm the existence and better characterise the properties of this counter-rotating component.

\begin{figure}[!ht]
    \centering
    \includegraphics[width=0.9\linewidth] {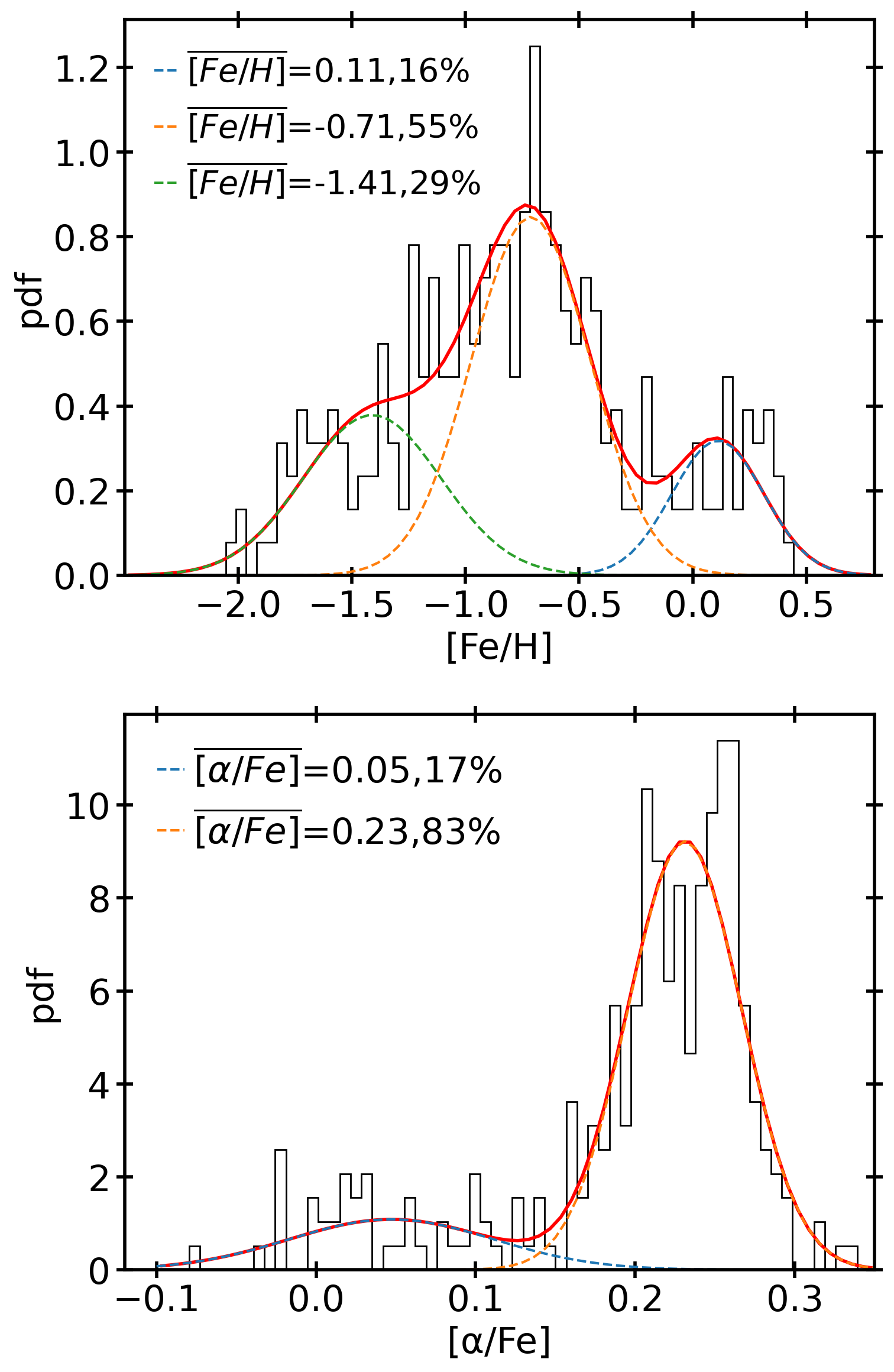}
    \caption{MDF (top) and \alphafe\ (bottom) distribution for the 287 $X_4$ stars, along with the main components identified using GMM. For each GMM component, the mean and the respective weights are listed.}
    \label{fig:gmm_x4}
\end{figure}

We now investigate the chemical properties of this counter-rotating stellar population. In Fig. \ref{fig:gmm_x4} we present the MDF and \alphafe\ distribution for the $X_4$ family along with the respective components identified via Gaussian mixture modelling (GMM). The MDF reveals a predominantly metal-poor population requiring a three-component fit with peaks at $\feh\approx -1.40, -0.70, \mathrm{and} +0.10$\ dex. The metal-poor peak at $-$0.70 has the highest contribution of 55\%, followed by 29\% for the very metal-poor peak at $-$1.40 dex, and finally a low fraction (16\%) for the metal-rich stars. Both low-metallicity peaks are at lower values compared to the (chemical-old) thick disc \citep{Miglio2021, Queiroz2023}. The \alphafe\ distribution shows that most stars (83\%) have high-\alphafe\ with a low fraction (17\%) of low-\alphafe\ stars. The fractions of low-\alphafe\ and the metal-rich component are consistent in both distributions. The two metal-poor components of the MDF have high-\alphafe\ abundances, and upon visual inspection we find two high-\alphafe\ peaks $\sim$0.05 dex apart. However, the GMM only fits one solution owing to a small difference in the \alphafe. 

Overall, we find that the metallicity and \alphafe\ abundance of stars on $X_4$ orbits do not match those of the bar-supporting stars discussed in Sec. \ref{sec:bar_chem}, suggesting that they originate from a different stellar population. Fig. \ref{fig:orbit_x4} complements this picture showing that, although both components have apocenter below 4 kpc, the $X_4$ orbits are more concentrated towards the inner parts but extend to larger $\mathrm{Z_{max}}$ (up to around 3 kpc).

The presence of $X_4$ orbits is tied to the existence of the Galactic bar, and their retrograde motion is a direct consequence of the bar’s dynamical influence \citep{Voglis2007, Contopoulos2013, Valluri2016, Abbott2017}. Although stable periodic $X_4$ orbits can trap stars for extended times and form a perpendicularly oriented orbital family \citep{Contopoulos2013}, simulations consistently show that the region around these orbits is sparsely populated. This is mainly because their geometry does not align with the elongated structure of the bar, making them inefficient in supporting it. As a result, they typically account for only a low fraction—about 1.5\% to 2\%—of the total orbital population in pure bar models \citep{Valluri2016, Abbott2017}. Due to their retrograde nature and misalignment with the bar's prograde motion, stars trapped in $X_4$ orbits are more likely to come from hotter, dynamically less organised populations such as the Galactic halo, a pressure-supported bulge, and the thick disc (more metal-poor), rather than from the colder disc stars that form the bar's backbone \citep{Contopoulos2013, Sellwood2014RvMP, Valluri2016}.

The counter-rotating population suggested by \citetalias{Queiroz2021} is now more evident thanks to the complementary information provided by the RVS sample (better sampling low metallicities and still reaching the $X_4$ orbits of stars in the nearby side of the bar, as shown in Fig. \ref{fig:freq_distros}). In summary, our analysis shows that this counter-rotation population is largely associated with stars on $X_4$ orbits in the innermost regions of the Galaxy. Importantly, this population does not require an external origin such as accretion; rather, it naturally arises from bar-driven dynamics in the presence of an early, spheroidal bulge component. To the best of our knowledge, this is the first time a connection between the MW’s bar and its spheroidal bulge has been demonstrated in this manner with spectroscopic observations.

\subsection{Evidence for the pressure-supported spheroidal bulge of the MW}\label{sec:SpB}

\begin{figure}[!ht]
    \centering
    \includegraphics[width=0.8\linewidth] {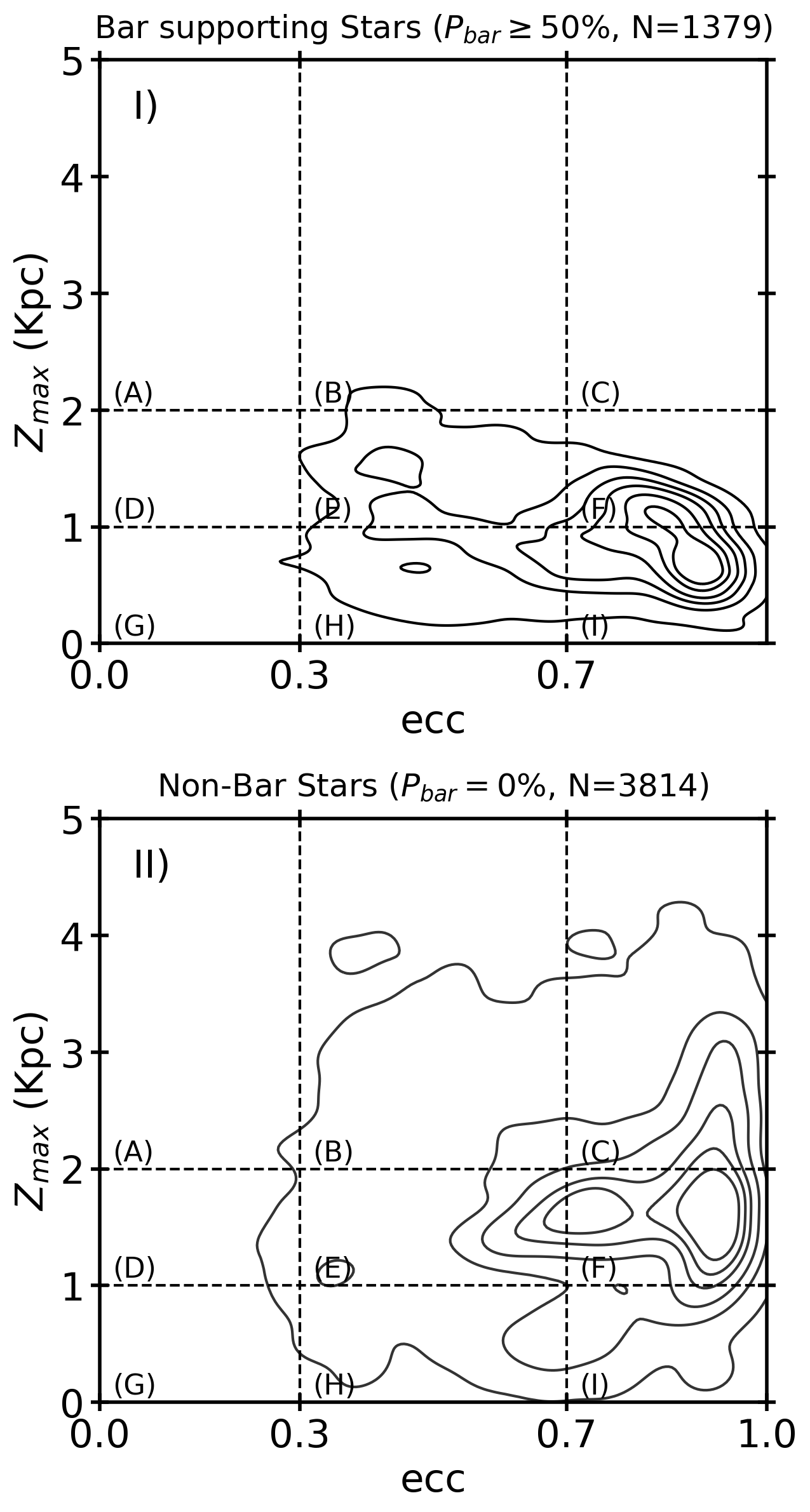}
    \caption{Distribution of bar (top) and non-bar (bottom) stars in the $\mathrm{Z_{max}}$--ecc plane, shown as KDEs. The KDE for the non-bar stars clearly shows the presence of at least two density groups. The distributions are very different and reflect the different kinematic groups co-existing in the Bulge regions.}
    \label{fig:bar_nb_ecc_zmax}
\end{figure}

\begin{figure*}[!ht]
    \centering
    \includegraphics[width=0.99\linewidth] {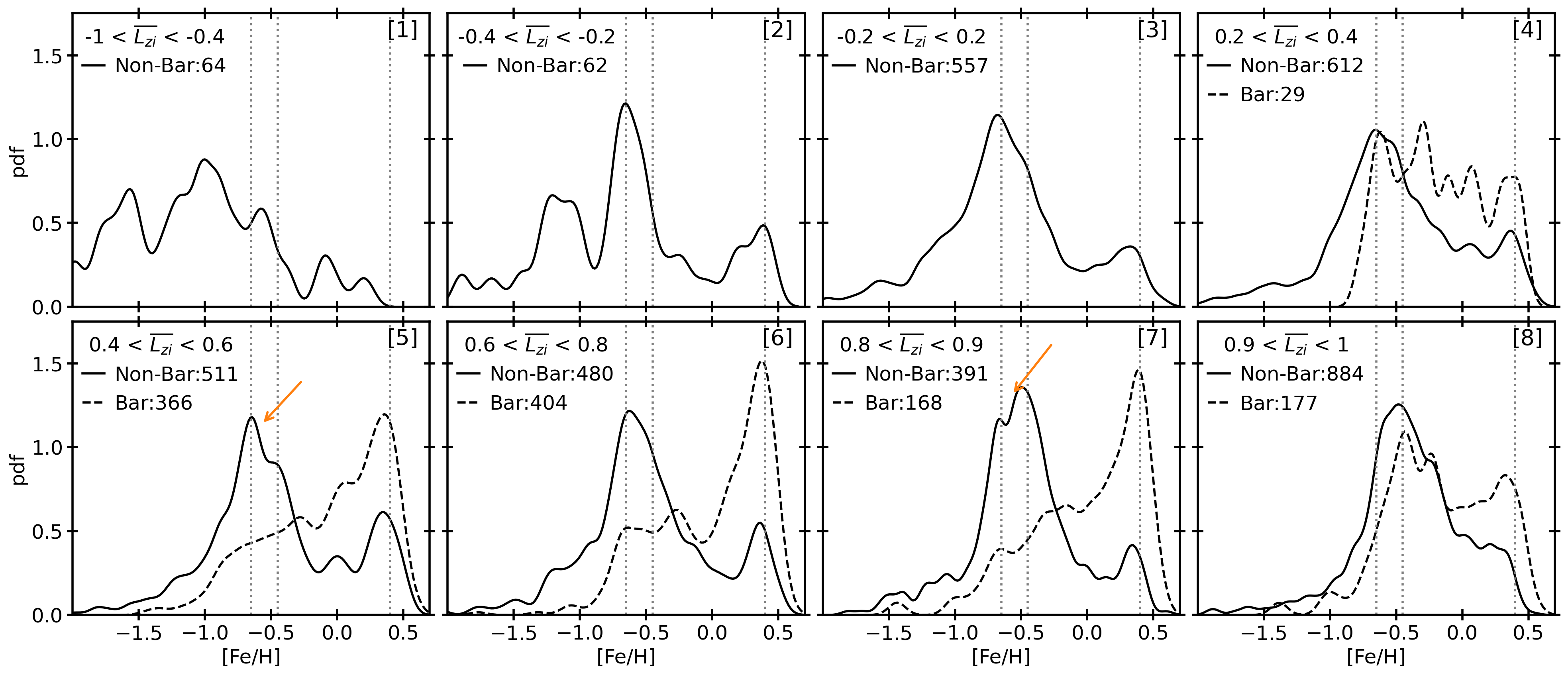}
    \caption{MDFs for bar (dashed line) and non-bar (solid line) stars in bins of the net angular momentum parameter in inertial frame ($\overline{L_{\mathrm{zi}}}$). The three vertical red lines, at \feh of $-$0.65, $-$0.5, $-$0.07, and $+$0.35 dex, are drawn to guide the location of main peaks in the MDFs, while the arrows in panel 5 and panel 7 highlight the change in the MDF peaks. For the non-bar population, the main MDF peak shifts from $-$0.65 dex, for the slowly non-rotating old bulge population, to $-$0.5 dex for the rotating thick-disc-dominated population.}
    \label{fig:mdfs_Lzni}
\end{figure*}

\begin{figure}
    \centering
    \includegraphics[width=0.8\linewidth] {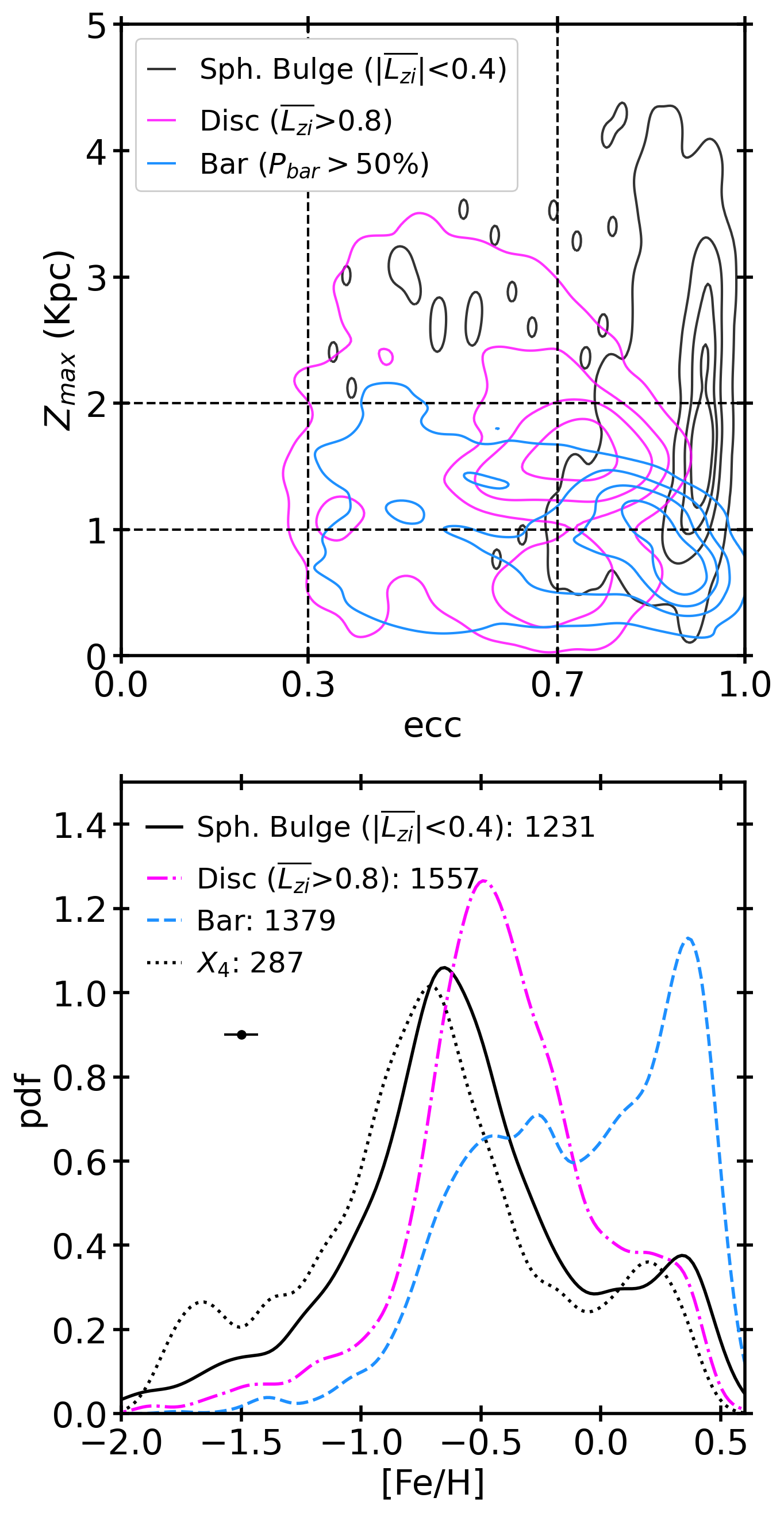}
    \caption{Dissection of the ‘non-bar’ stars into inner-thick disc and spheroidal bulge populations. Top: $\mathrm{Z_{max}}$--ecc plane showing distribution of the spheroidal bulge (black), disc (magenta), and bar (blue). Bottom: MDFs of the spheroidal bulge, thick disc, and bar components. Note the very similar MDF of the spheroidal bulge and the population of stars in $X_4$ orbits, and how distinct these are from the bar population. All non-bar populations seem to be contaminated by stars in bar-shaped orbits at high \feh.}
    \label{fig:3components}
\end{figure}

In the previous section, we used orbital frequency analysis to identify the bar-supporting stars, classified them into different orbital groups, and studied their chemodynamical properties. Additionally, we identified stars in the $X_4$ family of orbits, which exist due to the Galactic bar but have a different chemistry compared to the bar stars. In this section, we now study the rest of the stars in our bulge sample, which we call `Non-Bar' stars, defined with $\mathrm{P_{bar}}=0\%$\footnote{For a discussion of stars with a low bar-support probability, i.e. $0\%<\mathrm{P_{bar}}<50\%$, please refer to Appendix \ref{sec:lowPbar}}.

In Figure \ref{fig:bar_nb_ecc_zmax}, we present the distribution of the `bar' and `non-bar' stars in the $\mathrm{Z_{max}}$--ecc plane. We see that the bar-supporting stars mostly have high eccentricity and are confined to $\sim1.5$\,kpc from the galactic plane. These bar stars populate the cells E, F, H, and I of the $\mathrm{Z_{max}}$--ecc plane, as discussed in previous Sec. \ref{sec:zmaxecc}. In panel II), the kernel density estimate plot reveals the existence of two main density groups among the non-bar stars. One group includes the highest eccentricity orbits ($ecc>$0.9) and extends to larger distances ($\approx3$\,kpc) from the galactic plane compared to the bar. The second group has a peak in ecc $\approx0.7$ and is confined to $1 \lesssim  Z_{\mathrm{max}}\lesssim 2$\, kpc. The first group maps to the high velocity-dispersion dominated cells C and F, and the second group overlaps at cell F to continue towards cell E with a thick disc-like velocity distribution (see Figs. \ref{fig:mdf_ecc_zmax} and \ref{fig:vz_vphi} in Sec. \ref{sec:zmaxecc}). We have seen that cell C has a bi-modal MDF with both very metal-rich and metal-poor stars, while cells C and E contain mostly metal-poor stars.

Using these lines of evidence, we can deduce the existence of two additional components in the bulge region besides the bar, i.e. an extended pressure-supported component and a thick disc-like component. To further characterise these two additional components and to confirm if they are also chemically and kinematically distinct, we now study these stars in bins of the net rotation parameter $\overline{L_{\mathrm{zi}}}$ in the inertial frame.

In Figure \ref{fig:mdfs_Lzni} we present the MDFs for the non-bar (solid) and the bar (dashed) stars in bins of the net rotation parameter $\overline{L_{\mathrm{zi}}}$ in the inertial frame. A star with $\overline{L_{\mathrm{zi}}} \sim +1$ would indicate a fully prograde motion around the z-axis in the inertial frame at every sampling during orbit integration, and $\overline{L_{\mathrm{zi}}} \sim -1$ represents a fully retrograde orbit. The panels are arranged in increasing order of $\overline{L_{\mathrm{zi}}}$ ranges and are labelled [1] to [8]. The three vertical lines in the MDF plots at $-$0.65, $-$0.45, and $+$0.40 dex are shown to guide the readers. Panel [1] includes stars with retrograde orbits, which are mostly metal-poor ($\feh<-0.5$) with multiple peaks prominently at $\approx-1.0$ and $\approx-1.5$ dex. Panel [2], for $-0.4<\overline{L_{\mathrm{zi}}}<-0.2$, shows a prominent peak in the MDF at $-0.65$ dex, along with two smaller peaks at $\approx-1.0$ and $+0.4$ dex, respectively. Panel [3], with $-0.2<\overline{L_{\mathrm{zi}}}<+0.2$, includes the most non-rotating stars and shows a sharp increase in the number of stars, with the MDF having a large peak at $\approx-0.65$, a tail towards lower metallicity, and a small bump at $+0.4$ dex. The panel [4] MDF appears similar to that of panel [3] for the non-bar stars, with a major peak at $\approx-0.65$ dex, and a low number of mostly metal-poor bar stars are present. From panels [5] to [7], for the non-bar stars, we observe a gradual change in the metal-poor peak of the MDF from $\approx-0.65$ to $\approx-0.45$ dex, leading to the peak at $\approx-0.45$ for the most prograde bin in panel [8], along with a larger fraction of sub-solar metallicity stars. Additionally, in panel [8], we find that the metal-poor peak of the bar MDF coincides with that of the non-bar stars at $\approx-0.45$ dex. Interestingly, in panels [5] to [7], the MDFs for the bar-supporting stars are strikingly different compared to the non-bar stars, supporting a scenario of origin from different stellar populations for the bar and the non-bar stars.

Using the net rotation parameter, we find that the MDF gradually changes with a metal-poor peak at $\approx-0.65$ dex for the stars with no net rotation in the inertial frame (i.e. $|\overline{L_{\mathrm{zi}}}|<0.4$) towards a peak at $\approx-0.45$ dex for the stars rotating as a disc. This $\feh \approx-0.45$ also corresponds to the local thick disc MDF peak. The two peaks differ by $\sim 0.2$ dex, which is about $2\sigma$ larger than the combined measurement uncertainty for our \feh\ (see Fig. \ref{fig:3components}). This suggests that the difference is statistically significant and robust against measurement uncertainty, providing evidence for the existence of a chemically distinct non-rotating component in the galactic bulge region. For $\overline{L_{\mathrm{zi}}}\gtrsim0.4$, we find the bar-supporting stars. To robustly classify the non-rotating component and the disc components and study their properties, we adopted $|\overline{L_{\mathrm{zi}}}|<0.4$ and $\overline{L_{\mathrm{zi}}}>0.8$, respectively.

In the top panel of Fig. \ref{fig:3components}, we show the position of the bar, the disc, and the non-rotating component in the $\mathrm{Z_{max}}$--eccentricity plane. They occupy distinct locations in the plane owing to their distinct orbital properties. In the bottom panel, we present the combined MDFs of the three components, which are distinct. Following these lines of evidence, we confirm the existence of a non-rotating component (i.e. the spheroidal pressure-supported Bulge) in the bulge region, which is distinct from the bar and the inner-thick disc. To guide the readers through this intricate exploration and the subsequent characterisation, in Appendix \ref{sec:flowchart} we present the scheme summarising the classification of the stars into the three inner galaxy components. In the next section, we summarise the chemodynamical properties of the three components. This work represents the first statistically robust separation of bulge stellar populations for field stars (noting that the classification remains probabilistic, given the continuous nature of the underlying distributions, and therefore some contamination is unavoidable).

\subsection{Chemodynamical properties of the spheroidal bulge, the inner-thick disc, and the Galactic bar}\label{sec:SpBDB}

\begin{figure*}[!ht]
    \centering
    \includegraphics[width=0.99\linewidth] {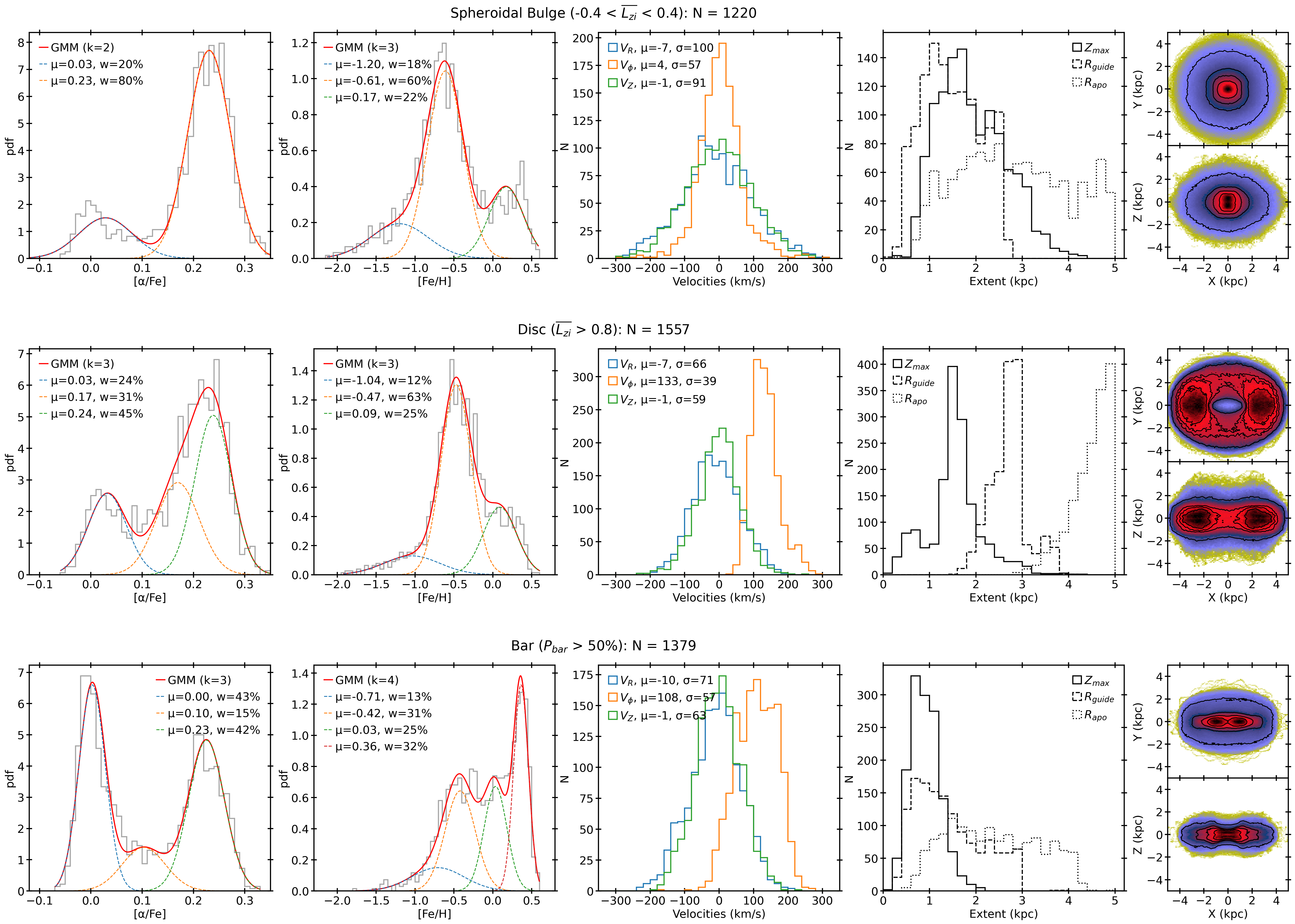}
    \caption{Spheroidal bulge, inner-thick disc, and bar. Top row: From left to right, the \alphafe\ distribution and the MDF, both decomposed using GMM; the distribution of the three galactocentric velocity components; the spatial extent derived from the orbital parameters $\mathrm{Z_{max}}$, $\mathrm{R_{apo}}$, and $R_{\mathrm{guide}}$ (the guiding centre radius); and the orbital density projections in face-on (XY) and edge-on (XZ) views. The mean of the GMM components and their respective weights are listed. The mean of the velocities and the respective dispersions are also listed. Middle and Bottom row: Same as above, but for the inner-thick disc and bar, respectively.}
    \label{fig:bulge_disc}
\end{figure*}

\begin{figure*}[!ht]
    \centering
    \includegraphics[width=0.9\linewidth] {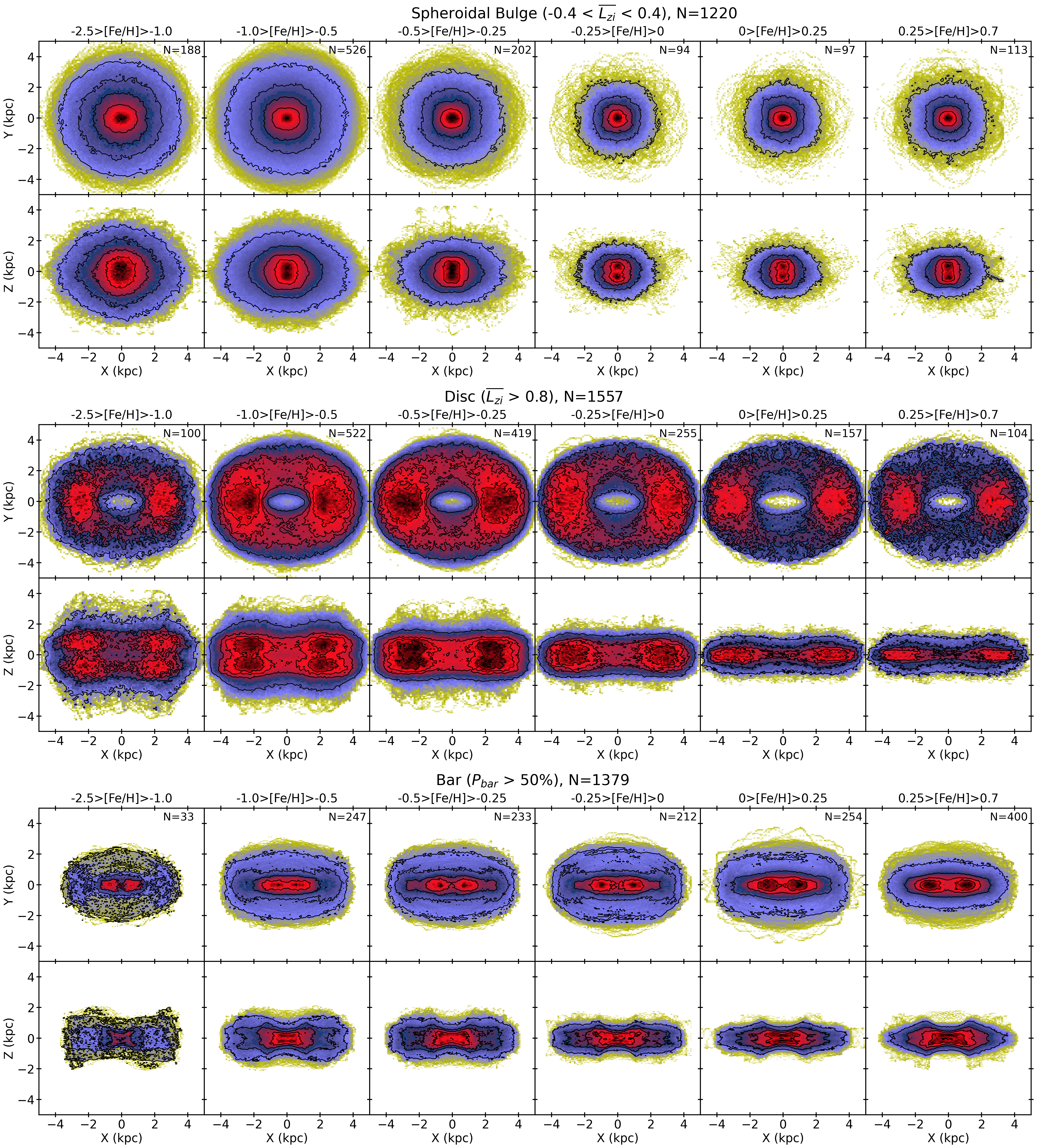}
    \caption{Face-on (XY) and edge-on (XZ) orbital density projections for the spheroidal bulge (top row), inner thick disc (middle row), and bar (bottom row), shown across increasing metallicity bins from left to right. Each panel includes contours marking the 25\%, 50\%, 75\%, and 95\% orbital density levels.}
    \label{fig:orbits_in_feh}
\end{figure*}

In Figure \ref{fig:bulge_disc}, we present the chemical, kinematic, and spatial properties of the three Galactic bulge components. For each component, a row of panels (from left to right) shows the \alphafe\ distribution and the MDF, both decomposed using GMM; the distribution of the three galactocentric velocity components; the spatial extent derived from the orbital parameters $\mathrm{Z_{max}}$, $\mathrm{R_{apo}}$, and $R_{\mathrm{guide}}$; and the orbital density projections in face-on (XY) and edge-on (XZ) views.

\paragraph{Spheroidal bulge:} The spheroidal bulge component primarily consists of moderately metal-poor stars with high-\alphafe\ abundances. About 80\% of the stars have high-\alphafe\ ($\mu\approx0.25$), and the rest have low-\alphafe\ ($\mu\approx0.0$) abundances. The MDF can be fitted by three Gaussian components. The main component with 60\% weight exhibits a peak at $\sim-0.60$ dex. The other two MDF components, each with $\approx20\%$ weight, have GMM means at $-1.20$ dex and $\approx0.20$ dex, respectively\footnote{These are similar to the halo and the SMR component of the bar and hint at a possible contamination from the halo and the bar.}. The stars show very high radial and vertical velocity dispersions with $\sigma_{V_{R}} \approx \sigma_{V_{R}}\approx100$\kms\,, which support our classification of this group as the pressure-supported spheroidal bulge (slightly triaxial; see discussion below) component and are comparable to literature values (e.g. \citealt{Arentsen2024} and references therein). The azimuthal velocity shows a narrow distribution centred at 0 \kms\ with a smaller dispersion $\approx60$\kms\,, due to the steps we adopted, which include: a cut in the apocenter radius to remove disc and halo contaminants; the removal of the bar and the inner-thick disc stars; and a limit of the net angular momentum parameter, favouring the exclusion of stars with high positive and negative rotational velocities from the parent population. The distribution of the orbital parameters $\mathrm{Z_{max}}$, $\mathrm{R_{apo}}$, and $R_{\mathrm{guide}}$ and the orbital density projections reveal a slightly oblate spheroidal shape, which extends $\approx3$ kpc from the galactic plane. The average radii (i.e. $R_{\mathrm{guide}}$) also show a similar distribution to $\approx3$ kpc. We measure overall vertical and radial metallicity gradients of $\nabla[Fe/H]/\nabla{Z_{\mathrm{max}}} = -0.35\pm0.02$ dex/kpc and $\nabla[Fe/H]/\nabla{R_{\mathrm{apo}}} = -0.19\pm0.01$ dex/kpc\footnote{These gradient measurements, however, require larger samples to be robust and will be the topic of a forthcoming paper.\label{fn:grad}}. The $\mathrm{R_{apo}}$ show an extension in the radial direction, typical of triaxial components (see discussion in the next sections).

\paragraph{Inner-thick disc:} The inner-thick disc, shown in the middle row of Fig. \ref{fig:bulge_disc}, also contains primarily metal-poor, high-\alphafe\ stars but with notable differences in its chemical makeup. The \alphafe\ distribution is best fitted by three components, including a distinct intermediate-\alphafe\ group centred at $\sim$0.15 dex, in addition to peaks at $\sim$0.0 and $\sim$0.25 dex, similar to the spheroidal bulge. The disc MDF shows a higher number of stars with sub-solar metallicity (i.e. $-0.5<\feh<0.0$), and the GMM decomposition yields peaks at [Fe/H] $\sim -1.0$, $\sim -0.45$, and $\sim +0.10$ dex, with the intermediate component ($\sim -0.45$ dex) dominating ($\sim60\%$ weight). This main peak is at a lower value than that of the spheroidal bulge. The most metal-poor MDF component centred at $\approx-1.0$ dex is at slightly higher values and of lower weight compared to the spheroidal bulge. The appearance of the sub-solar \feh\ stars with intermediate \alphafe\ abundance hints at the slowdown of the early high and bursty star-formation, which had contributed to the sustained Supernovae (SNe) II production of alpha elements as the disc saw increased \feh\ enrichment. This gives rise to the steeper decrease of \alphafe\ with \feh. The disc has a mean $V_{\phi}$ of 133 \kms\ and velocity dispersions $\sigma_{V_R}=66$, $\sigma_{V_{\phi}}=39$, and $\sigma_{V_{Z}}=59$\,\kms\,, which are significantly lower than the spheroidal bulge. Compared to the bar (see the next paragraph), the disc has $\overline{V_{\phi}}$ higher by $\approx25$\kms\,, and $\sigma_{V_{\phi}}$ lower by $\approx20$\kms\ , while the vertical and the radial dispersions are similar. Most disc stars lie within $Z_{\mathrm{max}} < 2$\ kpc, with a concentration in the range $1 \lesssim Z_{\mathrm{max}} \lesssim 2$\ kpc, and extend radially beyond the bar, with $R_{\mathrm{apo}} > 4$\ kpc. A small group of stars are confined to $Z_{\mathrm{max}}<1$\ kpc, which could indicate contamination from the bar or could belong to the thin disc. We measure an overall vertical metallicity gradient of $\nabla [Fe/H]/\nabla Z_{\mathrm{max}} = -0.35\pm0.01$ dex/kpc\footref{fn:grad}. The orbital projections confirm a thick disc-like morphology. The kinematic properties and the high eccentricities $\sim0.7$ for the orbits (see Fig. \ref{fig:3components}) inform us that the inner-thick disc is kinematically hotter than the local thick disc.
   
\paragraph{Bar:} The galactic bar, shown in the last row of Fig. \ref{fig:bulge_disc}, has a starkly different chemical makeup. The \alphafe\ distribution is best fitted by three components with two equally weighted ($\approx40\%$ ) low and high-\alphafe\ components centred at 0.0 dex and 0.23 dex, respectively. There is an intermediate component, similar to the inner-thick disc, centred at 0.10 dex with 15\% weight. In contrast to the bulge and the disc, the bar has a high fraction of high metallicity ($\feh>0.0$) stars, and its MDF is best described by a four-component GMM with peaks approximately at $+0.40$, $0.0$, $-0.40$, and $-0.70$\ dex, contributing roughly 30\%, 25\%, 30\%, and 15\% of the total weight, respectively. The two GMM peaks at $-0.40$ and $-0.70$\ dex are strikingly close to the MDF peaks for the inner-thick disc and the spheroidal bulge populations. This hints at the possible contributions of these stellar populations to the galactic bar. The most metal-poor stars with $\feh\lesssim -1.0$\ dex contribute least to the bar, more to the inner thick disc, and most significantly to the spheroidal bulge component. The bar shows a wide spread in azimuthal velocity with a mean of about 110\kms\,, which is about 25\kms\ slower than the disc. However, the different orbit family members cover different radii along the bar and provide separate contributions to the overall bar morphology. Moreover, for the different orbit family members, we find a varying trend: the $\overline{V_{\phi}}$ ($\sigma_{V_{\phi}}$) increases (decreases) with an increasing $f_{z}/f_{z}$ ratio (see Table \ref{tab:orb_kinematics}). The bar, elongated in face-on view and showing the boxy-peanut exterior with central X morphology, appears as the most compact among the three components with its stars confined within 4 kpc radially and $\approx$1.5 kpc from the galactic plane.

The absolute heights of the MDF components are affected by the selection effects. The APOGEE (RPM) and RVS-CNN samples each have distinct selection functions, analysis pipelines, and inhomogeneities (See Fig. \ref{fig:lb_extinction} and Sec.\ref{sec:rvs_apo}). Hence, while the GMM reliably identifies the number and location of the MDF peaks, the absolute normalisation should be interpreted with caution. Importantly, the dips between the main metallicity components are significant relative to the measurement uncertainties.

\begingroup
\renewcommand{\arraystretch}{1.5}
\begin{table*}[ht]
\caption{Orbital and kinematic properties for the three Galactic components in bins of metallicity. For $\mathrm{Z_{max}}$ and $\mathrm{R_{apo}}$ the median together with the first and third quartiles of the distribution are listed. The mean azimuthal velocity ($\overline{V_{\phi}}$), as well as the azimuthal ($\sigma_{V_{\phi}}$), radial ($\sigma_{V_{R}}$), and vertical ($\sigma_{V_{Z}}$) velocity dispersions and their respective uncertainties, estimated via a bootstrap sampling, are also shown.}
\centering
\resizebox{0.75\textwidth}{!}{
\begin{tabular}{|l|lllllll|}
\hline
 & [Fe/H] $\in$ & [$-$2.5, $-$1.0] & [$-$1.0, $-$0.5] & [$-$0.5, $-$0.25] & [$-$0.25, 0.0] & [0.0, 0.25] & [0.25, 0.7] \\
\hline\hline
\multirow{6}{2em}{Sph. Bulge} & ${Z_{\mathrm{max}}}^{Q3}_{Q1}$ &  $2.6^{2.9}_{1.9}$ &  $2.0^{2.5}_{1.6}$ &  $1.7^{2.2}_{1.3}$ &  $1.4^{1.8}_{1.1}$ &  $1.3^{1.7}_{1.1}$ &  $1.2^{1.6}_{1.0}$ \\
& ${R_{\mathrm{apo}}}^{Q3}_{Q1}$ &  $3.8^{4.4}_{2.9}$ &  $3.1^{4.1}_{2.1}$ &  $2.6^{3.3}_{1.8}$ &  $2.0^{2.8}_{1.5}$ &  $1.9^{2.4}_{1.4}$ &  $1.8^{2.5}_{1.2}$ \\
& $\overline{V_{\phi}}$ & $3\pm4$ & $9\pm3$ & $3\pm4$ & $4\pm7$ & $-5\pm7$ & $-7\pm5$ \\
& $\sigma_{V_{\phi}}$ & $57\pm5$ & $62\pm3$ & $63\pm4$ & $69\pm5$ & $72\pm7$ & $59\pm7$ \\
& $\sigma_{V_{R}}$ & $85\pm4$ & $101\pm3$ & $111\pm5$ & $109\pm7$ & $111\pm8$ & $110\pm7$ \\
& $\sigma_{V_{Z}}$  & $89\pm5$ & $92\pm3$ & $95\pm4$ & $98\pm6$ & $109\pm7$ & $96\pm6$ \\
\hline
\multirow{6}{2em}{Disc} & ${Z_{\mathrm{max}}}^{Q3}_{Q1}$ &  $2.0^{2.3}_{1.7}$ &  $1.7^{2.1}_{1.5}$ &  $1.5^{1.8}_{1.4}$ &  $1.4^{1.6}_{1.0}$ &  $1.0^{1.4}_{0.5}$ &  $1.3^{1.5}_{0.6}$ \\
& ${R_{\mathrm{apo}}}^{Q3}_{Q1}$      &  $4.5^{4.8}_{4.1}$ &  $4.5^{4.8}_{4.2}$ &  $4.6^{4.8}_{4.3}$ &  $4.6^{4.8}_{4.3}$ &  $4.7^{4.9}_{4.4}$ &  $4.5^{4.8}_{4.0}$ \\
& $\overline{V_{\phi}}$  & $127\pm4$ & $131\pm2$ & $135\pm2$ & $139\pm2$ & $141\pm3$ & $127\pm4$ \\
& $\sigma_{V_{\phi}}$ & $41\pm3$ & $40\pm1$ & $43\pm2$ & $39\pm2$ & $38\pm2$ & $36\pm3$ \\
& $\sigma_{V_{R}}$ &  $80\pm5$ &  $68\pm2$ &  $65\pm2$ &  $63\pm3$ &  $65\pm4$ &  $86\pm5$ \\
& $\sigma_{V_{Z}}$ &  $76\pm5$ &  $67\pm2$ &  $63\pm2$ &  $53\pm3$ &  $43\pm3$ &  $56\pm5$ \\
\hline
\multirow{6}{3em}{Bar} & ${Z_{\mathrm{max}}}^{Q3}_{Q1}$ & $1.0^{1.5}_{0.8}$ & $1.0^{1.2}_{0.7}$ & $0.9^{1.2}_{0.7}$ & $0.9^{1.1}_{0.7}$ &  $0.8^{1.0}_{0.6}$ & $0.9^{1.1}_{0.7}$ \\
& ${R_{\mathrm{apo}}}^{Q3}_{Q1}$ & $2.5^{3.4}_{1.4}$ & $2.3^{3.3}_{1.5}$ & $2.7^{3.4}_{1.7}$ & $2.6^{3.6}_{1.7}$ &  $2.2^{3.1}_{1.5}$ & $2.2^{2.7}_{1.5}$ \\
& $\overline{V_{\phi}}$ & $100\pm10$ & $107\pm4$ & $118\pm4$ & $120\pm4$ & $107\pm4$ & $98\pm3$ \\
& $\sigma_{V_{\phi}}$ & $55\pm6$ & $62\pm2$ & $58\pm2$ & $58\pm3$ & $58\pm3$ & $55\pm2$ \\
& $\sigma_{V_{R}}$ &  $67\pm9$ & $72\pm3$ & $66\pm3$ & $71\pm4$ & $79\pm3$ & $76\pm2$ \\
& $\sigma_{V_{Z}}$ &  $73\pm7$ & $68\pm3$ & $65\pm3$ & $64\pm3$ & $60\pm3$ & $65\pm2$ \\
\hline
\end{tabular}
}
\label{tab:orb_metallicitybins}
\end{table*}
\endgroup

We now investigate the three inner Galactic components—spheroidal bulge, inner-thick disc, and bar—by separating their member stars into bins of metallicity to examine their metallicity-dependent properties in greater detail. In Figure \ref{fig:orbits_in_feh}, we show the face-on (XY) and edge-on (XZ) orbital density projections for the spheroidal bulge (top row), inner thick Disc (middle row), and Bar (bottom row). Corresponding orbital and kinematic statistics for each component and metallicity bin are summarised in Table \ref{tab:orb_metallicitybins}.

The spheroidal bulge appears nearly circular and symmetric in the face-on view across all metallicity bins, while its radial extent decreases with increasing [Fe/H]. The edge-on projections show a slightly oblate structure that similarly contracts at higher metallicities. In all metallicity bins, a centrally concentrated structure is evident, highlighted by the 50\% density contour and the black and red colour. This is quantitatively reflected in the statistics: the median $\mathrm{Z_{max}}$ and $\mathrm{R_{apo}}$ decrease steadily with [Fe/H], from 2.6 kpc and 3.8 kpc, respectively, in the most metal-poor bin to 1.2 kpc and 1.8 kpc in the most metal-rich bin. This trend illustrates the metallicity gradient of the spheroidal bulge. In terms of kinematics, the spheroidal bulge shows broadly similar mean velocities and dispersions across all metallicity bins, with the exception of the most metal-poor bin, where the radial velocity dispersion $\sigma_{V_{R}}$ is lower by 10–25 km/s compared to the other bins.

In the orbital density projections, the inner thick disc appears as a flat, circular, donut-like structure across all metallicity bins, showing a radially uniform distribution. We note that this uniformity is influenced by our selection as the disc obviously extends outwards. However, we clearly see that the disc stars tend to avoid the very central region. The disc gets vertically thinner with increasing metallicity, showing that the highest metallicity stars prefer to stay close to the galactic midplane. Quantitatively, as shown in Table \ref{tab:orb_metallicitybins}, the median $\mathrm{Z_{max}}$ gradually decreases with [Fe/H], from 2.0 kpc for the most metal-poor stars to 1.0 kpc for the [0.0,0.25] bin. For the [0.25,0.70] bin, a marginal increase to 1.3 kpc is seen. These trends, again, reflect the vertical metallicity gradient for the inner-thick disc stars, which we estimated to be \(\nabla{[Fe/H]}/\nabla Z_{\mathrm{max}}=-0.35\pm0.01\)  dex/kpc.
We note again that for the disc the $\mathrm{R_{apo}}$ statistics is uninformative, as the disc extends outwards beyond the 5 kpc limit and is only presented here for completeness. Kinematically, the disc shows nearly constant mean azimuthal velocities, increasing slightly from 127 \kms\ in the lowest \feh\ bin to 141 \kms\ in the [0.0,0.25] bin, before decreasing again to 127 \kms\ in the most metal-rich bin. The azimuthal velocity dispersion $\sigma_{V_{\phi}}$ remains fairly uniform across metallicities, ranging from 36 to 43 \kms. The radial velocity dispersion is $\sim$65 \kms\ between $-1.0<\feh<0.25$, while for the lowest and the highest \feh\ bins it is higher by 15-20 \kms. The vertical velocity dispersion shows the strongest metallicity dependence, declining from 76 \kms\ in the most metal-poor bin to 43 \kms\ in [0.0,0.25] bin, with a subsequent increase to 56 \kms\ in the [0.25, 0.70] bin. These reversals in $\mathrm{Z_{max}}$, $\overline{V_{\phi}}$, $\sigma_{V_{R}}$, and $\sigma_{V_{Z}}$ in the most metal-rich bin may reflect contamination from bar stars. 

\begin{figure*}[!ht]
    \centering
    \includegraphics[width=0.99\linewidth] {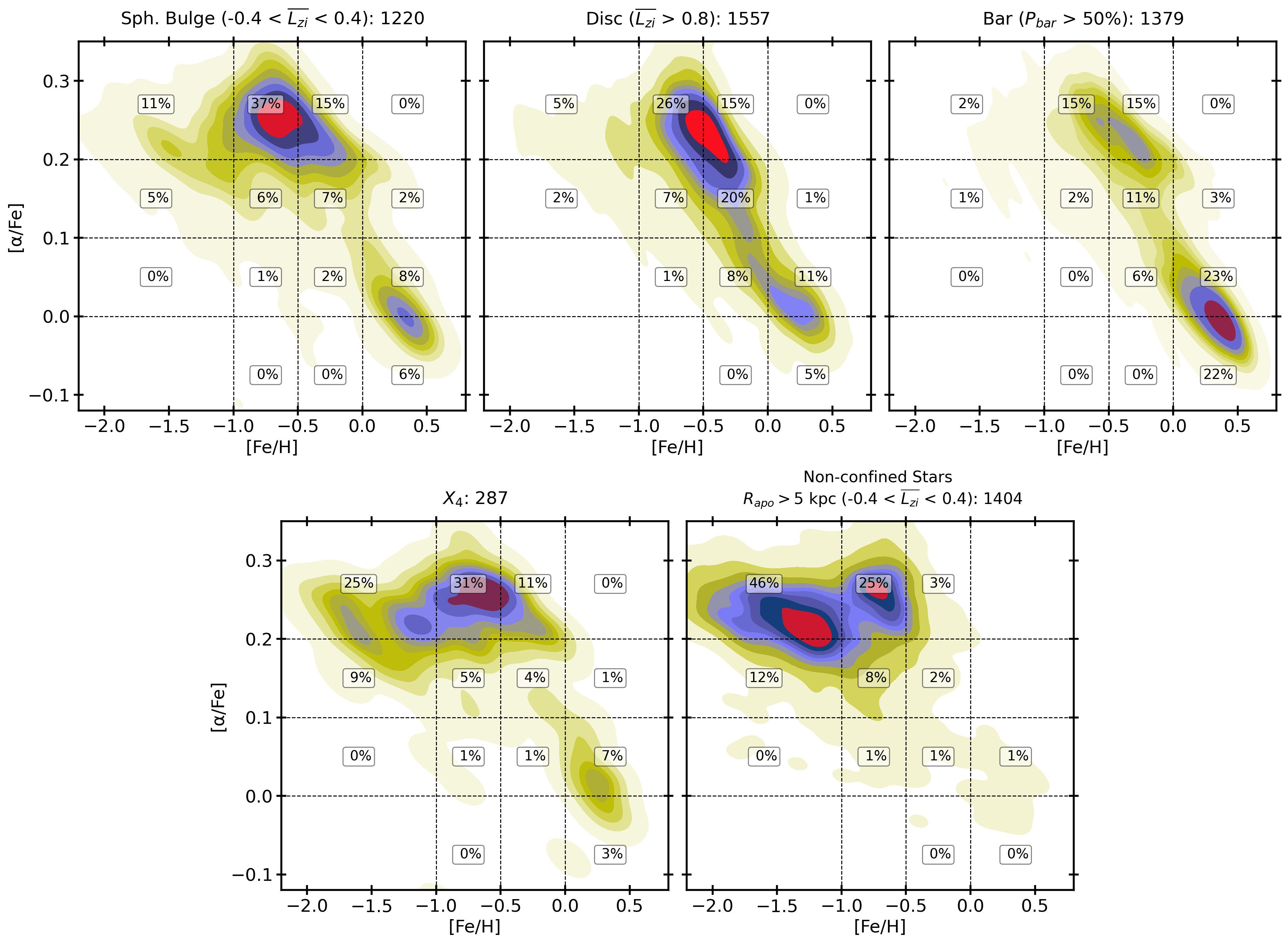}
    \caption{\alphafe\ vs \feh\ relation for the spheroidal bulge (top-left), the inner-thick disc (top-middle), the bar (top-right), the $X_4$ orbit family of stars (bottom-left), and the non-confined stars with ($-0.4<\overline{L_{\mathrm{zi}}}<0.4$) (bottom-right). Each panel shows the KDE -- overlaid with contours highlighting regions of high stellar density. Colours indicate density levels, with red being the highest, followed by blue and green, while white indicates regions with no stars. The \alphafe\ vs \feh\ plane is further divided into rectangular regions with lines at \feh=[-1.0,-0.5,0.0] and \alphafe=[0.0,0.1,0.2]. The fraction of stars in each region is provided.}
    \label{fig:afeh_5_components}
\end{figure*}

The overall morphology of the Galactic bar, as seen in the orbital density projections, remains largely consistent across all metallicity bins. In the face-on (XY) plane, the bar is clearly elongated along the X-direction, while the edge-on (XZ) view displays the characteristic boxy-peanut shape. The central X-shaped structure is visible at all metallicities, becoming most prominent in the highest \feh\ bin. We remind readers that the bar’s overall morphology is a composite of various orbital families. In particular, the brezel and banana orbits—which contribute most to the X shape—contain a higher fraction of metal-rich stars compared to metal-poor ones (see Sec. \ref{sec:bar_chem} and Fig. \ref{fig:chem_orbit_families}). The $\mathrm{Z_{max}}$, contrary to the spheroidal bulge and the disc, shows no apparent trend with metallicity, as the median remains nearly constant at $\sim1.0$ kpc and the median apocenter varies only slightly between 2.7 kpc to 2.2 kpc. The intermediate \feh\ bins of [$-$0.5,$-$0.25] and [$-$0.25,0.0] have the highest median $\mathrm{Z_{max}}$ values. Kinematically, the bar stars have $\overline{V_{\phi}}$ lower by $\sim$20 \kms\ compared to the disc stars in the same metallicity bin. Moreover, the intermediate \feh\ bins of [$-$0.5,$-$0.25] and [$-$0.25,0.0] have the highest $\overline{V_{\phi}}$. These differences for the sub-solar \feh\ regime may reflect contamination from the thick disc stars. The velocity dispersions show no apparent trend. These chemical, orbital, and kinematic properties reveal that the galactic bar is a very well-mixed dynamical structure. 

In addition to the orbital-density projections, the complementary configuration-space distributions presented in the Appendix (Figs. \ref{fig:XYZ}–\ref{fig:XYZ_mr}) confirm these trends: spheroidal-bulge stars appear centrally concentrated with metal-rich stars tightly clustered towards the centre and metal-poor stars spanning the full volume; the bar shows a pronounced elongated and boxy-peanut morphology across metallicities; and the inner-thick disc exhibits a vertically extended, radially broad structure in which metal-rich stars remain near the midplane, while metal-poor stars reach higher $\mathrm{|Z|}$. These maps visually reinforce the metallicity-dependent structural differences inferred from the orbital analysis.

\subsection{Evolution of the bulge stellar populations and the halo}\label{sec:bulge_halo}

To understand and constrain the formation and evolution scenario of the galactic bulge components and the inner MW halo, we present in Fig. \ref{fig:afeh_5_components} the \alphafe\ vs \feh\ diagrams. In the top panels, the spheroidal bulge (left), the inner thick disc (centre), and the bar-supporting stars (right) are shown, and in bottom-left the $X_4$ stars and in bottom-right the non-confined stars (i.e. $\mathrm{R_{apo}}$>5 kpc) with no significant rotation about the z-axis in inertial frame, similar to the spheroidal bulge, are shown. The latter is used to characterise the halo component. Each panel shows the \alphafe-\feh\ diagram as a density plot (KDE) with the red, blue, and green colours representing the stellar density in decreasing order. The \alphafe\ versus \feh\ planes are further subdivided, and the fraction of stars contained are labelled to facilitate the analysis. 

The spheroidal bulge, the inner thick disc, and the bar reflect distinct abundance patterns, as characterised in Sec. \ref{sec:SpBDB} and Figure \ref{fig:bulge_disc}. As a novelty of this figure, the predominantly high-\alphafe ($\gtrsim$ 0.2 dex) stars of the spheroidal bulge show a nearly flat distribution with a wide metallicity range and appear to extend to high metallicities reaching solar \feh\ with \alphafe\ around 0.2 dex. The peak of the spheroidal bulge MDF, at $\feh\sim-0.7$, also shows the highest \alphafe\ abundance values, followed by a gradual decrease towards higher and lower \feh. The small concentration of stars with \feh>0.0 and low-\alphafe\ appears detached from the rest of the high-\alphafe\ stars of the spheroidal bulge and is possibly a contamination of the more numerous low-\alphafe\ stars of the bar. We also observe intermediate- and high-\alphafe\ stars at metallicities below $-$1.0 dex. This is about 16\% of the total number in the spheroidal bulge. This metal-poor group is less prevalent in the thick disc and the bar.

Moving from the spheroidal bulge to the inner-thick disc, we clearly see an evolution across the \alphafe\ versus \feh\ diagram. The disc, in addition to stars with \alphafe\ above 0.2 dex, has a high fraction of sub-solar metallicity ($-0.5<\feh<0.0$) and intermediate \alphafe ($\sim0.1$) stars, which are sparse in the spheroidal bulge. The main concentration of high-\alphafe\ stars for the thick disc lies at a higher \feh\ compared to the spheroidal bulge, and a steep decrease in \alphafe\ with \feh\ is observed in contrast to the spheroid. This is expected for a chemical evolution model in which the spheroidal bulge forms with a higher star formation efficiency than the thick disc, occurring on extremely short timescales. This scenario is consistent with models such as those of \citet{Friaca2017} and the discussions in the \citet{Barbuy2018} review. See also \citet{Matteucci2019} and \citet{Molero2024}. However, it is worth noting that these previous models rely on observational constraints that mix different stellar populations. This problem was alleviated in \citet{Razera2022} and \citet{Barbuy2023, Barbuy2024, Barbuy2025}, where the models of \citet{Friaca2017} were compared to a sample of bona fide spheroidal bulge (selected from the sample of \citepalias{Queiroz2021}, and concentrating on the most metal-poor objects for the study of detailed chemical properties \citep[see also][]{Souza2023}).

Stars supporting the bar have both high- and low-\alphafe\,, as discussed in Section \ref{sec:bar_chem}, but the high-\alphafe\ population occupies a wider metallicity range than the more metal-rich low-\alphafe\ stars. The high-\alphafe stars in the bar show an abundance footprint similar to those in the thick disc, with possibly a smaller contamination from the spheroidal bulge. The high metallicity (>0.0 dex) stars constitute nearly half of the bar population, while they contribute only about 15\% to the thick disc and the spheroidal bulge. These metal-rich stars could have been previously in an inner-metal-rich thin disc prior to the bar formation. This further supports the claim that the bar structure is able to trap older stars, from the inner thick disc (and possibly the spheroidal bulge) with high-\alphafe\ to the old inner thin disc with lower-\alphafe during its formation. The galactic bar is not a stellar population but a very well-mixed dynamical structure.

In the bottom row of Fig. \ref{fig:afeh_5_components}, we present the \alphafe\ versus \feh\ diagram for the $X_4$ orbit stars (left panel). Earlier in Sec. \ref{sec:X4}, we confirmed, via GMM decomposition of the MDF and \alphafe\ distributions, that the $X_4$ orbits trace the spheroidal bulge component well. Here, the \alphafe\ versus \feh\ diagram also reflects a chemical abundance pattern similar to the spheroidal bulge, i.e. a high-\alphafe\ population ($\gtrsim0.2$\ dex) showing a flat distribution for a wide range of metallicities from $\sim-$2.0 to $\sim$0.0 dex. The metal-rich and low-\alphafe\ stars ($\sim10\%$) again are similarly placed as in the bar. However, the most metal-poor stars show an interesting difference—the stars with $\feh<-1.0$\ dex account for about 34\% of $X_4$ by number. This is significantly higher than in the spheroidal bulge, which has about 16\% of the same. Given the main component of both the spheroidal bulge and the $X_4$ peak at $\sim-0.70$, these more metal-poor stars could belong to a separate population or populations.

\begin{figure}[!ht]
    \centering
    \includegraphics[width=0.99\linewidth] {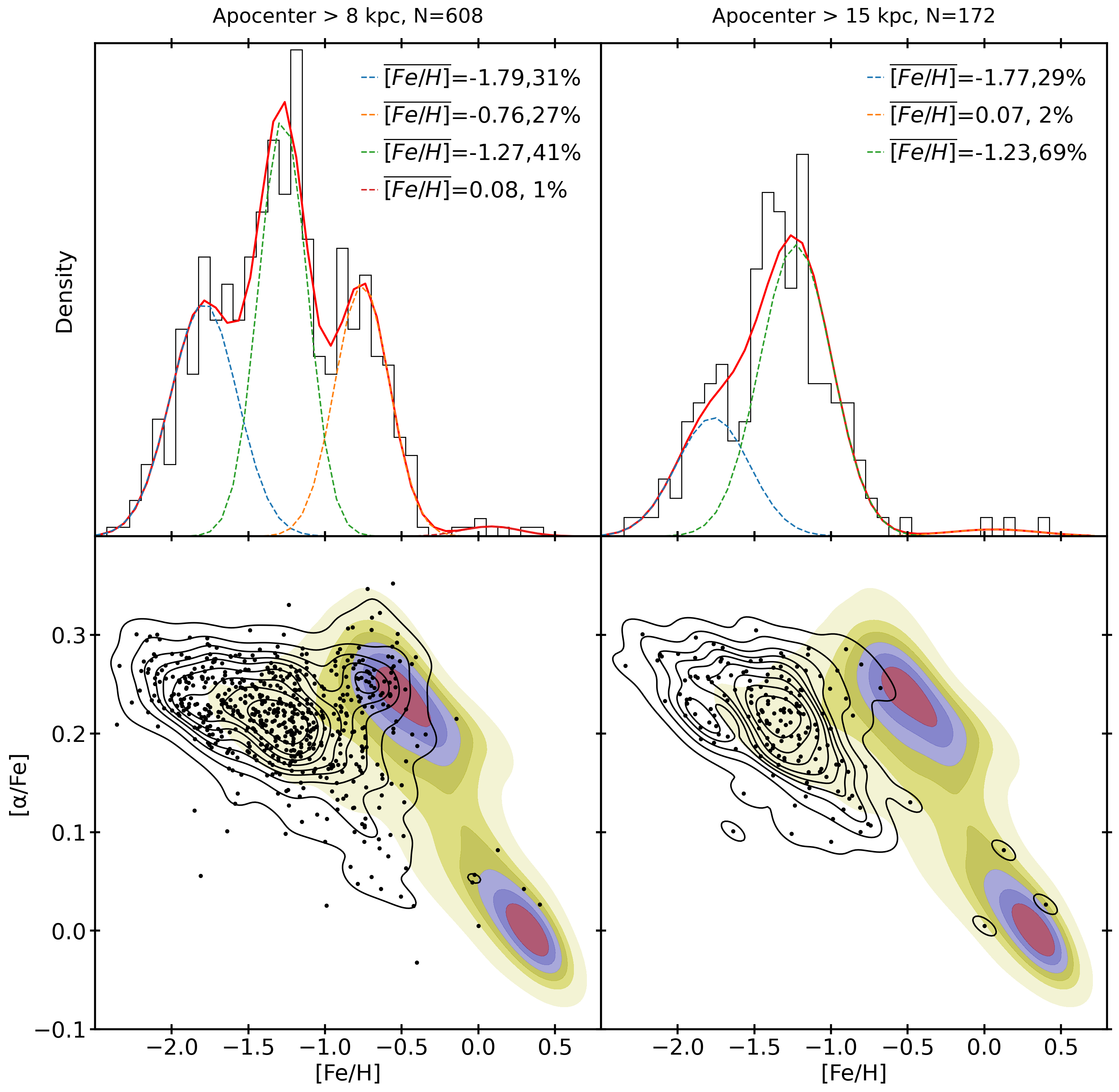}
    \caption{Identification of the GES in the galactic bulge. Left column: MDF with GMM decomposition for the non-confined stars with apocenter larger than 8 kpc. Similar to the spheroidal bulge, these stars exhibit no significant rotation about the z-axis in the inertial frame (i.e. $-0.4<\overline{L_{\mathrm{zi}}}<0.4$). For each GMM component, the respective mean and weight are shown. Right column: Same as the left, but for stars with apocenter larger than 15 kpc. Bottom: \alphafe\ vs \feh\ for the respective non-confined stars (black dots). The contours (KDE) highlight the main density features. In the background, the abundances for the bulge sample are shown.}
    \label{fig:gse}
\end{figure}

Considering that $X_4$ orbits are efficient at trapping stars with low angular momentum ($L_Z$)—such as those originating from a pressure-supported system (see Sec. \ref{sec:X4})—they would indiscriminately capture both the older, more metal-poor halo stars and stars from the central spheroidal bulge component. To test this hypothesis, in Fig. \ref{fig:afeh_5_components} (bottom-right panel), we analyse stars currently located in the bulge but with apocenters extending beyond 5 kpc, and whose orbits resemble those of the spheroidal bulge, i.e. showing no net rotation about the $z$-axis. The \alphafe\ versus \feh\ diagram reveals that nearly all such stars have metallicities below $\sim -0.50$ dex and fall into two main groups: a smaller, compact group with $\feh \approx -0.70$ dex and \alphafe $\geq 0.2$, consistent with the spheroidal bulge peak, and a larger group (comprising about 70\% of the stars) at lower metallicities and with a broader \alphafe\ distribution, including $\sim$20\% of stars with $0.1 <$ \alphafe\ $< 0.2$. Interestingly, we find similar stars in the inner thick disc and the bar, but at only $\sim$1\% contribution. These metal-poor, low-\alphafe\ stars could also include merger remnants \citep[e.g.][]{HortaSchiavon2024}. The chemical similarity between the metal-poor component of the spheroidal bulge or the $X_4$ stars and this larger population of metal-poor stars supports our hypothesis. It also reinforces the conclusion that the spheroidal bulge contains only a minor contribution ($\lesssim$16\%) from the more metal-poor (\feh\ $< -1.0$) genuine bulge stars or halo interlopers, including merger debris.

The dynamically selected halo-like group from the inner-galaxy sample, shown in the bottom-right panel of Fig. \ref{fig:afeh_5_components}, exhibits a distinct abundance pattern and a higher fraction of metal-poor stars compared to the confined bulge populations. To further characterise these halo-like stars, we examine their MDFs and \alphafe–[Fe/H] distributions in Fig. \ref{fig:gse}. To minimise contamination from genuine bulge stars, we applied two additional apocenter-based selections: one with $R_{\mathrm{apo}} > 8$ kpc and another with $R_{\mathrm{apo}} > 15$ kpc. 

For the $R_{\mathrm{apo}} > 8$ kpc selection, the MDF is best described by three Gaussian components: a dominant peak at [Fe/H] $\sim -1.30$ dex contributing $\sim$40\% of the stars, and two secondary peaks at $\sim -1.80$ and $\sim -0.75$ dex, each contributing $\sim$30\%. In contrast, the more stringent $R_{\mathrm{apo}} > 15$ kpc cut yields a cleaner sample, where the MDF is best fitted by two components at $\sim -1.80$ and $\sim -1.30$ dex, with $\sim$70\% and $\sim$30\% relative weights, respectively. The component peaking at [Fe/H] $\sim -1.30$ stands out as the dominant inner-halo contribution.  

In the \alphafe–[Fe/H] plane, these stars occupy loci clearly distinct from the confined bulge populations. While the $R_{\mathrm{apo}} > 8$ kpc sample shows some overlap with the spheroidal bulge, the $R_{\mathrm{apo}} > 15$ kpc selection reveals a clear sequence of declining \alphafe\ with increasing [Fe/H], including an $\alpha$-knee at [Fe/H] $\sim -1.10$ dex. Compared to the in situ bulge populations, these stars exhibit systematically lower \alphafe\ values at a given metallicity, reaching solar \alphafe\ near [Fe/H] $\sim -0.5$. 

The chemodynamical properties of these stars closely resemble those of the Gaia-Enceladus/Sausage (GES) merger remnant, which dominates the local stellar halo (e.g. \citealt{Helmi2018, Belokurov2018}; see also \citealt{Nissen2010, Feuillet2021, Buder2022, Limberg2022, Horta2023} and references therein). The secondary metallicity peak around [Fe/H]~$\sim -1.80$ is similar to the Sequoia event proposed by \citet{Myeong2019}, who associates the existence of retrograde stars with metallicities lower by $\sim$0.3 dex compared to the GES (see also e.g. \citealt{Koppelman2019, Feuillet2021, Limberg2022, Horta2023}). However, this peak has also been linked to GES itself and may point to episodic star formation within the GES system \citep[see][]{AguadoAgelet2025, GonzalezKoda2025}. See also \citet{Kruijssen2019, Massari2019, Forbes2020, Callingham2022} for possible contributions by other low-mass merger remnants. Overall, based on our analysis of the MDF and the \alphafe–[Fe/H] plane, we find evidence supporting a single dominant merger contribution to the inner galaxy, in agreement with the work by, for example, \citet{Rix2022}. 

We also note that the overall high-\alphafe\ and low-\alphafe, metal-poor population ([Fe/H] < $-1.0$)\footnote{The nature of the metal-poor stars in our “inner-galaxy sample” will be studied in detail in a separate work.} in the MW bulge region likely represents a mixture of both in situ stars and ex situ debris from early accretion events. Moreover, a low fraction of the merger remnants could also now belong to the inner thick disc or the bar (see Fig. \ref{fig:afeh_5_components}). Targeted studies of these metal-poor bulge stars, such as those by \citet{Anke2020a, Anke2020b, Arentsen2024}, together with follow-up studies to obtain detailed chemical abundance measurements, are crucial to disentangling their origins.

\section{Discussion and conclusion}\label{sec:conclusion}

In this paper, we explored the stellar populations in the inner 5 kpc of the MW with the goal of understanding the composition and gaining insights into the origin and evolution of the Galactic bulge region. To accomplish this, we used observational data comprising approximately 18,000 stars coming from two spectroscopic catalogues. The first sample was extracted from the catalogue of \citet{rvs_cnn_2023}, which provides spectroscopic parameters for the {\it Gaia}-RVS spectra using the {\tt hybrid-CNN} machine learning approach, with precision comparable to higher-resolution spectroscopic surveys such as APOGEE. The second is an updated version of the RPM sample of \citet{Queiroz2021} based on the spectroscopic parameters from DR17 of the APOGEE survey \citep{Abdurrouf2022}. We used the {\tt StarHorse} Bayesian isochrone fitting code \citep{queiroz2018, queiroz2020, Queiroz2023} to estimate accurate distances and extinctions for our spectroscopic samples. Using these distances together with 6D phase-space coordinates from {\it Gaia}-DR3, we performed full orbit integrations in a barred potential.  

We began our exploration of the bulge by first examining the overall MDF and kinematics, which initially revealed a bimodal MDF with rotation support and high velocity dispersions. The MDFs in bins of ($l, b$) confirmed previously reported changes in the nature of MDFs with sky positions. To differentiate overlapping populations, we proceeded with an analysis in the $\mathrm{Z_{max}}$--eccentricity plane, a diagnostic tool used to identify stellar populations based on orbital similarity, which revealed complex interplays of kinematics and chemistry across different regions of this parameter space. Most importantly, this revealed a transition for high eccentricity stars, from a metal-rich population that remains close to the galactic plane to a metal-poor population extending beyond 2 kpc in the $Z$-direction. We then applied a full orbital frequency analysis approach to robustly identify and characterise stars supporting the Galactic bar and study its detailed properties across various orbital families. Additionally, we identified the stars in the $X_4$ orbit family that are chemically distinct from the bar. After isolating the bar, we further classified the remaining stars, based on their distinct orbital and chemical signatures into the spheroidal bulge and the inner thick disc components. A summary of the classification scheme is presented in Appendix \ref{sec:flowchart}. Finally, we checked for the possible contribution from the halo-like stars to the spheroidal bulge population. Our main findings are summarised as follows:

\begin{enumerate}
    \item Identification of the spheroidal bulge: 
    \begin{itemize}
        \item The spheroidal bulge MDF peaks around $-$0.70 dex and extends to solar values. It is dominated by a high-\alphafe\ population with almost no metallicity dependency, consistent with a very rapid early formation, predating the thick disc and the bar. This suggests a high early star formation rate and efficiency.
        \item It exhibits high velocity dispersions, specifically $\sigma_{V_R}\approx\sigma_{V_Z} \approx100$ \kms, supporting its classification as a pressure-supported component.
        \item It extends approximately 3 kpc from the Galactic plane and shows mild triaxiality.
        \item We observe strong metallicity-dependent trends: as [Fe/H] increases, both the median apocenter and the $\mathrm{Z_{max}}$ of spheroidal bulge stars decrease, indicating that the most metal-rich stars are more centrally concentrated and the spheroidal bulge becomes increasingly compact with metallicity. This suggests the very early presence of a metallicity gradient.
        \item We identified a group of stars belonging to the retrograde $X_4$ orbit family. They have $\overline{V_{\phi}}\approx-85$ \kms\ with a dispersion of $\approx50$ \kms. Their fractional contribution and orbital characteristics match those predicted in N-body models. We find the $X_4$ stars predominantly reflect the chemistry of the spheroidal bulge with a main MDF peak at $-0.70$ dex, including a more metal-poor component.
        \item The metal-poor stars ([Fe/H] < $-$1.0 dex) provide only a minor contribution ($\lesssim16$\%) to the spheroidal bulge population, with a significant fraction likely from the merger remnants.
    \end{itemize}
    \item Characterisation of the inner-thick disc: \begin{itemize}
        \item This disc is predominantly metal-poor, with an MDF peak at $\feh\approx-0.45$, including a high fraction of stars with sub-solar \feh\ and intermediate \alphafe, similar to the local thick disc \citep[e.g.][]{bensby2014, Queiroz2023}.
        \item In contrast to the spheroidal bulge, the disc's \alphafe\ shows a steeper decline with [Fe/H], indicative of less intense star formation efficiency.
        \item The inner thick disc, with $\overline{V_{\phi}} = 133$ \kms, velocity dispersions [$\sigma_{V_{R}}, \sigma_{V_{\phi}}, \sigma_{V_{Z}}$] = [66, 39, 59] \kms, and mostly high-eccentricity orbits ($e\sim0.7$), is kinematically hotter than the local thick disc \citep[e.g.][]{bensby2014, Queiroz2023}.
        \item We measure an overall vertical metallicity gradient of \(\nabla [Fe/H]/\nabla Z_{\mathrm{max}} = -0.35\pm0.01\) dex/kpc, with most metal-rich stars confined to $\sim1$ kpc from the galactic plane.
    \end{itemize}

    \item Properties of the Galactic bar: \begin{itemize}
        \item We employed the orbital frequency analysis technique to identify bar-supporting stars and show that the overall morphology of the bar is a composite of various orbit family groups. The brezel and banana orbits significantly contribute to the X shape, while higher-order orbits build the boxy-peanut and the thin bar morphology.
        \item The bar shows equal contributions from the high- and low-\alphafe\ stars, while the MDF is fitted by four GMM components with means at $-$0.70, $-$0.40, 0.0, and $+$0.40 dex. The peaks at $-$0.70 and $-$0.40 dex closely match the spheroidal bulge and the thick disc MDF peaks.
        \item Across the orbital groups, we find significant chemical differences. For example, the brezel orbits show a strong central X shape and an 8\% higher contribution from high-\alphafe\ stars, while banana orbits, supporting the outer X shape, have 16\% more low-\alphafe\ stars and a massive contribution from super-solar metallicity stars.
        \item Similarly, across the orbital groups, we find significant kinematic differences. With the increasing frequency ratio, the velocity dispersions systematically decrease, while the $\overline{V_{\phi}}$ increases.
        \item The bar displays no strong metallicity trends in orbital extent ($\mathrm{R_{apo}}$ or $\mathrm{Z_{max}}$) or velocity dispersions and maintains a consistent elongated shape (with nearly identical face-on and edge-on orbital projections) across all metallicities, indicating a well-mixed dynamical structure.
    \end{itemize} 

    \item Identification of GES in the bulge: \begin{itemize}
        \item Chemodynamical analysis of the stars, currently in the bulge, with halo-like orbits reveal two main MDF peaks at around $-$1.80 and $-$1.30. The MDF peak at $-$1.30 is dominant ($\sim70\%$) for stars with large apocentres, and the peak at $-$1.80 has consistently around 30\% contribution.
        \item We identify the GES merger remnants in the bulge region with chemical characteristics (loci in the \alphafe-[Fe/H] diagram) similar to those from the local halo \citep[e.g.][]{Helmi2018, Horta2023}.
        \item The GES stars show an MDF peak at $\sim-$1.30 dex and a clear sequence of declining \alphafe\ with increasing [Fe/H], including an $\alpha$-knee at [Fe/H] $\sim-1.10$ dex and systematically lower \alphafe\ values at a given metallicity, compared to the in situ stars, reaching solar \alphafe\ near [Fe/H] $\sim -0.50$.
        \item Some of these merger remnants, remain confined within the bulge region and are now part of the spheroidal bulge, the inner thick disc, and the bar, in decreasing contribution, respectively, as observed from the metal-poor low-\alphafe\ stars in these components.
    \end{itemize}
\end{enumerate}

Our population classification is purely dynamical, so the observed chemical differences between the components and populations are unlikely to arise from systematics or selection functions. Our results demonstrate that the spheroidal bulge, thick disc, and bar—while clearly co-spatial in the inner Galaxy—remain well separated in phase-space and chemistry, despite the large overlaps in all properties. The addition of RVS data presented here was instrumental in complementing the RPM dataset of \citetalias{Queiroz2021} by providing access to a larger sample of metal-poor stars (specially \feh\ between $-$2 to $-$0.5). This expanded coverage now corroborates the presence of a spheroidal bulge component in the MW, which likely formed prior to the development of the disc and bar structures. Moreover, the combined datasets have allowed us, for the first time, to place tighter constraints on the properties of the inner thick disc, revealing its dynamical and chemical characteristics in unprecedented detail.

The physical overlap, coupled with a continuous range of kinematic, structural, and chemical properties across components, underscores the necessity of a probabilistic and multi-dimensional approach, such as the one employed in this study. By modelling these populations probabilistically, we are able to account for the inherent uncertainties and overlapping signatures in both configuration and velocity space. Importantly, we find that these distinct structures can co-exist and remain dynamically coherent despite their physical overlap. The hotter components, especially the spheroidal bulge, are more resistant to perturbations, consistent with their higher velocity dispersion.

Nonetheless, we do detect signs of mild triaxiality in our spheroidal bulge component, especially towards larger metallicities, which could reflect real dynamical influence from the bar. This is in line with theoretical work by, for instance, \citet{Saha2012}, who showed through high-resolution N-body simulations that a low-mass classical bulge can absorb angular momentum emitted by a growing bar via resonant interactions (notably through the Lagrange point and inner Lindblad resonances). This process can spin up an initially isotropic, non-rotating spheroidal bulge. In \citet{PerezVillegas2017}, the authors showed that the metal-poor component ([Fe/H] $\lesssim-$1.0) of the inner part of the galaxy has a triaxial shape and that it does not support the X-shaped structure. Using N-body simulations, they also showed that the initially oblate stellar halo evolves into triaxial under the influence of the Galactic bar. More recently, in their N-body simulation study of barred galaxies with classical bulges,\citet{McClure2025} found that up to 50\% of their initial bulge stars were trapped in bar resonances and only $\sim25\%$ of stars maintained their initial isotropic, dispersion-dominated character. These findings provide a compelling theoretical framework to interpret the mild triaxiality we observe in our spheroidal bulge component as a consequence of dynamical interaction with the bar, rather than as mere contamination or observational uncertainty. Also, the inner thick disc component—which we find to be slightly dynamically hotter than the thick disc at the solar vicinity—could itself be a result of interactions with the bar \citep[e.g.][]{Patsis2019}. Additionally, the low fraction of low-\alphafe\ and super-solar metallicity stars observed in the spheroidal bulge most probably belonged to the bar or disc and have lost angular momentum during bar growth to the halo and the bulge \citep[e.g.][]{Debattista1998, Athanassoula2003}. This also provides a natural explanation for the Knot of SMR stars as observed by \citet{Rix2024} in the galactic $l, b$ plane, whose origin can now be assigned to the bar.

Alternatively, this may be partially attributed to contamination between populations, stemming from residual uncertainties in distance measurements and the assumed Galactic potential. The fact that these different structures are also linked to distinct chemical properties—particularly the thick disc and spheroidal bulge, which, despite significant overlap in chemical abundances, exhibit subtle differences in the peak of their metallicity distributions and in the shape of the \alphafe\ versus \feh\ relation—gives us confidence that we have successfully disentangled these populations. These results emphasise the complexity of the inner 4-5 kpc of the MW and the value of a multi-dimensional analysis involving both chemistry and kinematics to interpret the structure and properties of the different stellar populations (the spheroidal -- old -- bulge and inner hotter chemical thick disc) and structures (such as the bar).

We are now in a position to provide initial answers to long-standing questions in this field. One such question is whether the bulge is entirely formed via thickened bar structures through either buckling instabilities or resonance trapping by a slowing bar. If this were the case, the stellar population within the spheroidal bulge component should exhibit chemical properties consistent with those of the bar. However, if the spheroidal component formed prior to the bar, its MDF would be expected to differ significantly, indicating a distinct formation pathway. In particular, a substantial difference in the MDF—and in the \alphafe\ distribution—between the spheroidal component and the bar would imply that the former formed during an earlier epoch. Our results suggest precisely this: the MDF and \alphafe\ trends of the spheroidal bulge component are clearly distinct from those of the bar population (and the inner thick disc).

Furthermore, the differences between the spheroidal bulge and the thick disc indicate that these two kinematically hot components are not chemically identical, with slightly different MDF peaks (and maybe other chemical differences (see discussion in \citet{Barbuy2018} and the detailed chemistry for bona fide spheroidal bulge stars in \citealt{Razera2022, Barbuy2023, Barbuy2024}). A detailed high-resolution follow-up of our inner thick disc sample—mainly dominated by RVS stars—would be necessary to map in detail the chemical abundance differences between these two important old stellar populations. Indeed, the thick disc is also expected to be old because we know that in the solar vicinity this component is coeval in age, peaking at around 11 Gyr (see \citealt{Miglio2021} using ages from asteroseismology and \citealt{Queiroz2023} using isochrone ages). The spheroidal bulge should be even older. Following the theoretical age-metallicity relationship (AMR) approach, which is widely used for Globular cluster studies, and with new constraints from more precise bulge GC's ages of \citet{Souza2024a, Ortolani2025}, places our metallicity peak of $\sim-0.70$ for MW spheroidal bulge at an age range of 12-13 Gyr (see Fig.\ref{fig:age}).

The metal-rich, low-\alphafe, stars -- with a wide age distribution and a prominent peak at $\sim$4 Gyr observed for the micro-lensed bulge dwarfs stars of \citet{Bensby2013,Bensby2017} or for the Baade's Window stars of \citet{Schultheis2017} -- appear mostly likely to belong to the galactic bar population. A strong star-formation peak at around 3--4 Gyr has also been observed in the solar-neighbourhood disc stars with a possible connection to the galactic bar \citep{Nepal2024_bar}.

Studies based on RR Lyrae stars (tracers of an old—$>$10 Gyr—stellar population) have provided compelling evidence for the existence of a pressure-supported, centrally concentrated stellar component in the MW bulge. The BRAVA-RR survey \citealt{Kunder2020} revealed that a subset of RR Lyrae stars in the inner Galaxy exhibit tight, eccentric orbits and a spatial distribution that does not align with the barred structure, suggesting they formed prior to the bar and belong to an ancient, spheroidal population.

We may be discussing age differences of less than 1-1.5 Gyr. Achieving a MW age map with this level of resolution—not only in terms of precision (at the ~10\% level), but also accuracy—will require large samples of bulge stars with well-determined ages. These could come from isochrone fitting near the MSTO or from asteroseismology of red giant stars \citep{Chaplin2013, Miglio2017AN}. Significant progress in this area will depend on future missions such as the proposed \textit{Haydn} mission \citealt{MiglioHAYDN2021}, as well as on the capabilities of large telescopes equipped with adaptive optics and near-infrared spectrographs \citep[e.g.][]{Magrini2023}. In the nearer term, smaller but valuable samples will be provided by instruments such as MOONS \citep{Cirasuolo2020Msngr} on the VLT and MOSAIC \citep{Hammer2021Msngr} on the ELT.

Our examination of the halo-like stars, in Fig. \ref{fig:afeh_5_components} and \ref{fig:gse}, provides interesting insights on the metal-poor ([Fe/H]<$-$1.0) stellar populations of the inner MW. While the spheroidal bulge does host metal-poor stars, they represent only a minor fraction compared to the dominant moderately metal-poor population ([Fe/H]$\approx-$0.7). Among stars with large apocentres, we clearly identify the GES merger remnant \citep{Helmi2018, Belokurov2018} as the primary contributor to the inner halo population, with a metallicity peak at [Fe/H] $\approx-$1.3 dex and a distinct locus in the \alphafe–[Fe/H] plane. A secondary peak at [Fe/H] $\approx-$1.8 may be associated with the Sequoia event \citep{Myeong2019}. Due to the lack of detailed chemical abundances, especially for the RVS stars, we cannot examine the high-\alphafe\ metal-poor stars with chemical selection as performed by \citet{Horta2021} for identification of the Heracles merger remnant. The Heracles MDF also peaks around [Fe/H] $\approx-$1.30 dex \citep{Horta2021}, similar to our halo-like population in the bulge, but it shows no clear $\alpha$-knee—consistent with our finding that a distinct high-\alphafe\ sequence is absent.

A complementary RR Lyrae study by \citet{Kunder2024} similarly finds that stars below [Fe/H]$\lesssim -$1.0 comprise both halo and accreted populations. While two RR Lyrae stars in their sample are linked to GES, a higher fraction (58\%) occupies the energy and eccentricity range associated with Heracles, Kraken, and Koala, which also overlaps with the in situ Aurora population \citep{Belokurov2022, Arentsen2024}. The high fraction of stars in this low-energy regime suggests that most are not accreted but are instead members of Aurora—thought to be a dense, old, in situ component similar to our spheroidal bulge. Although some contamination by accreted stars cannot be excluded for either the RR Lyrae or our spheroidal bulge sample, the lack of detailed elemental abundances prevents firm distinctions between remnants such as Heracles, Kraken, or Aurora. From the RR Lyrae perspective, Heracles is not clearly identified kinematically within the inner Galaxy; its identification rests mainly on chemical signatures, which also show overlap with the spheroidal bulge population, as demonstrated by \citet{Barbuy2024}. In summary, as also argued by \citet{Rix2022} in the context of metal-poor stars, labels such as Aurora, Heracles, and Knot may not be necessary when the observed structures can be explained by the co-existence of distinct but overlapping stellar populations in the inner Galaxy.
  
Finally, these findings (i.e. the detection of the spheroidal bulge and inner thick disc in the inner parts of the MW) pose a challenge to models, which posit that the entire inner Galaxy formed via the vertical heating of the disc, for example via buckling instabilities or the disc-fractionation scenario proposed by \citet{Debattista2017}. In particular, the observed chemical properties of the thick disc and the spheroidal bulge differ markedly from those of the bar stars, making it clear that a single evolutionary pathway cannot account for all the populations observed in the innermost Galactic regions.
On the other hand, clump models for the spheroidal bulge formation seem to be better aligned with what the data indicate \citep[see][]{Elmegreen2008, Debattista2023}. However, large samples are needed to more robustly constrain these models. This will be one of the goals of the \textit{4MIDABLE-LR} survey \citep{Chiappini2019} (one of the 4MOST surveys - \cite{deJong2019}) and its sub-surveys dedicated to the bulge and moderately metal-poor stars.

\bibliographystyle{aa}
\bibliography{bulge_biblio}

\begin{acknowledgements}
We thank the anonymous referee for insightful suggestions that have greatly improved this manuscript. SN and CC thank the e-Science \& IT team for COLAB service and research infrastructure at AIP. SN thanks Potsdam Graduate School (PoGS) for partial financial support for this publication, funding code $\mathrm{D\_2674}$. BB and CC acknowledges partial financial support from FAPESP, and BB also from CNPq, and CAPES- Finantial code 001. This work was partially funded by the Spanish MICIN/AEI/10.13039/501100011033 and by the "ERDF A way of making Europe" funds by the European Union through grant RTI2018-095076-B-C21 and PID2021-122842OB-C21, and the Institute of Cosmos Sciences University of Barcelona (ICCUB, Unidad de Excelencia ’Mar\'{\i}a de Maeztu’) through grant CEX2019-000918-M. FA acknowledges financial support from MCIN/AEI/10.13039/501100011033 through a RYC2021-031638-I grant co-funded by the European Union NextGenerationEU/PRTR. APV and SOS acknowledge the DGAPA–PAPIIT grant IA103224. SOS acknowledges the support from Dr. Nadine Neumayer's Lise Meitner grant from the Max Planck Society. GG acknowledges support by Deutsche Forschungs-gemeinschaft (DFG, German Research Foundation) – project-IDs: eBer-22-59652 (GU 2240/1-1 "Galactic Archaeology with Convolutional Neural-Networks: Realising the potential of Gaia and 4MOST"). This work has made use of data from the European Space Agency (ESA) mission {\it Gaia} (\url{http://www.cosmos.esa.int/gaia}), processed by the {\it Gaia} Data Processing and Analysis Consortium (DPAC, \url{http://www.cosmos.esa.int/web/gaia/dpac/consortium}). Funding for the DPAC has been provided by national institutions, in particular the institutions participating in the {\it Gaia} Multilateral Agreement. Funding for the Sloan Digital Sky Survey V has been provided by the Alfred P. Sloan Foundation, the Heising-Simons Foundation, the National Science Foundation, and the Participating Institutions. SDSS acknowledges support and resources from the Center for High-Performance Computing at the University of Utah. SDSS telescopes are located at Apache Point Observatory, funded by the Astrophysical Research Consortium and operated by New Mexico State University, and at Las Campanas Observatory, operated by the Carnegie Institution for Science. The SDSS web site is \url{www.sdss.org}. SDSS is managed by the Astrophysical Research Consortium for the Participating Institutions of the SDSS Collaboration, including Caltech, The Carnegie Institution for Science, Chilean National Time Allocation Committee (CNTAC) ratified researchers, The Flatiron Institute, the Gotham Participation Group, Harvard University, Heidelberg University, The Johns Hopkins University, L'Ecole polytechnique f\'{e}d\'{e}rale de Lausanne (EPFL), Leibniz-Institut f\"{u}r Astrophysik Potsdam (AIP), Max-Planck-Institut f\"{u}r Astronomie (MPIA Heidelberg), Max-Planck-Institut f\"{u}r Extraterrestrische Physik (MPE), Nanjing University, National Astronomical Observatories of China (NAOC), New Mexico State University, The Ohio State University, Pennsylvania State University, Smithsonian Astrophysical Observatory, Space Telescope Science Institute (STScI), the Stellar Astrophysics Participation Group, Universidad Nacional Aut\'{o}noma de M\'{e}xico, University of Arizona, University of Colorado Boulder, University of Illinois at Urbana-Champaign, University of Toronto, University of Utah, University of Virginia, Yale University, and Yunnan University. This work made use of \texttt{overleaf.com}, and of the
following \textsc{python} packages (not previously mentioned):
\textsc{matplotlib} \citep{Hunter2007}, \textsc{scikit-learn} \citep{scikit-learn}, \textsc{numpy}
\citep{Harris2020}, \textsc{pandas}
\citep{mckinney-proc-scipy-2010}, \textsc{seaborn}
\citep{Waskom2021}. This work also benefited from
\textsc{topcat} \citep{Taylor2005}.
\end{acknowledgements}

\begin{appendix}

\section{The non-confined stars ($\mathrm{R_{apo}}$\,>\,5\,kpc)} \label{sec:nonConfined}
\begin{figure}[!ht]
    \centering
    \includegraphics[width=0.6\linewidth]{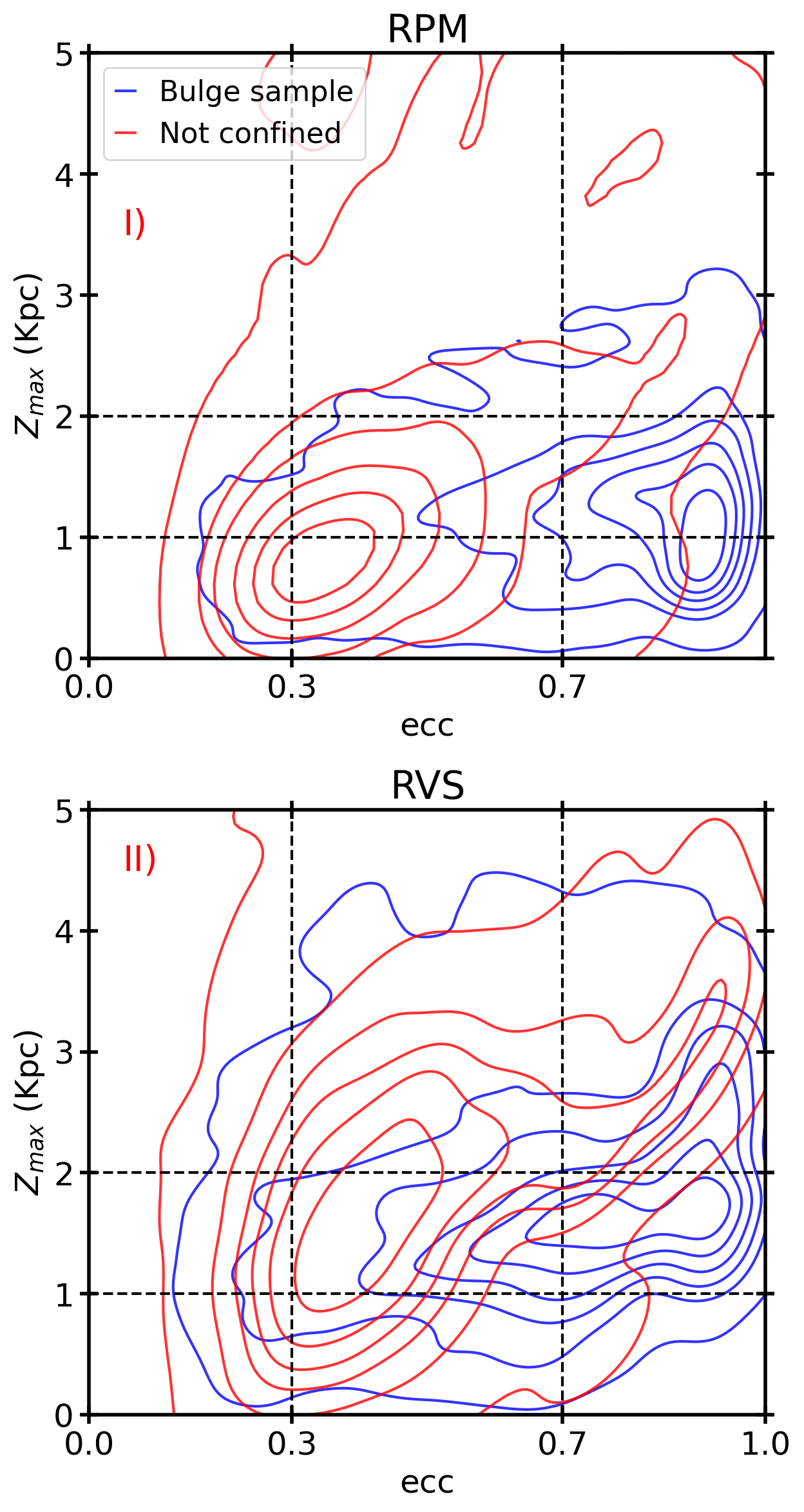}
    \caption{Distribution of bulge sample stars (blue) and non-confined stars (red) in the $\mathrm{Z_{max}}$--eccentricity plane for the RPM (top) and the RVS (bottom) stars.}
    \label{fig:temp}
\end{figure}

In Figure \ref{fig:temp} we present the distribution of the non-confined stars in the $\mathrm{Z_{max}}$--eccentricity plane. For the sample definition please refer to the Data section (see Sec. \ref{fig:data intro}) in the main text. For the RPM sample (red), we observe that the non-confined stars mostly have thin disc-like orbits with lower eccentricities ($\sim$0.3) and are confined to $\sim1$\ kpc from the galactic midplane. In \citepalias{Queiroz2021} the authors also characterise a significant thin disc-like population in the bulge region; however, we remind readers that in our current study we impose a limit of Apocenter < 5 kpc. For the RVS sample, we observe the non-confined stars also have lower eccentricities with a peak at ($\sim$0.4) and are confined to a $\mathrm{Z_{max}}$ of $\sim2$\ kpc. As expected, in the case of RVS these stars are dominated by the thick disc contribution. Among the non-confined stars, we also find a group that have very high eccentricities and $\mathrm{Z_{max}}$, which probably represent halo stars passing through the bulge region (see \ref{sec:bulge_halo} for further details). The non-confined sample will be studied in a forthcoming work. 

\section{Low probability bar stars}\label{sec:lowPbar}

\begin{figure}[!ht]
    \centering
    \includegraphics[width=0.99\linewidth]{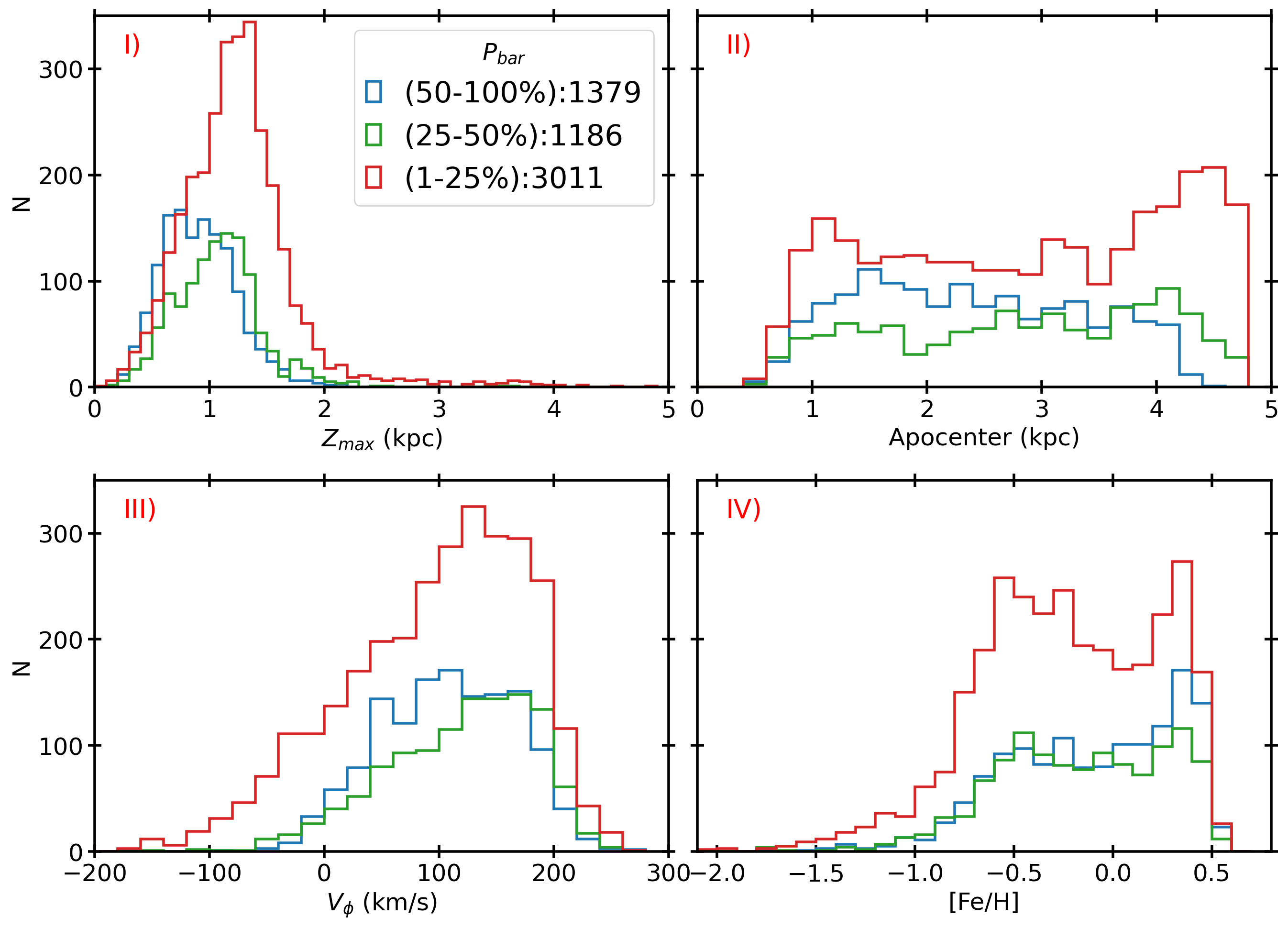}
    \caption{Distributions of $\mathrm{Z_{max}}$, $\mathrm{R_{apo}}$, and $\mathrm{V_{\phi}}$ and the metallicity in three bins of $\mathrm{P_{bar}}$.}
    \label{fig:lowBarP}
\end{figure}

In Figure \ref{fig:lowBarP} we present the orbital ($\mathrm{Z_{max}}$ and $\mathrm{R_{apo}}$), kinematic ($\mathrm{V_{\phi}}$), and chemical (MDF) properties of the stars according to their bar-support probability ($\mathrm{P_{bar}}$). Stars with 1 < $\mathrm{P_{bar}}$ < 50\%  stars have, in general, larger $\mathrm{Z_{max}}$ values than the higher stars with higher $\mathrm{P_{bar}}$. The low $\mathrm{P_{bar}}$ stars, comparatively, also have small apocentres (< 2 kpc) or $\mathrm{R_{apo}}$> 4 kpc. The velocity distribution also show most of the stars have 100<$V_{\phi}$<200 with a significant fraction of stars around $V_{\phi}\approx0$ \kms most probably belonging to the spheroidal bulge (see Sec. \ref{sec:SpB}). The MDFs show the wide range with a high fraction of metal-poor stars along with the peak at super-solar metallicity. Overall, the low $\mathrm{P_{bar}}$ appears to not only reflect contamination from both the disc and the spheroidal bulge but also include genuine bar stars that have been assigned low probabilities due to larger uncertainties in distance, radial velocities, or \textit{Gaia} proper motions. 

\section{Hints from globular clusters on the epoch of the spheroidal bulge formation.}

\begin{figure}[!ht]
    \centering
    \includegraphics[width=0.99\linewidth] {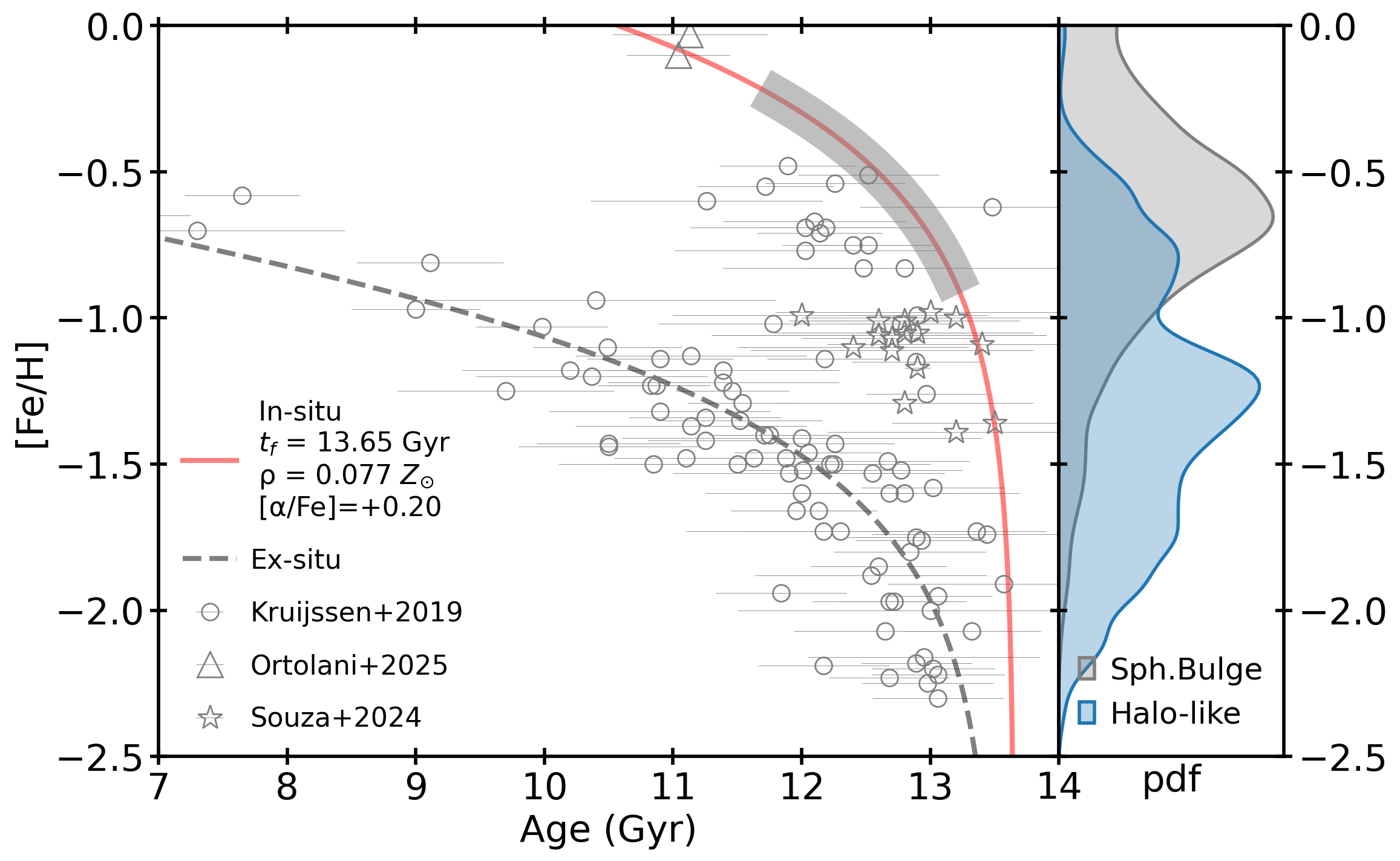}
    \caption{Age estimate for the spheroidal bulge of the MW using the AMR for the in situ (red line) GCs. The GCs are compiled from \citet{Kruijssen2019, Souza2024a, Ortolani2025}. For details on the method and the AMR fitted parameters, see \citet{Ortolani2025}. On the right, the MDFs for the spheroidal bulge (grey) and the halo-like stars (blue) from Fig. \ref{fig:afeh_5_components} are shown. The thick grey curve along the in situ AMR line corresponds to the central peak of the spheroidal bulge MDF and provides an estimate for the age. We note that, given the long tail of the spheroidal bulge, it also includes the bona fide metal-poor GCs, as shown in \citet{Souza2024a}.}
    \label{fig:age}
\end{figure}

\onecolumn
\section{Classification scheme}\label{sec:flowchart}
\begin{figure*}[!h]
    \centering
    \includegraphics[width=0.95\linewidth]{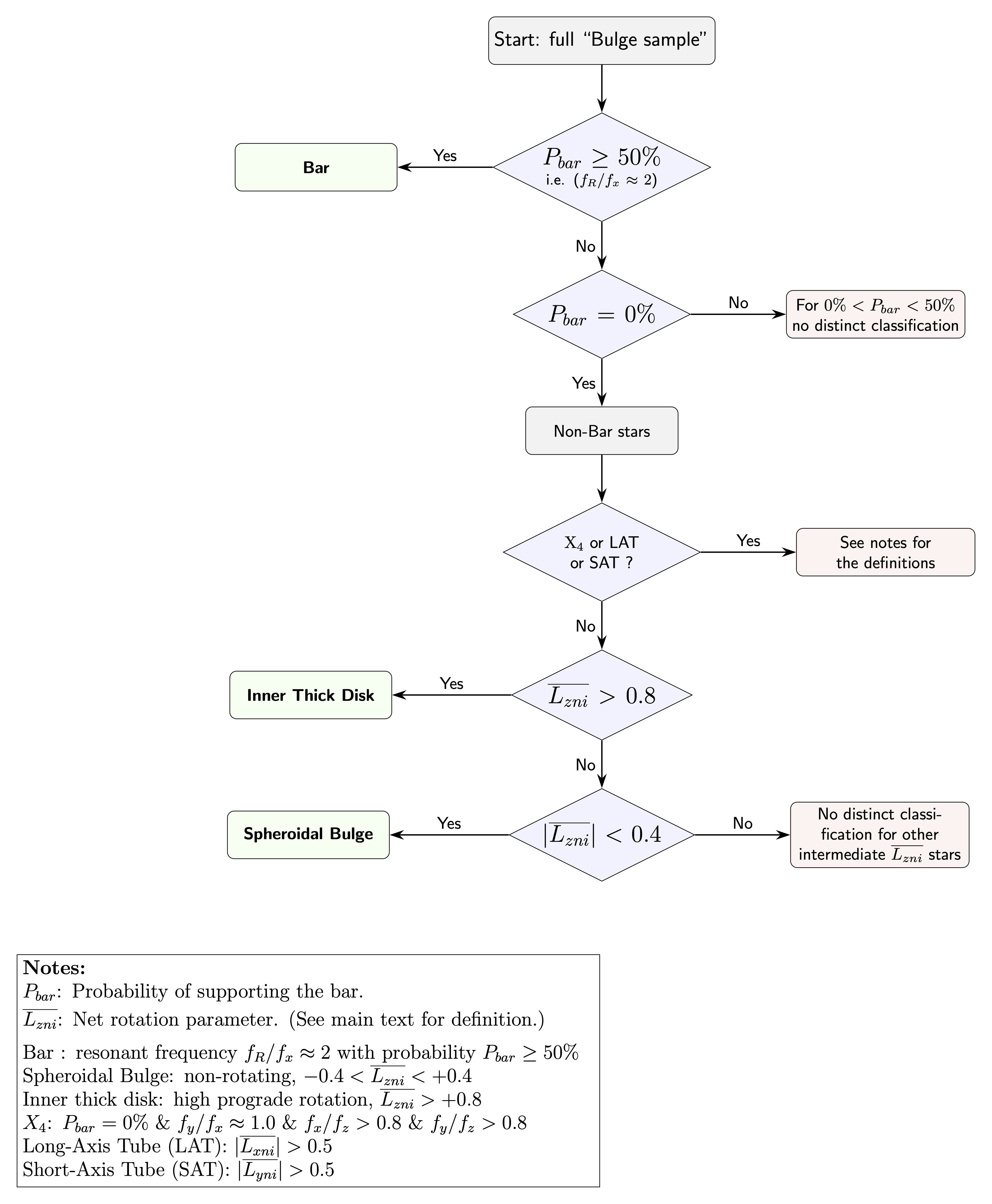}
    \caption{Flowchart of the classification scheme, summarizing the sequence of criteria used to separate the stars into the bar, spheroidal bulge, and inner-thick disc components. The classification scheme first relies on the robust identification of the bar-supporting stars with a probabilistic approach. Subsequently, the $X_4$ or the long- and short-axis tube orbits are extracted. Finally, the spheroidal bulge and the inner thick disc stars are classified based on the $\overline{L_{zni}}$, i.e. the net rotation parameter.}
    \label{fig:classification_flowchart}
\end{figure*}

\twocolumn
\section{Spatial projections}\label{sec:XYZ}

\begin{figure}[!h]
  \centering
  \includegraphics[width=0.99\linewidth]{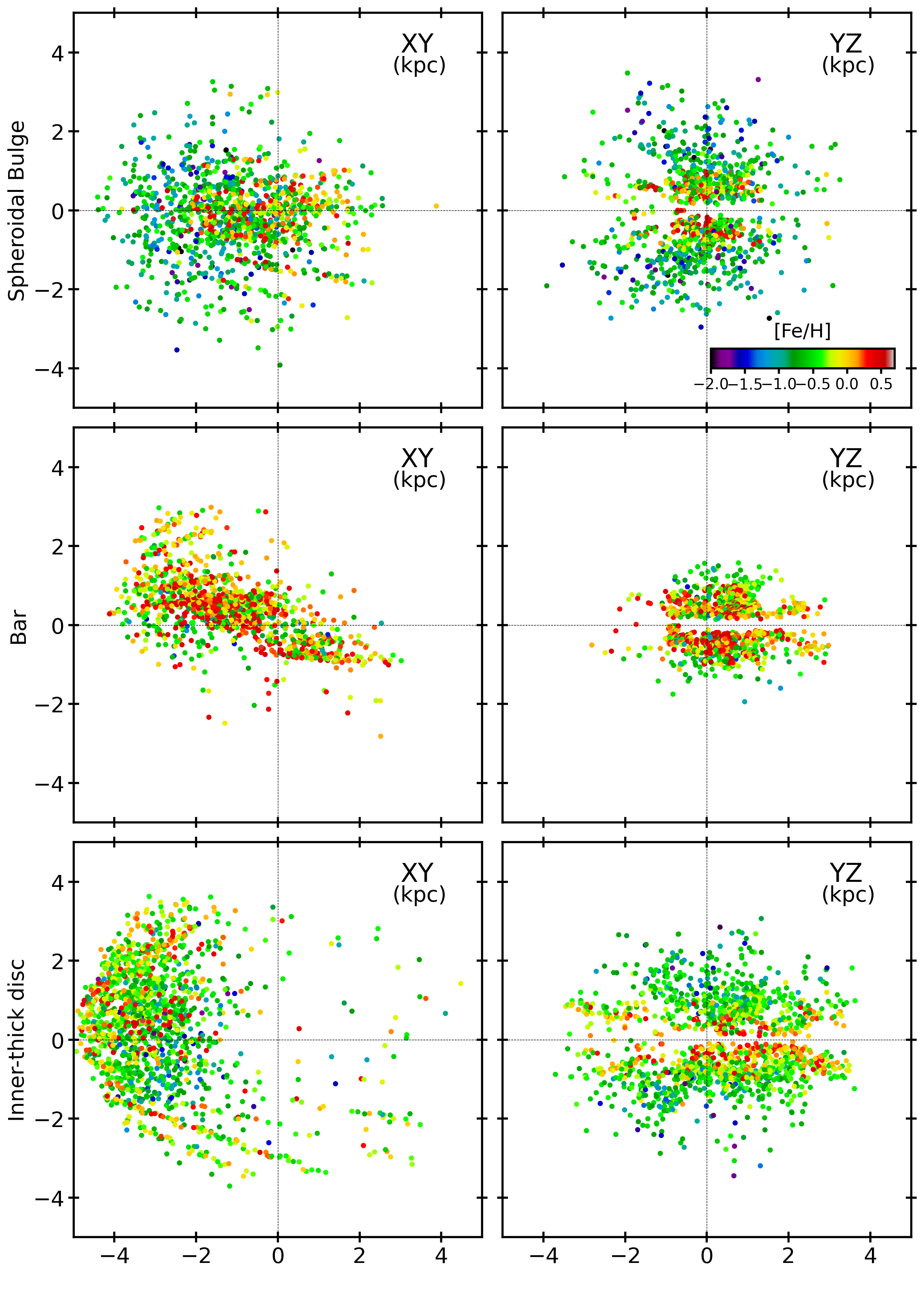}
  \caption{Spatial distributions colour-coded by [Fe/H], for the spheroidal bulge (top), bar (middle), and the inner thick disc (bottom).}
  \label{fig:XYZ}
\end{figure}

In Fig. \ref{fig:XYZ} we present the spatial distribution of the Spheroidal bulge (top), Bar (middle), and the Inner-thick disc (bottom) stars colour-coded by their metallicity. For all three components the spatial distributions of the current position of the stars clearly depict their geometric/orbital characteristics. The spheroidal bulge stars appear to be uniformly distributed, with metal-poor ([Fe/H]$<-$0.5) stars distributed across the volume, while the more metal-rich stars ([Fe/H]$>$0.0) remain centrally concentrated. The bar stars depict an elongated geometry with an angular orientation along the line of sight. The inner-thick disc stars, despite being mostly sampled towards the near side of the bulge region, clearly depict a “thick” disc with larger extent in both Y and Z directions compared to the bar. The metal-rich stars in the disc are located close to the mid plane, while the metal-poor stars extend out to $\sim$2 kpc.

\begin{figure}[!h]
  \centering
  \includegraphics[width=0.9\linewidth]{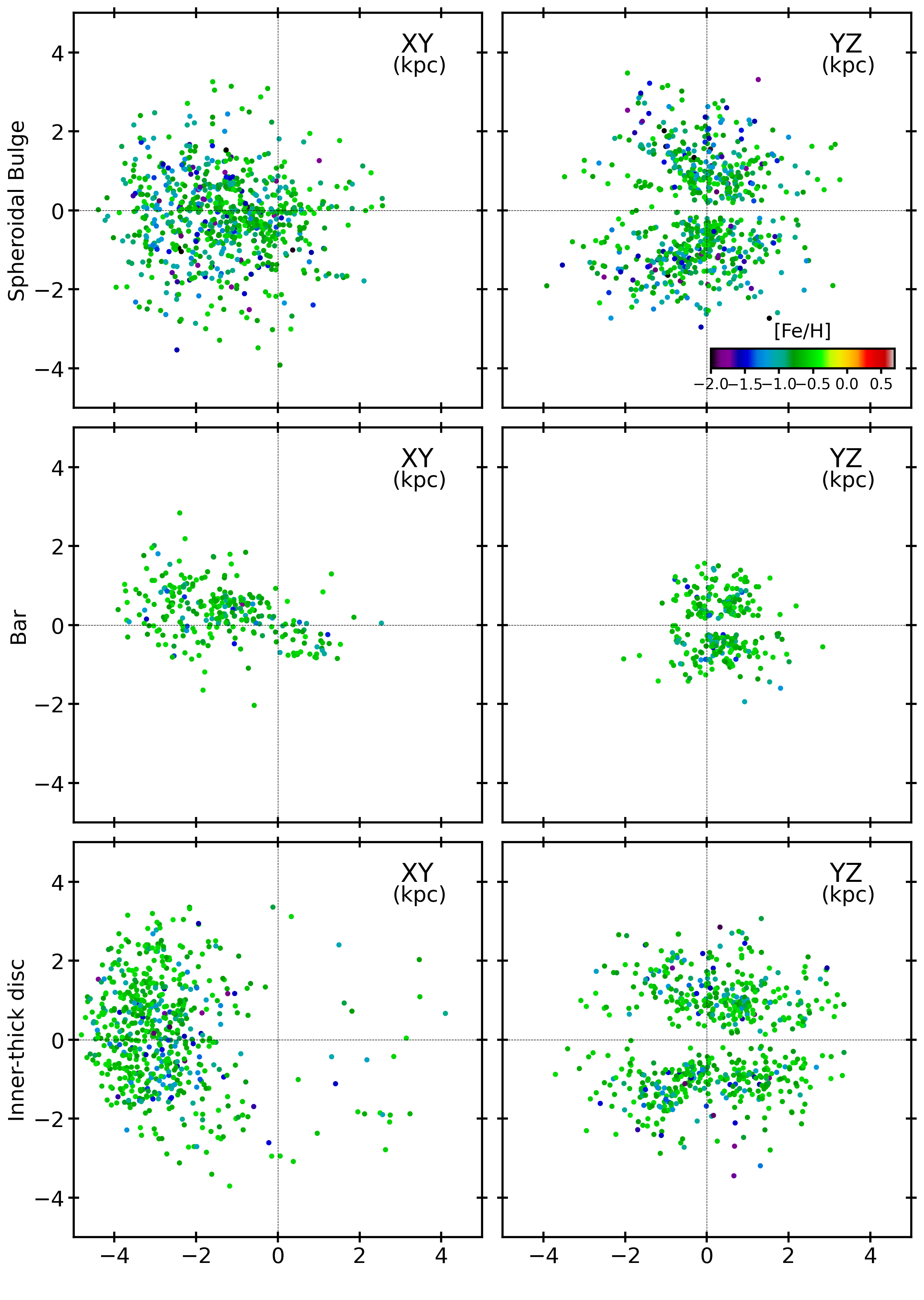}
  \caption{Same as Fig. \ref{fig:XYZ}, but for stars with [Fe/H]$<-$0.5.}
  \label{fig:XYZ_mp}
\end{figure}

\begin{figure}[!h]
  \centering
  \includegraphics[width=0.9\linewidth]{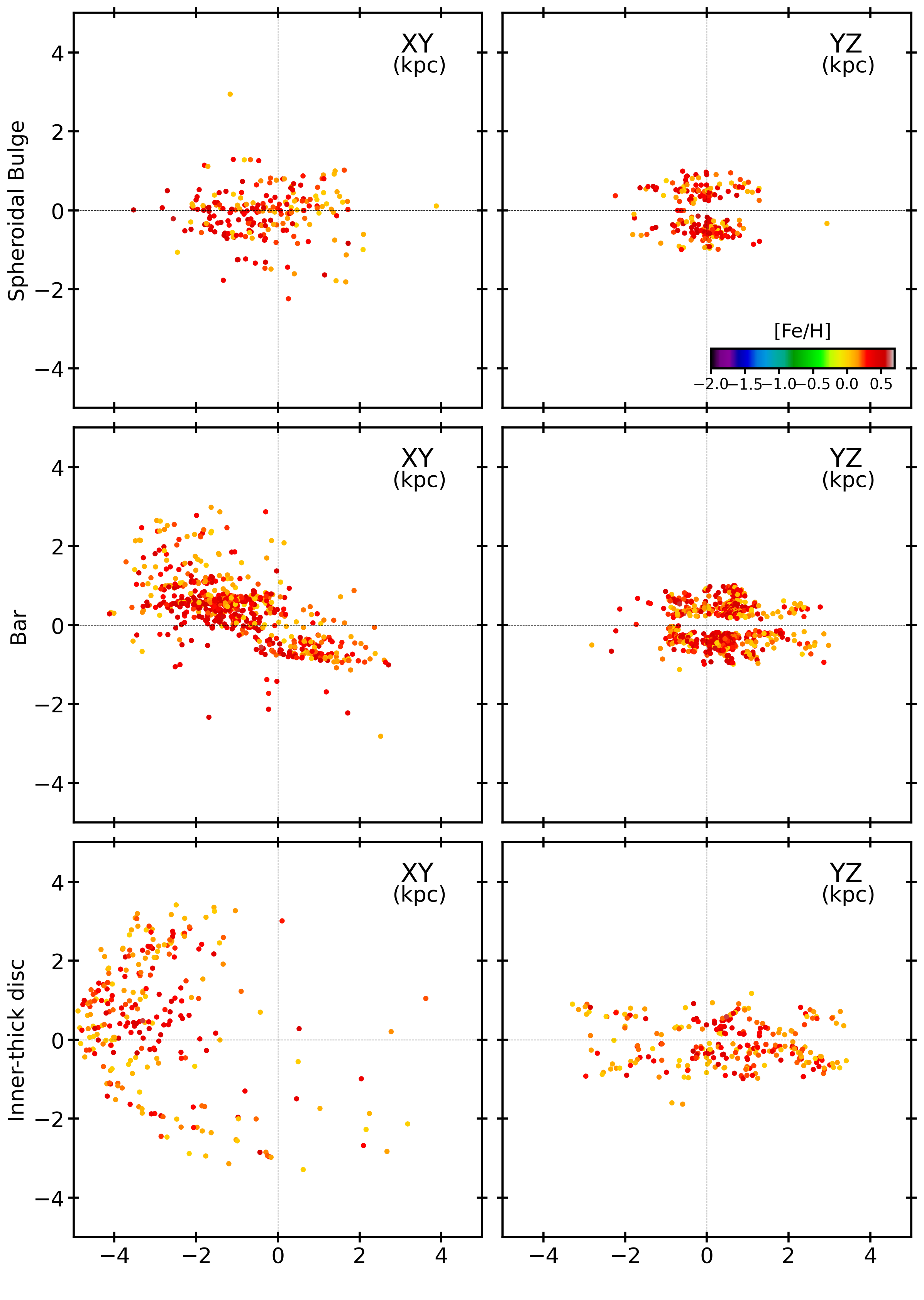}
  \caption{Same as Fig. \ref{fig:XYZ}, but for stars with [Fe/H]$>$0.0.}
  \label{fig:XYZ_mr}
\end{figure}

The spatial distribution plots (see also Fig. \ref{fig:XYZ_mp} and \ref{fig:XYZ_mr}) reveals a trend with the more metal-rich stars being centrally concentrated for the Spheroidal bulge and along the galactic mid plane and supporting the bar. However, the metal-poor stars, make significant contributions are present throughout the volume for for all three inner Galaxy components.

\end{appendix}
\end{document}